\documentclass{article}
\usepackage[utf8]{inputenc}

\usepackage{amsmath, amssymb, amsfonts,amsthm}
\usepackage{bm}
\usepackage{mathrsfs}
\usepackage{xr}
\usepackage{tablefootnote}
\usepackage{tabularx}
\usepackage{graphicx}
\usepackage{float}
\usepackage{color}
\usepackage{booktabs}
\usepackage{natbib}
\usepackage{color}
\usepackage{fullpage}
\usepackage{algorithm}
\usepackage{algpseudocode}

\newcommand{\E}{\mathbb{E}}
\newcommand{\pr}{\mathbb{P}}
\newcommand{\var}{\operatorname{Var}}
\newcommand{\cov}{\operatorname{Cov}}
\newcommand{\tr}{\operatorname{tr}}
\newcommand{\diag}{\operatorname{diag}}

\newcommand{\bms}{\bm{\Sigma}}
\newcommand{\bmu}{\bm{\mu}}

\newcommand{\bvarep}{\bm{\varepsilon}}
\newcommand{\bv}{\bm{v}}
\newcommand{\I}{\mathbf{I}}

\newcommand{\mA}{\mathbf{A}}
\newcommand{\mS}{\mathbf{S}}
\newcommand{\mO}{\mathbf{\Omega}}
\newcommand{\mM}{\mathbf{M}}

\def\X{\bm X}
\def\U{\bm U}
\def\x{\bm x}

\def\c{\rightarrow}
\def\cd{\xrightarrow{d}}

\allowdisplaybreaks[4]

\newtheorem{theorem}{Theorem}
\newtheorem{assu}{Assumption}
\newtheorem{lemma}{Lemma}
\newtheorem{corollary}{Corollary}
\title{High-Dimensional Hettmansperger-Randles Estimator and its Applications}
\author{Guowei Yan$^1$, Long Feng$^{1*}$ and Xiaoxu Zhang$^{2}$\thanks{Co-corresponding author and equally contributed.}\\
$^1$School of Statistics and Data Science, KLMDASR, LEBPS and LPMC, Nankai University\\
$^2$Academy of Mathematics and Systems Science, Chinese Academy of Science
}
\date{\today}

\begin{document}

\maketitle

\begin{abstract}
The classic Hettmansperger-Randles Estimator has found extensive use in robust statistical inference. However, it cannot be directly applied to high-dimensional data. In this paper, we propose a high-dimensional Hettmansperger-Randles Estimator for the location parameter and scatter matrix of elliptical distributions in high-dimensional scenarios. Subsequently, we apply these estimators to two prominent problems: the one-sample location test problem and quadratic discriminant analysis. We discover that the corresponding new methods exhibit high effectiveness across a broad range of distributions. Both simulation studies and real-data applications further illustrate the superiority of the newly proposed methods. 

\end{abstract}

\section{Introduction}

Estimating the mean vector and covariance matrix is a fundamental problem in statistics. In low-dimensional settings, when data follow a multivariate normal distribution, the sample mean and covariance matrix are known to be efficient \citep{hardle2007applied}. However, their performance often degrades under departures from normality, prompting the development of robust alternatives. For elliptical distributions, robust estimators such as the spatial median for location and Tyler’s scatter matrix for dispersion have gained prominence \citep{oja2010multivariate}. Moreover, \cite{hettmansperger2002practical} proposed a unified procedure for jointly estimating both location and scatter robustly.

With the rise of high-dimensional data in fields such as genomics and finance, new challenges have emerged. When the number of features (dimensions) becomes comparable to or even exceeds the number of observations (sample size), traditional estimators like the sample covariance matrix become singular and irreversible. In response,  extensive research has focused on high-dimensional covariance estimation, including thresholding, regularization, and shrinkage techniques \citep{bickel2008covariance, bickel2008regularized}; see \cite{fan2016overview} for a comprehensive review. Nevertheless, these methods largely rely on the sample covariance matrix, making them not robust for heavy-tailed distributions. 

To overcome this, robust methods under elliptical distributions, which encompass a broad range of heavy-tailed distributions such as the multivariate $t$-distribution and multivariate mixture normal distributions, have attracted increasing attention. Several studies have explored the properties of the sample spatial median and its use in high-dimensional sphericity testing and location parameter testing problems, including \cite{zou2014multivariate}, \cite{li2022asymptotic}, and \cite{cheng2023}. However, these estimators are not scalar-invariant. To address this issue, scalar-invariant spatial median estimators were proposed \citep{feng2016multivariate, feng2016, liu2024spatial} by extending the simultaneous estimation framework of \cite{hettmansperger2002practical}. However, these approaches are not affine-invariant with respect to scatter transformations, limiting their flexibility and applicability in practice. In parallel, robust scatter estimation has advanced through the study of spatial-sign covariance matrices, known for their affine equivariance. Recent works have developed linear shrinkage methods tailored for high-dimensional settings \citep{raninen2021linear, raninen2021bias, ollila2022regularized, ollila2024linear}, and sparse precision matrix estimation based on spatial-sign covariance \citep{lu2025}, extending earlier advances such as \cite{cai2011constrainedl1minimizationapproach} and \cite{YuanLin2007}. Nonetheless, most existing methods address location and scatter separately, lacking a coherent framework that unifies both aspects in high dimensions.

Motivated by these limitations, we propose a novel framework for robust high-dimensional inference. Specifically, we introduce the high-dimensional HR (Hettmansperger-Randles) estimator, which simultaneously estimates the spatial median and the scatter matrix in an affine-invariant manner. 
The resulting spatial median estimator is affine-invariant with respect to scatter transformations, thereby addressing certain limitations of previous approaches and enhancing robustness in high-dimensional inference under elliptical distributions. 
We demonstrate the practical utility of the HR estimator through its applications to two core problems in modern high-dimensional statistics: one-sample location testing and quadratic discriminant analysis.

For the high-dimensional one-sample location testing problem, substantial research has been conducted over the past two decades, leading to three main categories of testing procedures. The first category comprises sum-type tests, which aggregate statistics across all variables and are powerful against dense alternatives, as studied in \cite{bai1996effect, chen2010tests, wang2015high, ayyala2017mean, feng2015two, feng2016, feng2016multivariate, feng2021inverse}. The second category includes max-type tests, which focus on the maximum of individual statistics and excel under sparse alternatives, explored in works such as \cite{Zhong2013, CLX14, cheng2023, chang2017testing}. The third category consists of adaptive-type tests, which combine sum-type and max-type strategies to achieve robustness across diverse sparsity regimes, with important contributions from \cite{xu2016adaptive, he2021, feng2022a, Feng2022AsymptoticIO, chang2023testing, chen2024asymptotic, ma2024adaptive}. Comprehensive overviews are available in \cite{huang2022overview} and \cite{liu2024spatial}.

Since the seminal contribution of \cite{chernozhukov2013gaussian, chernozhukov2017central}, Gaussian approximations have become a cornerstone of high-dimensional statistical inference. Inspired by their theoretical framework, we first derive a Bahadur representation for the standardized spatial median estimator and establish its Gaussian approximation over a class of simple convex sets. This theoretical development provides a solid foundation for analyzing the limiting distributions of our proposed test statistics and facilitates the verification of the asymptotic independence between the max-type and sum-type statistics. Specifically, we introduce two types of test statistics based on the \(L_2\) and \(L_\infty\) norms of the corresponding standardized spatial-median estimator, which correspond to sum-type and max-type test procedures, respectively. We rigorously establish that these statistics are asymptotically independent. Leveraging this property, we develop a Cauchy combination test that integrates both sources of information. Notably, while the method of \cite{liu2024spatial} achieves robustness concerning the sparsity of $\bmu$, our approach attains robustness with respect to the sparsity of $\bms^{-1/2}\bmu$. Given that the true sparsity structure is generally unknown in practice, we further extend our procedure by combining four test statistics to achieve greater adaptability across various sparsity regimes. Simulation studies confirm that the proposed Cauchy combination tests consistently perform well across different distributional settings and sparsity levels, thereby demonstrating the practical robustness and wide applicability of our method in high-dimensional hypothesis testing.

We further apply the proposed HR estimator to improve quadratic discriminant analysis (QDA), which is a natural extension of linear discriminant analysis (LDA) \citep{Friedman1989, muirhead2009aspects}. When population parameters are known, QDA achieves optimal classification by comparing likelihood ratios. In low-dimensional settings, replacing population parameters with sample estimates generally preserves strong classification performance. However, in high-dimensional regimes, the singularity of the sample covariance matrix renders classical QDA infeasible. To address this, previous works have proposed sparse estimators for the covariance matrix \citep{Wu2019, Xiong2016} or its inverse \citep{cai2011constrainedl1minimizationapproach, YuanLin2007}. Nonetheless, these approaches fundamentally rely on the sample covariance matrix, which is highly sensitive to heavy-tailed distributions and thus undermines robustness. To overcome this limitation, we propose a robust QDA procedure by replacing the sample mean and precision matrix with the HR-based spatial median and scatter estimators. The resulting classifier retains high efficiency even under heavy-tailed distributions. We rigorously establish the asymptotic properties of the proposed method under mild moment conditions and demonstrate its superior performance through extensive simulations and real data applications. These results highlight the significant gains in robustness and classification accuracy offered by our framework in high-dimensional, non-Gaussian settings.

The remainder of this paper is structured as follows. In Section 2, we will introduce the high-dimensional HR estimator. Section 3 is dedicated to establishing the theoretical results of the corresponding estimators and applying them to the high-dimensional one-sample location problem. In Section 4, we will delve into high-dimensional quadratic discriminant analysis. The simulation studies are presented in Section 5, and the real data applications are shown in Section 6. Section 7 will cover some conclusions and suggest directions for further research. Finally, all the proofs of the theorems are provided in the appendix.   					

{\bf Notations:}  For $d$-dimensional $\boldsymbol{x}\in\mathbb{R}^d$, we use the notation $\|\boldsymbol{x}\|$ and $\|\boldsymbol{x}\|_\infty$ to denote its Euclidean norm and maximum-norm respectively. Denote $a_n\lesssim b_n$ if there exists constant $C$, $a_{n}\leq$ $Cb_{n}$ and $a_n\asymp b_n$ if both $a_n\lesssim b_n$ and $b_n\lesssim a_n$ hold. Let $\psi_\alpha(x)=\exp(x^\alpha)-1$ be a function defined on $[0,\infty)$ for $\alpha>0.$ Then the Orlicz norm $\|\cdot\|_{\psi_{\alpha}}$ of a random variable $ X$ is defined as $\| X\| _{\psi _{\alpha }}= \inf \left \{ t> 0, \mathbb{E}  \left \{ \psi _{\alpha }( | X| / t) \right \} \leqslant 1\right \} .$ Let $\tr(\cdot)$ be a trace of matrix, $\lambda_{\min}(\cdot)$ and $\lambda_{\max}(\cdot)$ be the minimum and maximum eigenvalue for symmetric matrix. For a matrix $\mA=(a_{ij})\in\mathbb{R}^{p\times q}$, we define the elementwise $\ell_{\infty}$ norm $\|\mA\|_{\infty} = \max_{1 \leq i \leq p, 1 \leq j \leq q} |a_{ij}|,$ the operation norm $\|\mA\|_{\text{op}} = \sup_{\|\x\| \leq 1} \|\mA \x\|,$ the matrix $\ell_1$ norm $\|\mA\|_{L_1} = \max_{1 \leq j \leq q} \sum_{i=1}^{p} |a_{ij}|,$ the Frobenius norm $\|\mA\|_F = \sqrt{\sum_{i,j} a_{ij}^2},$ and the elementwise $\ell_1$ norm $\|\mA\|_1 = \sum_{i=1}^{p} \sum_{j=1}^{q} |a_{ij}|.$ $\I_p$ represents a $p$-dimensional identity matrix, $\diag(v_1,\dots,v_p)$ represents the diagonal matrix with entries $\bv=(v_1,\dots,v_p)$. And $\mathbb{S}^{d-1}$ represents the unit sphere in $\mathbb{R}^d$.

\section{High-dimensional HR estimator}
Let $\X_1, \ldots, \X_n$ be an independently and 
identically distributed (i.i.d) observations from $p$-variate
elliptical distribution with density functions $|\bms|^{-1/2}g\{\|{\bf 
 \bms}^{-1/2}({\x}-\bmu)\|\}$,
where $\bmu$ is the location parameter, $\bms$ is a positive definite symmetric $p\times p$ scatter matrices, and $g(\cdot)$ is a scale function.
The spatial sign function is defined as $U({\bf x})=\|{\bf
x}\|^{-1}{\bf x}\mathbb{I}({\bf x}\neq {\bf 0})$. Denote
$\bvarep_i=\bms^{-1/2}(\X_i-\bmu)$. 
The modulus $\|\bvarep_i\|$ and the direction
$\U_i=U(\bvarep_i)$ are independent, and the direction 
vector $\U_i$ is uniformly distributed on $\mathbb{S}^{p-1}$. It is then well known that $\E(\U_i)=0$ and
$\cov(\U_i)=p^{-1}\I_p$. Without loss of generality, we assume that the scatter matrix satisfies $\tr(\bms)=p$.

The Hettmansperger-Randles (HR) (\cite{hettmansperger2002practical}) estimates for the location and scatter matrix are the values that simultaneously satisfy the following two equations:
$$
\frac{1}{n}\sum_{i=1}^n U\left(\hat{\bvarep}_i\right)=\mathbf{0} \quad \text { and } \quad \frac{p}{n}\sum_{i=1}^n\left\{U\left(\hat{\bvarep}_i\right)U\left(\hat{\bvarep}_i\right)^{\top}\right\}=\mathbf{I}_p,
$$
where $\hat{\bvarep}_i=\hat\bms^{-1/2}(\X_i-\hat\bmu)$.
These estimators are affine equivariant and provide robust estimates of both location and scatter. \cite{hettmansperger2002practical} further established their asymptotic distributions and showed that the HR estimators possess bounded influence functions and a positive breakdown point.

The HR estimator can be computed via the iterative procedure summarized in Algorithm \ref{alg:HR1}, which alternately updates the residuals, location, and scatter matrix until convergence. However, this procedure cannot be directly applied in high-dimensional settings, as the sample spatial-sign covariance matrix $\hat{\mS}\doteq n^{-1}\sum_{i=1}^n U(\hat{\bvarep}_i)U(\hat{\bvarep}_i)^{\top}$ becomes singular, making Step 3 inapplicable. To address this, \cite{feng2016multivariate} proposed a high-dimensional extension of Algorithm~\ref{alg:HR1} by restricting $\bms$ to be diagonal. While this allows estimation of the location and scale, it does not recover the full scatter structure.

\begin{algorithm}
\caption{HR estimator}
\label{alg:HR1}
\begin{algorithmic}[1]
\Procedure{Update}{$\X_1, \ldots, \X_n, \hat\bmu^{(k)}, \hat\bms^{(k)}, p$}
   \vspace{0.08cm} 
    \State \textbf{Step 1}: $\hat{\bvarep}_i^{(k)}\gets \{\hat\bms^{(k)}\}^{-1 / 2}\{\X_i - \hat\bmu^{(k)}\}$
    \vspace{0.08cm} 
    \State \textbf{Step 2}: $\hat\bmu^{(k+1)} \gets \hat\bmu^{(k)}+ \frac{\{\hat\bms^{(k)}\}^{1 / 2}\sum_{i=1}^nU\{\hat{\bvarep}_i^{(k)}\}}{\sum_{i=1}^n\|\hat{\bvarep}_i^{(k)}\|^{-1}}$
    \vspace{0.08cm}  
    \State \textbf{Step 3}: $\hat\bms^{(k+1)} \gets p \{\hat\bms^{(k)}\}^{1 / 2} \big[n^{-1}\sum_{i=1}^nU\{\hat{\bvarep}_i^{(k)}\}U\{\hat{\bvarep}_i^{(k)}\}^\top\big] \{\hat\bms^{(k)}\}^{1 / 2}$
    \vspace{0.08cm}  
    \State \textbf{Step 4}: Repeat Steps 1 - 3 until convergence.
    \vspace{0.08cm}  
    \State \Return $\hat\bmu, \hat\bms$
\EndProcedure
\end{algorithmic}
\end{algorithm}
The main challenge in Algorithm~\ref{alg:HR1} lies in the non-invertibility of $\hat{\mS}$. However, if a reliable initial estimator of the precision matrix ${\bf \Sigma}^{-1}$ is available, $\hat{\mS}$ can be approximated by the diagonal matrix $p^{-1}{\bf I}_p$, so it is unnecessary to consider all entries of $\hat{\mS}$ in Step 3. Motivated by \cite{bickel2008regularized}, we apply banding to  $\hat{\mS}$. For any matrix $\mathbf{M} = (m_{ij})_{p \times p}$ and any $0 \leq h < p$, define:
$$
\mathcal{B}_h(\mathbf{M})=\left\{m_{i j}\mathbb{I}(|i-j| \leq h)\right\}.
$$
In this paper, we set the bandwidth to $h=3$.
The remaining problem is to obtain good initial estimators for $\bmu$ and $\bms$. For location parameter, we use the spatial median:
\begin{align}
\hat\bmu_0=\arg\min_{\bmu\in \mathbb{R}^p} \sum_{i=1}^n\|\X_i-\bmu\|,
\end{align}
which has been shown to be consistent under high-dimensional settings (see \cite{zou2014multivariate}, \cite{feng2016multivariate}, \cite{feng2024spatial}).
For the precision matrix, we adopt the sparse graphical Lasso (SGLASSO) estimator proposed by \cite{lu2025}, which solves:
\begin{align}\label{sglasso}
\hat{\bf \Omega}_{0}=\arg\min_{\bf \Theta\succ 0} \tr(p{\bf \Theta}\hat{\mS}_0)-\log (|{\bf \Theta}|)+\lambda_n \|{\bf \Theta}\|_1,
\end{align}
where $\bf \Theta\succ 0$ indicates $\bf \Theta$ is positive define, $\hat{\mS}_0 = \frac{1}{n} \sum_{i=1}^n U(\X_i - \hat{\bmu}_0) U(\X_i - \hat{\bmu}_0)^\top$ is the sample spatial-sign covariance matrix based on the initial location estimate.

Combining these components, we propose a high-dimensional HR estimator for both the location and scatter matrix in Algorithm \ref{alg:HR2}.
\begin{algorithm}
\caption{High Dimensional HR estimator}
\label{alg:HR2}
\begin{algorithmic}[1]
\Procedure{Update}{$\bm X_1, \ldots, \bm X_n, \hat\bmu^{(k)}, \hat\bms^{(k)}, p$}
\vspace{0.08cm} 
\State Initial estimator $\hat\bmu^{(0)}=\hat{\bm \mu}_0, \hat\bms^{(0)}=\hat{{\bf \Omega}}_0^{-1}$
\vspace{0.08cm} 
    \State \textbf{Step 1}: $\hat{\bvarep}_i^{(k)} \gets \{\hat\bms^{(k)}\}^{-1 / 2}\{\bm X_i -\hat\bmu^{(k)}\}$
    \vspace{0.08cm} 
    \State \textbf{Step 2}: $\hat\bmu^{(k+1)} \gets \hat\bmu^{(k)} + \frac{\{\hat\bms^{(k)}\}^{1 / 2} n^{-1}\sum_{i=1}^n\mathbf{U}\{\hat{\bvarep}_i^{(k)}\}}{n^{-1}\sum_{i=1}^n\|\hat{\bvarep}_i^{(k)}\|^{-1}}$
    \vspace{0.08cm} 
    \State \textbf{Step 3}: $\hat\bms^{(k+1)} \gets p \{\hat\bms^{(k)}\}^{1 / 2} \mathcal{B}_h\left[n^{-1}\sum_{i=1}^nU\{\hat{\bvarep}_i^{(k)}\} U\{\hat{\bvarep}_i^{(k)}\}\top\right] \{\hat\bms^{(k)}\}^{1 / 2}$, $\hat\bms^{(k+1)} \gets \frac{p\hat\bms^{(k+1)} }{\tr\{\hat\bms^{(k+1)} \}}$
    \vspace{0.08cm} 
    \State \textbf{Step 4}: Repeat Steps 1 - 3 until convergence.
    \vspace{0.08cm} 
    \State \Return $ \hat\bmu, \hat\bms$
\EndProcedure
\end{algorithmic}
\end{algorithm}

In next section, we will show the consistency of the above high dimensional HR estimators $\hat{\bm \mu}$ and $\hat{\mathbf{\Sigma}}$ and apply it in one-sample location test problem.

\section{High-dimensional one-sample location problem}
In this section, we consider the following one-sample hypothesis testing problem:  
\[
H_0:\bm{\mu}=0 \quad \text{versus} \quad H_1:\bm{\mu}\neq0.
\]  
When the dimension \( p \) is fixed and the observations $\X_1,\dots,\X_n \overset{\text{i.i.d.}}{\sim} N(\bm 0, \bms_X)$, the classical Hotelling's \( T^2 \) test statistic commonly used:  
\[
T^2=n\bar{\bm X}^\top \hat{\mathbf{\Sigma}}_X^{-1} \bar{\bm X},
\]  
where \( \bar{\bm X} \) and \( \hat{\mathbf{\Sigma}}_X \) represent the sample mean vector and the sample covariance matrix, respectively. However, when the dimension \( p \) exceeds the sample size \( n \), the sample covariance matrix \( \hat{\mathbf{\Sigma}}_X \) becomes singular, rendering Hotelling’s \( T^2 \) test inapplicable.  

To overcome the limitation, \citet{Fan2015p} proposed replacing the sample covariance matrix with a sparse estimator $\bms_{\tau}$ and introduced the following test statistic:
$$
T_{FLY}=\frac{n\bar{\bm X}^\top \hat{\mathbf{\Sigma}}_{\tau}^{-1} \bar{\bm X}-p}{\sqrt{2p}}.
$$
Under the null, they showed that as $(n,p)\c\infty$, $T_{FLY} \cd N(0,1)$.
As a sum-type test, \( T_{FLY} \) is effective under dense alternatives but deteriorates in performance under sparse ones. To better handle sparse alternatives, \cite{chen2024asymptotic} introduced a max-type test statistic:  
$$
T_{CFL}=\max_{1\le i\le p} W_i^2-2\log p+\log\log p,
$$ 
where \( \bm W=(W_1,\cdots, W_p)^\top= n^{1/2} \hat{\mathbf{\Sigma}}_\tau^{-1/2} \bar{\bm X}\). They show that under the null, \( T_{CFL} \) follows a Gumbel distribution.   

Both \( T_{FLY} \) and \( T_{CFL} \) rely on the assumption of a multivariate normal distribution or an independent component model. However, these methods are not robust under heavy-tailed distributions, such as the multivariate \( t \)-distribution or mixture multivariate normal distributions. To address this limitation, we aim to develop new test procedures that are more effective under heavy-tailed distributions.  

For elliptical distributions, spatial-sign methods have been widely studied in the literature; see \cite{oja2010multivariate} for a comprehensive review. When the dimension \( p \) is fixed, the spatial-sign test with inner standardization, as proposed by \cite{Randles2000}, is given by:  
\[
Q^2 = np\bar{\U}^\top_T\bar{\U}_T,
\]  
where  
\[
\bar{\U}_T=\frac{1}{n}\sum_{i=1}^n \hat{\U}_{i,T}, \quad \hat{\U}_{i,T}=U(\hat{\mathbf{\Sigma}}_T^{-1/2}\bm X_i).
\]  
Here, \( \hat{\mathbf{\Sigma}}_T \) denotes Tyler's scatter matrix \citep{Tyler1987}, a widely used robust covariance estimator. However, in high-dimensional settings where \( p > n \), Tyler’s scatter matrix is no longer well-defined, making \( Q^2 \) inapplicable.  

To overcome this challenge, this paper proposes novel test procedures based on high-dimensional HR estimators, aiming to improve robustness and efficiency in heavy-tailed distributions.

First, we investigate some theoretical properties of the high dimensional HR estimator $\hat{\bm\mu}$. Let $\U_i=U(\bm \varepsilon_i)$, $r_i=\|\bm \varepsilon_i\|$, $\mathbf{S}=\E\{U(\bm X_i-\bm \mu)U(\bm X_i-\bm \mu)^\top\}$ and $\zeta_k=\E(r_i^{-k})$ for $i=1,\dots,n$.
\begin{assu}\label{assu1}
    {\bf (Assumptions on $\bm{\varepsilon}$)} 
    There exist constants $\underline{b},\bar{B}>0$ such that $\underline{b}\le \allowbreak\lim\sup_p \E\{(r_i/\sqrt{p})^{-k}\}\le \bar{B}$ for $k\in \{-1,1,2,3,4\}$. And $\zeta_1^{-1}r_1^{-1}$ is sub-Gaussian distributed, i.e. $\|\zeta_1^{-1}r_1^{-1}\|_{\psi_2}\le K_1 <\infty$.
\end{assu}
\begin{assu}\label{assu2}
    {\bf (Assumptions on $\bm{\Sigma}$)} $\exists\,\eta,h>0,$ s.t. $\eta<\lambda_{\min}(\bm{\Sigma})\leq\lambda_{\max}(\bm{\Sigma})<\eta^{-1}$ , $\tr(\bm{\Sigma})=p$ and $\|\mathbf{\Sigma}\|_{L_1}\le h $. The diagonal matrix of $\bm{\Sigma}$ is denoted as $\mathbf{D}=\mathrm{diag}\{d_1^2,d_2^2,\dots,d_p^2\}$, $\liminf_{p\rightarrow\infty}\min_{j=1,\dots,p}d_j>\underline{d}$ for some constant $\underline{d}>0$ and $\limsup_{p\rightarrow\infty}\max_{j=1,\dots,p}d_j<\bar{D}$ for some constant $\bar{D}>0$.
\end{assu}
\begin{assu}\label{assu3}
    {\bf (Assumptions on $\mathbf{\Omega}$)} $\exists\, T>0,\;0\leq q<1,\;s_{0}(p)>0,\text{ s.t. }\left(1\right)\|\mathbf{\Omega}\|_{L_{1}}\leq T,\;\left(2\right)\max_{1\leq i\leq p}\allowbreak\sum_{j=1}^{p}|\omega_{ij}|^{q}\leq s_{0}(p).$
\end{assu}
\begin{assu}\label{assu4}
{\bf (Assumptions on $\mathbf{S}$)} $\limsup_{p}\|\mathbf{S}\|_{op}<1-\psi<1$ for some positive constant $\psi.$
\end{assu}

Assumption \ref{assu1} aligns with Assumptions 1-2 in \cite{liu2024spatial}, which state that $\bm X_i$ follows an elliptically symmetric distribution and that $\zeta_k\asymp p^{-k/2}$. Assumptions \ref{assu2} and \ref{assu3} are standard conditions in high-dimensional data analysis, as seen in \cite{bickel2008regularized} and \cite{cai2011constrainedl1minimizationapproach}, ensuring the sparsity of the covariance and precision matrices. Assumption \ref{assu4} corresponds to Assumption (A2) in \cite{feng2024spatial}, guaranteeing the consistency of the initial sample spatial median.

The following lemma provides a Bahadur-type representation of the standardized estimator $\hat{\bm\mu}$, which lays the foundation for the Gaussian approximation in Lemma~\ref{lemma2}. This approximation result, in turn, is instrumental for establishing the main theorems of this paper, including Theorems~\ref{thm1}, \ref{thm3}, and \ref{thm6}.
\begin{lemma}\label{lemma1}
    (Bahadur representation) Assume Assumptions \ref{assu1}-\ref{assu4} hold and $\log p=o(n^{1/3})$, There exist constants $C_{\eta,T}$ and $C$, such that if we pick 
    $$
    \lambda_{n}=T\{\sqrt{2}C(8+\eta^{2}C_{\eta,T})\eta^{-2}n^{-1/2}\log^{1/2}p+p^{-1/2}C_{\eta,T}\},
    $$ 
    and $\lambda_n^{1-q}s_0(p)(\log p)^{1/2}=o(1)$ then
    $$
    n^{1/2}\hat{\mathbf \Omega}^{1/2}(\hat{\boldsymbol \mu}-\boldsymbol \mu)=n^{-1/2}\zeta_1^{-1}\sum_{i=1}^n \boldsymbol U_i+C_n,
    $$
    where
$$
\begin{aligned}
\Vert C_n\Vert_\infty=O_{p}\{n^{-1/4}\log^{1/2}(np)+n^{-(1-q)/2}(\log p)^{(1-q)/2}\log^{1/2}(np)s_0(p)+p^{-(1-q)/2}\log^{1/2}(np)s_0(p)\}.
\end{aligned}
$$

\end{lemma}
Let $\mathcal{A}^{\operatorname{si}}$ be the class of simple convex sets (\cite{chernozhukov2017central}) in $\mathbb{R}^p.$ Based on the Bahadur representation of $\hat{\bm \mu}$, we acquire the following Gaussian approximation of $\hat{\mathbf{\Omega}}^{1/2}\left(\hat{\boldsymbol{\mu}}-\boldsymbol{\mu}\right)$ in $\mathcal{A}^\mathrm{si}$, where $\hat{\mathbf{\Omega}}=\hat{\bms}^{-1}$. 
\begin{lemma}\label{lemma2}
(Gaussian approximation) Assume Lemma \ref{lemma1} holds. If $\log p=o\left(n^{1 / 5}\right)$ ,then
\begin{align*}
\rho_n\left(\mathcal{A}^{\mathrm{si}}\right)=\sup _{A \in \mathcal{A}^{\mathrm{si}}}\left|\mathbb{P}\left\{n^{1 / 2}\hat{\mathbf \Omega}^{1/2}\left(\hat{\boldsymbol{\mu}}-\boldsymbol{\mu}\right) \in A\right\}-\mathbb{P}(\boldsymbol Z \in A)\right| \rightarrow 0,
\end{align*}
as $n \rightarrow \infty$, where $\boldsymbol Z \sim N\left(0, p^{-1}\zeta_1^{-2} \I_p\right)$.
\end{lemma}

Consequently, we derive the following corollary, which establishes the limiting distribution of the $L_2$- and $L_\infty$-norms of \( n^{1 / 2} \hat{\mathbf{\Omega}}^{1/2} (\hat{\boldsymbol{\mu}} - \boldsymbol{\mu}) \).

\begin{corollary}\label{cor1}
    Assume Assumptions Lemma \ref{lemma2} holds. Set $A$ to $\{\x\big|\|\x\|_\infty\le t\}$, $\{\x\big|\|\x\|\le t\}$ and $\{\x\big|\|\x\|_\infty\le t_1,\|\x\|\le t_2\}$  we have
$$\begin{aligned}
     \tilde\rho_{n,\infty}&=\sup _{t \in \mathbb{R}}\left|\mathbb{P}\left(n^{1 / 2}\Vert\hat{\mathbf \Omega}^{1/2}(\hat{\boldsymbol{\mu}}-\boldsymbol{\mu})\Vert_{\infty} \leqslant t\right)-\mathbb{P}\left(\Vert \boldsymbol Z\Vert_{\infty} \leqslant t\right)\right| \rightarrow 0 ,\\
     \tilde\rho_{n,2}&=\sup _{t\in \mathbb{R}}\left|\mathbb{P}\left(n^{1 / 2}\Vert\hat{\mathbf \Omega}^{1/2}(\hat{\boldsymbol{\mu}}-\boldsymbol{\mu})\Vert \leqslant t\right)-\mathbb{P}\left(\Vert \boldsymbol Z\Vert \leqslant t\right)\right| \rightarrow 0 ,\\
     \tilde\rho_{n,comb}&=\sup _{t_1,t_2 \in \mathbb{R}}\left|\mathbb{P}\left(n^{1 / 2}\Vert\hat{\mathbf \Omega}^{1/2}(\hat{\boldsymbol{\mu}}-\boldsymbol{\mu})\Vert_{\infty} \leqslant t_1, n^{1 / 2}\Vert\hat{\mathbf \Omega}^{1/2}(\hat{\boldsymbol{\mu}}-\boldsymbol{\mu})\Vert \leqslant t_2\right)-\mathbb{P}\left(\Vert \boldsymbol Z\Vert_{\infty} \leqslant t_1, \Vert \boldsymbol Z\Vert \leqslant t_2\right)\right| \rightarrow 0 ,
\end{aligned}$$
as $n\to \infty$, where $\boldsymbol Z \sim N\left(0, \zeta_1^{-2}p^{-1}\I_p \right)$.
\end{corollary}
We know that $\{\x\big|\|\x\|_\infty\le t\}$ and $\{\x\big|\|\x\|\le t\}$ are simple convex sets. The third equation holds because the intersection of a finite number of simple convex sets is still simply convex. 

From \cite{CLX14}, we can see that $p\zeta_1^2\max _{1 \leq i \leq p} Z_i^2-2 \log p+\log \log p$ converges to a Gumbel distribution with the cumulative distribution function (cdf) $F(x)=\exp (-\frac{1}{\sqrt{\pi}} e^{-x / 2})$ as $p \rightarrow \infty$. In combining with the Corollary \ref{cor1} we can conclude that,
\begin{equation}\label{eq:Tmax0}
    \mathbb P\left(n\left\|\hat{\mathbf \Omega}^{1 / 2}(\hat{\boldsymbol{\mu}}-\boldsymbol{\mu})\right\|_{\infty}^2 p \zeta_1^2-2 \log p+\log \log p \leq x\right) \rightarrow \exp \left(-\frac{1}{\sqrt{\pi}} e^{-x / 2}\right).
\end{equation}
Next we replace $E\left(r^{-1}\right)$ with its estimators. We denote $\Tilde{r}_i=\Vert\hat{\mathbf{\Omega}}^{-1 / 2}(\mathbf{X}_i-\hat{\boldsymbol{\mu}})\Vert$. Then the estimator is defined as $\hat \zeta_{1}:=n^{-1}\sum_{i=1}^n \Tilde{r}_i^{-1}$, and the proof of consistency is shown in Lemma \ref{lemma6}. Next, we propose the following max-type test statistic 
$$
T_{MAX}=n\left\|\hat{\mathbf{\Omega}}^{1 / 2} \hat{\boldsymbol{\mu}}\right\|_{\infty}^2  \hat{\zeta}_{1}^2p-2 \log p+\log \log p.
$$
Obviously, the new proposed test statistic $T_{MAX}$ is affine invariant.

\begin{theorem}\label{thm1}
Suppose the Assumptions \ref{assu1}-\ref{assu4} hold. Under the null hypothesis, as $(n,p)\to \infty$, we have
$$
\mathbb P\left(T_{M A X} \leq x\right) \rightarrow \exp \left(-\frac{1}{\sqrt{\pi}} e^{-x / 2}\right).
$$
\end{theorem}
According to Theorem \ref{thm1}, $H_0$ will be rejected when our proposed statistic $T_{MAX}$ is larger than the $(1-\alpha)$ quantile $q_{1-\alpha}=-\log\pi -2\log\log(1-\alpha)^{-1}$ of the Gumbel distribution $F(x)$. Next, we give the following theorem to demonstrate the consistency of our test.
\begin{theorem}\label{thm2}
    Suppose the conditions assumed in Theorem \ref{thm1} holds, for any given $\alpha\in (0,1)$, if $\|\mathbf{\Omega}^{1/2}\bm{\mu}\|_\infty\ge \widetilde{C}n^{-1/2}(\log p+q_{1-\alpha})^{1/2}$, for some large enough constant $\widetilde{C}$, then
    \begin{equation*}
        \mathbb{P}(T_{MAX}>q_{1-\alpha}|H_1)\to 1,
    \end{equation*}
    as $(n,p)\to \infty$.
\end{theorem}

Next, we consider a special case of alternative hypothesis: 
\begin{align}
\mathbf{\Omega}^{1/2}\bm{\mu}=(\mu_1,0,\cdots,0),\mu_1>0,
\end{align}
which means there are only one variable with nonzero mean. Similar to the calculation in \cite{liu2024spatial}, we can easily show the power function of new proposed $T_{MAX}$ test is
\begin{align*}
\beta_{MAX}(\bm \mu)\in \left(\Phi\left\{-x_\alpha^{1/2}+(n p)^{1 / 2} d_1^{-1} \mu_1 \zeta_1\right\}, \Phi\left\{-x_\alpha^{1/2}+(n p)^{1 / 2} d_1^{-1} \mu_1 \zeta_1\right\}+\alpha\right),
\end{align*}
where $x_\alpha=2 \log p-\log \log p+q_{1-\alpha}$.
Similarly, the power function of \cite{chen2024asymptotic}'s test is
$$
\beta_{CFL}(\boldsymbol{\theta}) \in\left(\Phi\left(-\sqrt{x_\alpha}+n^{1 / 2} \varsigma_1^{-1} \mu_1\right), \Phi\left(-\sqrt{x_\alpha}+n^{1 / 2} \varsigma_1^{-1} \mu_1\right)+\alpha\right),
$$
where $\varsigma_i^2$ is the variance of $X_{k i}, i=1, \cdots, p$. Thus, the asymptotic relative efficiency of $T_{M A X}$ with respective to \cite{CLX14}'s test could be approximated as 
$$\operatorname{ARE}\left(T_{M A X}, T_{CFL}\right)=\left\{\E\left(r_i^{-1}\right)\right\}^2 \E\left(r_i^2\right)\ge 1,$$ 
which has been widely observed by many literature to show the efficiency of the spatial sign based methods with respect to the least-square based methods, such as \cite{feng2016}, \cite{feng2016multivariate}, \cite{liu2024spatial}. If $\boldsymbol{X}_i$ are generated from standard multivariate $t$-distribution with $\nu$ degrees of freedom $(\nu>2)$,
$$
\operatorname{ARE}\left(T_{M A X}, T_{CFL}\right)=\frac{2}{\nu-2}\left[\frac{\Gamma\{(\nu+1) / 2\}}{\Gamma(\nu / 2)}\right]^2.
$$
For different $\nu=3,4,5,6$, the above ARE are $2.54,1.76,1.51,1.38$, respectively. Under the multivariate normal distribution $(\nu=\infty)$, our $T_{M A X}$ test is the same powerful as \cite{chen2024asymptotic}'s test. However, our $T_{M A X}$ test is much more powerful under the heavy-tailed distributions.

Similarly, we can see that \((2p)^{-1/2}\left(\sum\limits_{i=1}^p p\zeta_1^2 Z_i^2-p\right)\) converges to a standard Gaussian distribution with cdf $\Phi(x)$. In combining with the Corollary \ref{cor1} we can conclude that, 
\begin{equation}\label{eq:Tsum0}
    \mathbb P\left\{(2p)^{-1/2}\left(n\left\|\hat{\mathbf \Omega}^{1 / 2}(\hat{\boldsymbol{\mu}}-\boldsymbol{\mu})\right\|^2 p \zeta_1^2- p\right)\leq x\right\}\rightarrow \Phi(x).
\end{equation}
Then we propose the sum-type test statistic
\begin{align}
T_{SUM}=\frac{\sqrt{2p}}{2}\left(n\hat{\zeta}_1^{2}\hat{\bm \mu}^\top \mathbf{\hat{\Omega}} \hat{\bm \mu}-1\right).
\end{align}

\begin{theorem}\label{thm3}
    Suppose the Assumptions \ref{assu1}-\ref{assu4} hold. Under $H_0:\bm{\mu}=\bm{0}$, as $(n,p)\to \infty$, we have
    \begin{align*}
    T_{SUM}\xrightarrow{d} N(0,1),
    \end{align*}
    where $\cd$ stands for convergence in distribution.
    Furthermore, under $H_1:\bm{\mu}^\top\mathbf{\Omega}\bm{\mu}=o(pn^{-1})$, as $(n,p)\to \infty$, we have
    \begin{align*}
    T_{SUM}-\frac{\sqrt{2}}{2}np^{1/2}\zeta_1^2\bm{\mu}^\top\mathbf{\Omega}\bm{\mu}\xrightarrow{d} N(0,1).
    \end{align*}
\end{theorem}
By Theorem \ref{thm3}, the asymptotic power function of $T_{SUM}$ is
\begin{equation*}
    \beta_{SUM}(\bm{\mu})=\Phi\left(-z_{1-\alpha}+\frac{\sqrt{2}}{2}np^{1/2}\zeta_1^2\bm{\mu}^\top\mathbf{\Omega}\bm{\mu}\right).
\end{equation*}

After some simply calculations, we can obtain the power function of $T_{FLY}$ is
\begin{equation*}
    \beta_{FLY}(\bm{\mu})=\Phi\left(-z_{1-\alpha}+\frac{1}{\sqrt{2p}}n\bm{\mu}^\top\mathbf{\Sigma}_s^{-1}\bm{\mu}\right).
\end{equation*}
where $\mathbf{\Sigma}_s=\E(\bm X_i\bm X_i^\top)$ is the covariance matrix and $\mathbf{\Sigma}_s=p^{-1}\E(r_i^{2})\mathbf{\Sigma}$. So the asymptotic relative efficiency (ARE) of $T_{SUM}$ with repective to $T_{FLY}$ is
\begin{align*}
\textrm{ARE}(T_{SUM}, T_{FLY})=\{\E(r_i^{-1})\}^2\E(r_i^2)\ge 1,
\end{align*}
which is the same as $\textrm{ARE}(T_{MAX},T_{CFL})$.

However, when the dimension gets larger, there would be a non-negligible bias term in $T_{SUM}$ and $T_{MAX}$. To use the above sum-type and max-type test procedure, we adopt the bootstrap method to calculate the bias term. We simply generate $n$ samples $z_1,\cdots,z_n$ from the multivariate normal distribution $N(0,\hat{\mathbf{\Omega}}^{-1})$. Then, based on the random sample $z_1,\cdots,z_n$, we calculate the sum-type test statistic $T_{SUM}^*$  and max-type test statistic $T_{MAX}^*$. Repeat this procedure $M$ times, we could get a bootstrap sample of $T_{SUM}$ and $T_{MAX}$. Then, we calculate the sample mean and the sample variance of these bootstrap samples, denoted as $\mu_S^*$ and $\sigma_S^{2*}$ for $T_{SUM}^*$ and $\mu_M^*$ and $\sigma_M^{2*}$ for $T_{MAX}^*$. The corresponding $p$-values of $T_{SUM}$ and $T_{MAX}$ are
\begin{align*}
p_{SUM}&=1-\Phi\{(T_{SUM}-\mu^*)/\sigma^*\},\\
p_{MAX}&=1-F\left\{\sigma_0(T_{M A X}-\mu_M^*)/\sigma_M^*+\mu_0\right\},
\end{align*}
where $\mu_0=-\log(\pi)+2\gamma$ and $\sigma_0^2=3^{-1}2\pi^2$ are the expectation and variance of the Gumbel distribution $F(x)$. Here $\gamma$ is the Euler constant. Because we only need the mean and variance of the bootstrap samples, so the bootstrap size $M=50$ is always enough for controlling the empirical sizes.

It is well known that sum-type and max-type test statistics are powerful against dense and sparse alternatives, respectively. To adapt to unknown alternative structures, Cauchy combination procedures are widely used, as they remain valid even under dependent $p$-values. In our setting, we further establish the asymptotic independence between $T_{SUM}$ and $T_{MAX}$, highlighting their complementary detection power.

\begin{theorem} \label{thm5}
    Under Assumptions \ref{assu1}-\ref{assu4}, if $\|\bm{\mu}\|_\infty=o(n^{-1/2})$ and $\|\bm{\mu}\|=o(p^{1/4}n^{-1/2})$, as $n,p \to \infty$, $T_{MAX}$ and $T_{SUM}$ are asymptotic independent.
\end{theorem}

Based on Theorem \ref{thm5}, we construct a Cauchy combination test procedure as follow:
$$
\begin{aligned}
p_{C C1} & =1-G\left[0.5 \tan \left\{\left(0.5-p_{M A X}\right) \pi\right\}+0.5 \tan \left\{\left(0.5-p_{S U M}\right) \pi\right\}\right],
\end{aligned}
$$
where $G(\cdot)$ is the CDF of the the standard Cauchy distribution. If the final $p$-value is less than some pre-specified significant level $\alpha \in(0,1)$, then we reject $H_0$.

For comparison, we also consider the test procedures proposed by \cite{feng2016} and \cite{liu2024spatial}, which are designed for the sparsity structure of the location parameter $\bmu$. 
\begin{align*}
T_{S U M2}=&\frac{2}{n(n-1)} \sum \sum_{i<j} U\left(\tilde{\mathbf{D}}_{i j}^{-1 / 2} \boldsymbol{X}_i\right)^T U\left(\tilde{\mathbf{D}}_{i j}^{-1 / 2} \boldsymbol{X}_j\right),\\
T_{M A X2}=&n\hat{\zeta}_1^2 p\left\|\tilde{\mathbf{D}}^{-1 / 2} \tilde{\boldsymbol{\mu}}\right\|_{\infty}^2  \left(1-n^{-1 / 2}\right),\\
p_{C C2} =&1-G\left[0.5 \tan \left\{\left(0.5-p_{M A X2}\right) \pi\right\}+0.5 \tan \left\{\left(0.5-p_{S U M2}\right) \pi\right\}\right],
\end{align*}
where $p_{MAX2}$ and $p_{SUM2}$ are the p-values of $T_{MAX2}$ and $T_{SUM2}$, respectively. Here $\tilde{\bm \mu}$ and $\tilde{\mathbf{D}}_{ij}$ are the estimator of spatial-median and diagonal matrix of $\bms$ by the following algorithm:
\begin{itemize}
\item[(i)] $\tilde{\bvarep}_i \leftarrow \tilde{\mathbf{D}}^{-1 / 2}\left(\mathbf{X}_i-\tilde{\bm \mu}\right), \quad i=1, \cdots, n$;
\item[(ii)] $\tilde{\bm \mu} \leftarrow \tilde{\bm \mu}+\frac{\tilde{\mathbf{D}}^{1 / 2} \sum_{j=1}^n U\left(\tilde{\bvarep}_i\right)}{\sum_{j=1}^n\left\|\tilde{\bvarep}_i\right\|^{-1}}$;
\item[(iii)] $\tilde{\mathbf{D}} \leftarrow p \tilde{\mathbf{D}}^{1 / 2} \operatorname{diag}\left\{n^{-1} \sum_{j=1}^n U\left(\tilde{\bvarep}_i\right) U\left(\tilde{\bvarep}_i\right)^{\top}\right\} \tilde{\mathbf{D}}^{1 / 2}$.
\end{itemize}

Next, we demonstrate that, under mild regularity conditions, the sum-type test statistic \( T_{\text{SUM2}} \) is asymptotically independent of the max-type test statistic \( T_{\text{MAX}} \). Furthermore, the max-type test statistic \( T_{\text{MAX2}} \) is also asymptotically independent of the sum-type test statistic \( T_{\text{SUM1}} \).

\begin{theorem}\label{thm6}
  Under Assumptions \ref{assu1}-\ref{assu4}, if $\|\bm{\mu}\|_\infty=o(n^{-1/2})$ and $\|\bm{\mu}\|=o(p^{1/4}n^{-1/2})$, as $n,p \to \infty$, and Theorem 7 in \cite{liu2024spatial} holds, $T_{SUM2}/\sigma_n$ is asymptotically independent with $T_{MAX}$, $T_{SUM}$ is asymptotically independent with $T_{MAX2}-2\log p+\log\log p$. 
\end{theorem}

In practice, we could not know the sparsity level of the alternative, either $\mathbf{\Omega}^{1/2}\bm \mu$ or $\bm \mu$, so we suggest to use Cauchy combination test to combine all the four test procedures as follow:
\begin{align}
T_{CC3}=&1-G\Big[\frac{1}{4} \tan \left\{\left(0.5-p_{M A X}\right) \pi\right\}+\frac{1}{4}\tan \left\{\left(0.5-p_{S U M}\right) \pi\right\} \nonumber\\
&+\frac{1}{4} \tan \left\{\left(0.5-p_{M A X2}\right) \pi\right\}+\frac{1}{4}\tan \left\{\left(0.5-p_{S U M2}\right) \pi\right\}\Big].
\end{align}

In the following simulation studies, we demonstrate that the proposed Cauchy combination test $T_{CC3}$ performs robustly across a wide range of distributions and varying levels of sparsity in the alternatives.

\section{Quadratic Discriminant Analysis}\label{sec:QDA}
Consider the problem of classifying a p-dimensional normally distributed vector $\boldsymbol{x}$ into one of two classes represented by two $p$-dimensional normal distributions, $N_p\left(\boldsymbol{\mu}_1, \mathbf{\Xi}_1\right)$ and $N_p\left(\boldsymbol{\mu}_2, \mathbf{\Xi}_2\right)$, where $\boldsymbol{\mu}_k$'s are mean vectors and $\mathbf{\Xi}_k$ 's are positive definite covariance matrices. If $\boldsymbol{\mu}_k$ and $\mathbf{\Xi}_k, k=1,2$, are known, then an optimal classification rule having the smallest possible misclassification rate can be constructed. However, $\boldsymbol{\mu}_k$ and $\mathbf{\Xi}_k, k=1,2$, are usually unknown and the optimal classification rule, the Bayes rule, classifies $\boldsymbol{x}$ to class 2 if and only if
\begin{align}\label{qda0}
\left(\boldsymbol{x}-\boldsymbol{\mu}_1\right)^{\top} (\mathbf{\Xi}_2^{-1}-\mathbf{\Xi}_1^{-1})\left(\boldsymbol{x}-\boldsymbol{\mu}_1\right)-2 \boldsymbol{\delta}^{\top} \mathbf{\Xi}_2^{-1}\left(\boldsymbol{x}-\boldsymbol{\mu}_1\right)+\boldsymbol{\delta}^{\top} \mathbf{\Xi}_2^{-1} \boldsymbol{\delta}-\log \left(|\mathbf{\Xi}_1| /|\mathbf{\Xi}_2|\right)<0,
\end{align}
where $\boldsymbol{\delta}=\boldsymbol{\mu}_2-\boldsymbol{\mu}_1.$ In practical applications, when the dimension is lower than the sample size, we substitute the mean and covariance matrix in (\ref{qda0}) with their respective sample mean and covariance matrix. Nevertheless, when the dimension exceeds the sample size, the sample covariance matrix becomes non - invertible. As a result, a common approach, as described in \cite{Li2015} and \cite{Wu2019}, involves replacing the sample covariance matrix with various sparse covariance matrix estimators \citep{bickel2008covariance,bickel2008regularized}. However, it should be noted that these methods relying on the sample covariance matrix may not be highly efficient when the underlying distribution diverges from the normal distribution.

In fact it has been shown by \cite{Bose2015} that, for the class of elliptically symmetric distributions with the probability density function having the form
$$
f(\boldsymbol{x}; \bm \mu, \mathbf{\Xi})=|\mathbf{\Xi}|^{-1/2} g\left\{(\boldsymbol{x}-\boldsymbol{\mu}\}^{\top} \mathbf{\Xi}^{-1}(\boldsymbol{x}-\boldsymbol{\mu})\right\},
$$
the Bayes rule leads to the partition
$$
R_1=\left\{\boldsymbol{x}: \frac{1}{2} \log \left(\frac{|\mathbf{\Xi}_2|}{|\mathbf{\Xi}_1|}\right)+k \Delta_d^2 \geq 0\right\},
$$
where $\Delta_d^2(\bm x)=\left\{\left(\boldsymbol{x}-\boldsymbol{\mu}_{\mathbf{2}}\right)^{\top} \mathbf{\Xi}_2^{-1}\left(\boldsymbol{x}-\boldsymbol{\mu}_{\mathbf{2}}\right)-\left(\boldsymbol{x}-\boldsymbol{\mu}_{\mathbf{1}}\right)^{\top} \mathbf{\Xi}_1^{-1}\left(\boldsymbol{x}-\boldsymbol{\mu}_{\mathbf{1}}\right)\right\}$ and $k$ may depend on $\boldsymbol{x}$.
Therefore,  denoting $\varsigma_p\doteq\log(|\mathbf{\Xi}_1| /|\mathbf{\Xi}_2|)$, a general classification rule/classifier, proposed by \cite{Bose2015}, is given by
\begin{equation}\label{eq:qda}
    \begin{aligned}
& \boldsymbol{x} \in R_1 \quad \text { if } \Delta_d^2(\bm x) \geq c \varsigma_p, \\
& \boldsymbol{x} \in R_2 \quad \text { otherwise },
\end{aligned}
\end{equation}
for some constant $c \geq 0$. Clearly, this classifier boils down to the MMD and the QDA classifiers whenever $c$ is chosen to be 0 and 1 , respectively. It has a misclassification rate of
$$R_{QDA}=\frac{R_{QDA}^1+R_{QDA}^2}{2},\qquad R_{QDA}^k=\mathbb{P}(\text{incorrectly~classify~$x$~to~class~$k$}).$$

In practice, the parameters in the classifier (\ref{eq:qda}) are unknown and need to be estimated from the training set. Suppose we observe two independent samples $\{\bm X_{ik}\}_{k=1}^{n_i}, i=1,2$ from $f(\bm x; \bm \mu_i, \mathbf{\Xi}_i)$, respectively. 
Under the elliptical symmetric distribution assumption, we have $\mathbf{\Xi}=p^{-1}\tr(\mathbf{\Xi})\mathbf{\Sigma}$. So the inverse of the covariance matrix $\hat{\mathbf{\Xi}}^{-1}=p\tr^{-1}(\mathbf{\Xi})\mathbf{\Omega}$ could be estimated by 
$\tilde{\mathbf{\Omega}}_i=p\{\widehat{\tr(\mathbf{\Xi}_i)}\}^{-1}\hat{\mathbf{\Omega}}_i$ where 
\begin{align*}
\widehat{\tr(\mathbf{\Xi}_i)}=\frac{1}{n_i-1}\sum_{l=1}^{n_i} \bm X_{il}^T  \bm X_{il}-\frac{n_i}{n_i-1}\bar{\bm X}_i^T\bar{\bm X}_i,
\end{align*}
and $\bar{\bm X}_i=\frac{1}{n_i}\sum_{l=1}^{n_i}\bm X_{il}$. Then, we replace the parameters with its high dimensional HR estimators, i.e.
\begin{align*}
\hat{\Delta}_d^2(\bm x)=\left(\boldsymbol{x}-\hat{\boldsymbol{\mu}}_{{2}}\right)^{\top} \tilde{\mathbf{\Omega}}_2\left(\boldsymbol{x}-\hat{\boldsymbol{\mu}}_{{2}}\right)-\left(\boldsymbol{x}-\hat{\boldsymbol{\mu}}_{{1}}\right)^{\top} \tilde{\mathbf{\Omega}}_1\left(\boldsymbol{x}-\hat{\boldsymbol{\mu}}_{{1}}\right), \hat{\varsigma}_p=\log \left(|\tilde{\mathbf{\Omega}}_2|/|\tilde{\mathbf{\Omega}}_1|\right),
\end{align*}
where 
the parameter $c$ is estimated the same as Subsection 2.1 in \cite{Bose2015}, denoted as $\hat{c}$. So the final classification rule is
\begin{equation}\label{eq:qda2}
    \begin{aligned}
& \boldsymbol{x} \in R_1 \quad \text { if } \hat\Delta_d^2(\bm x) \geq \hat c 
\hat\varsigma_p, \\
& \boldsymbol{x} \in R_2 \quad \text { otherwise }.
\end{aligned}
\end{equation}
It has a misclassification rate of
$$R_{HRQDA}=\frac{R_{HRQDA}^1+R_{HRQDA}^2}{2},\qquad R_{HRQDA}^k=\mathbb{P}(\text{incorrectly~classify~$x$~to~class~$k$}).$$

To show the consistency of the misclassification rate of our proposed HRQDA method, we need the following additional assumptions.
\begin{assu}\label{assu5}
 $\tr \bm{\Xi}_k\asymp t_0(p)$ for each $k=1,2$. $\sigma_Q(p):=\sqrt{\|\bm{\Sigma}^{1/2}_1\bm{\Omega}_2\bm{\Sigma}_1^{1/2}-\mathbf{I}_p\|_F^2+t_0(p)^{-1}p\|\bm{\mu}_2-\bm{\mu}_1\|^2}\asymp p$.
\end{assu}
\begin{assu}\label{assu6}
     $r_k=\|\mathbf{\Xi}_k^{-1/2}(\bm{x}-\bm{\mu}_k)\|$ satisfies $\mathrm{Var}(r_k^2)\lesssim p\sqrt{p}$ and $\mathrm{Var}(r_k)\lesssim \sqrt{p}$, for $k=1,2$.
\end{assu}

Assumption \ref{assu5} assume the signal of the difference between the two distribution is larger enough. Assumption \ref{assu6} is needed to show the consistency of the trace estimator $\widehat{\tr(\mathbf{\Xi}_i)}$.

\begin{theorem}\label{thm7}
    If $\bm{\varepsilon}_k$, $\bm{\Sigma}_k$, $\bm{\Omega}_k$, $\mathbf{S}_k$ for $k=1,2$ satisfy Assumptions \ref{assu1}-\ref{assu4} and Assumptions \ref{assu5},\ref{assu6} hold. Assume that $n_1\asymp n_2$ and $n:=\min \{n_1,n_2\}$, we have
    \begin{equation*}
        |R_{HRQDA}-{R}_{QDA}|=O_p \{\lambda_n^{1-q/2}s_0(p)^{1/2}+\lambda_n^{1-q}s_0(p)\}.
    \end{equation*}
\end{theorem}
The result in Theorem \ref{thm7} show that HRQDA is able to mimic the optimal Bayes rule consistently under some mild assumptions, which is similar to Theorem 4.2 in \cite{cai2021convex}.

\section{Simulation}
\subsection{Location Parameter Test Problem}
First, we consider the empirical sizes of our proposed new test procedure. We consider the following three elliptical distributions:
\begin{itemize}
\item[(i)] Multivariate Normal Distribution: $\X_{i}\sim N(\bmu,\bms)$;
\item[(ii)] Multivariate $t$-distribution:  $\X_{i}\sim t(\bmu,\bms,3)/\sqrt{3}$;
\item[(iii)] Mixture of multivariate Normal distribution: $\X_{i}\sim MN(\bmu,\bms,10,0.8)/\sqrt{22.8}$;
\end{itemize}
Four covariance matrices are considered:
\begin{itemize}
\item[(I)] $\bms=(0.6^{|i-j|})_{1\le i,j\le p}$;
\item[(II)] $\bms=0.5\I_p+0.5\bm 1 \bm 1^\top$;
\item[(III)] $\mO=(0.6^{|i-j|})_{1\le i,j\le p}$, $\bms=\mO^{-1}$.
\item[(IV)] $\mO=\left(\omega_{i, j}\right)$ where $\omega_{i, i}=2$ for $i=1, \ldots, p, \omega_{i, i+1}=0.8$ for $i=1, \ldots, p-1, \omega_{i, i+2}=0.4$ for $i=1, \ldots, p-2, \omega_{i, i+3}=0.4$ for $i=1, \ldots, p-3, \omega_{i, i+4}=0.2$ for $i=1, \ldots, p-4, \omega_{i, j}=\omega_{j, i}$ for $i, j=1, \ldots, p$ and $\omega_{i, j}=0$ otherwise.
\end{itemize}
Table \ref{tab1} reports the empirical sizes of the new proposed test procedures $T_{SUM}$, $T_{MAX}$, $T_{CC1}$ and $T_{CC3}$ with $n=100$, $p=120,240$. We found that all the test could control the empirical sizes very well.

\begin{table}[htbp]
	\centering
 \caption{The empirical sizes of new proposed four test procedures with $n=100$.}
	\begin{tabular}{c|cc|cc|cc|cc} \hline \hline
Models &\multicolumn{2}{c}{(I)}&\multicolumn{2}{c}{(II)}&\multicolumn{2}{c}{(III)}&\multicolumn{2}{c}{(IV)}\\ \hline
$p$& 120&240& 120&240& 120&240& 120&240\\ \hline
\multicolumn{9}{c}{Normal distribution}\\ \hline
$T_{SUM}$&4.3&4.9&4.2&5.7&4.4&5.5&4.7&5.2\\
$T_{MAX}$&5.2&4.8&5.1&5.9&4.1&4.6&4.3&5.9\\
$T_{CC1}$&4.7&5.3&5.6&4.5&4.9&5.4&4.5&5.5\\
$T_{CC3}$&5.8&4.3&4.7&5.2&5.0&4.4&4.1&5.6\\ \hline
\multicolumn{9}{c}{$t_3$ distribution}\\ \hline
$T_{SUM}$&4.5&4.1&5.8&4.6&5.1&5.3&5.3&4.8\\
$T_{MAX}$&4.8&4.2&5.7&5.0&4.9&4.4&4.3&5.6\\
$T_{CC1}$&5.5&4.7&4.3&5.6&4.0&5.2&4.8&5.2\\
$T_{CC3}$&5.9&4.6&5.4&4.2&5.0&4.7&5.9&4.5\\ \hline
\multicolumn{9}{c}{Mixture normal distribution}\\ \hline
$T_{SUM}$&4.2&5.6&4.9&4.4&5.8&4.3&4.1&5.7\\
$T_{MAX}$&5.2&4.7&5.5&4.1&5.0&4.6&5.1&4.7\\
$T_{CC1}$&4.8&5.3&4.5&5.7&4.0&5.4&4.4&5.8\\
$T_{CC3}$&5.1&4.9&4.2&5.6&4.7&5.9&4.9&5.5\\ \hline \hline
	\end{tabular}
	\label{tab1}
\end{table}

Next, we conduct a comparison between our proposed methods and the test procedures founded on the sample covariance matrix. Specifically, \cite{chen2024asymptotic} put forward a max-type test procedure that hinges on the sample mean and a sparse precision matrix estimator, which we denote as $T_{CFL}$. Meanwhile, \cite{Fan2015p} introduced a sum-type test procedure relying on the sample mean and a sparse covariance matrix estimator, denoted as $T_{FLY}$. To ensure a fair comparison, we uniformly adopt the graphic lasso method to estimate the precision matrix in both $T_{CFL}$ and $T_{FLY}$. In addition, we also take into account a Cauchy combination test procedure that combines $T_{CFL}$ and $T_{FLY}$, denoted as $T_{CCF}$. Similar to the case of $T_{SUM}$, directly utilizing the asymptotic distribution to control the empirical sizes of $T_{FLY}$ would lead to substantial size distortion. Therefore, for the sake of a fair comparison, we propose a size-corrected power comparison approach. This approach involves first simulating the empirical distribution of all the test statistics under the null hypothesis. Subsequently, we determine the critical values for each test based on the simulated samples. This ensures that all the tests exhibit the same empirical sizes.

\begin{figure}[htbp]
    \caption{Power curves of each methods  with different sparsity under Model II and $n=100,p=120$.}
    \centering
    \begin{minipage}{0.3\textwidth}
        \includegraphics[width=\textwidth]{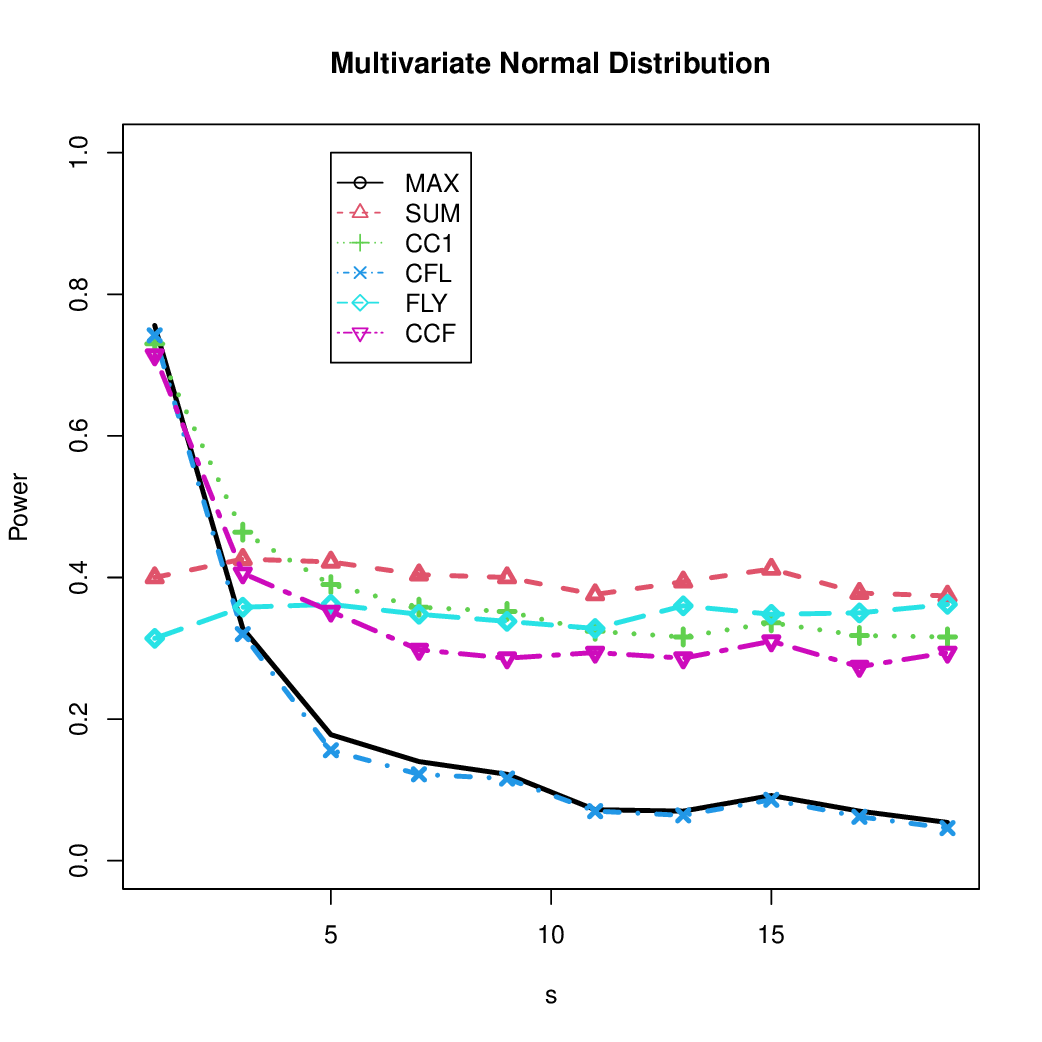}
        \parbox[t]{\textwidth}{\centering (a) Normal Distribution}
        \label{fig:n}
    \end{minipage}
    \begin{minipage}{0.3\textwidth}
        \includegraphics[width=\textwidth]{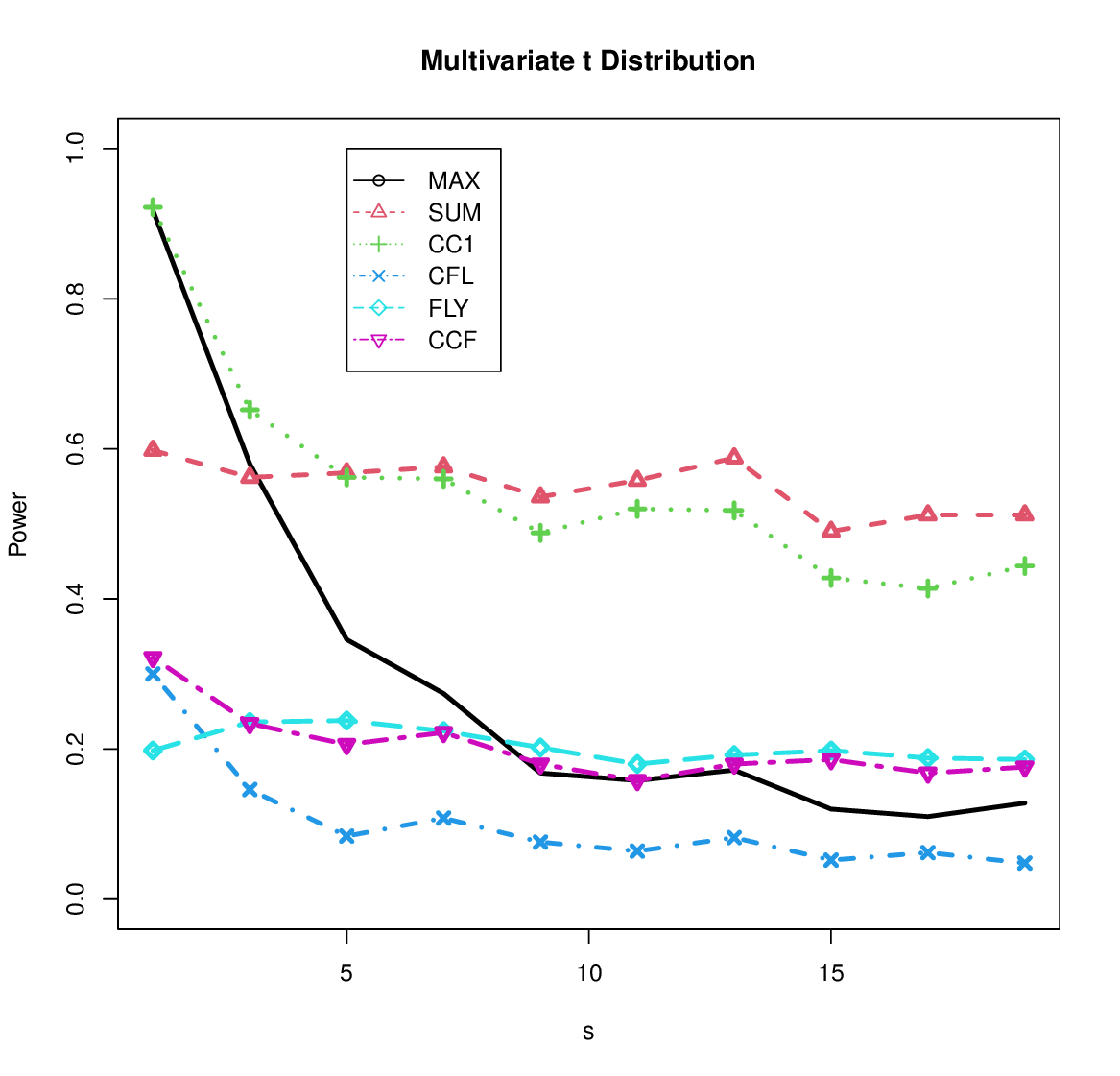}
       \parbox[t]{\textwidth}{\centering (b) Multivariate $t_3$ Distribution}
        \label{fig:t}
    \end{minipage}
    \begin{minipage}{0.3\textwidth}
        \includegraphics[width=\textwidth]{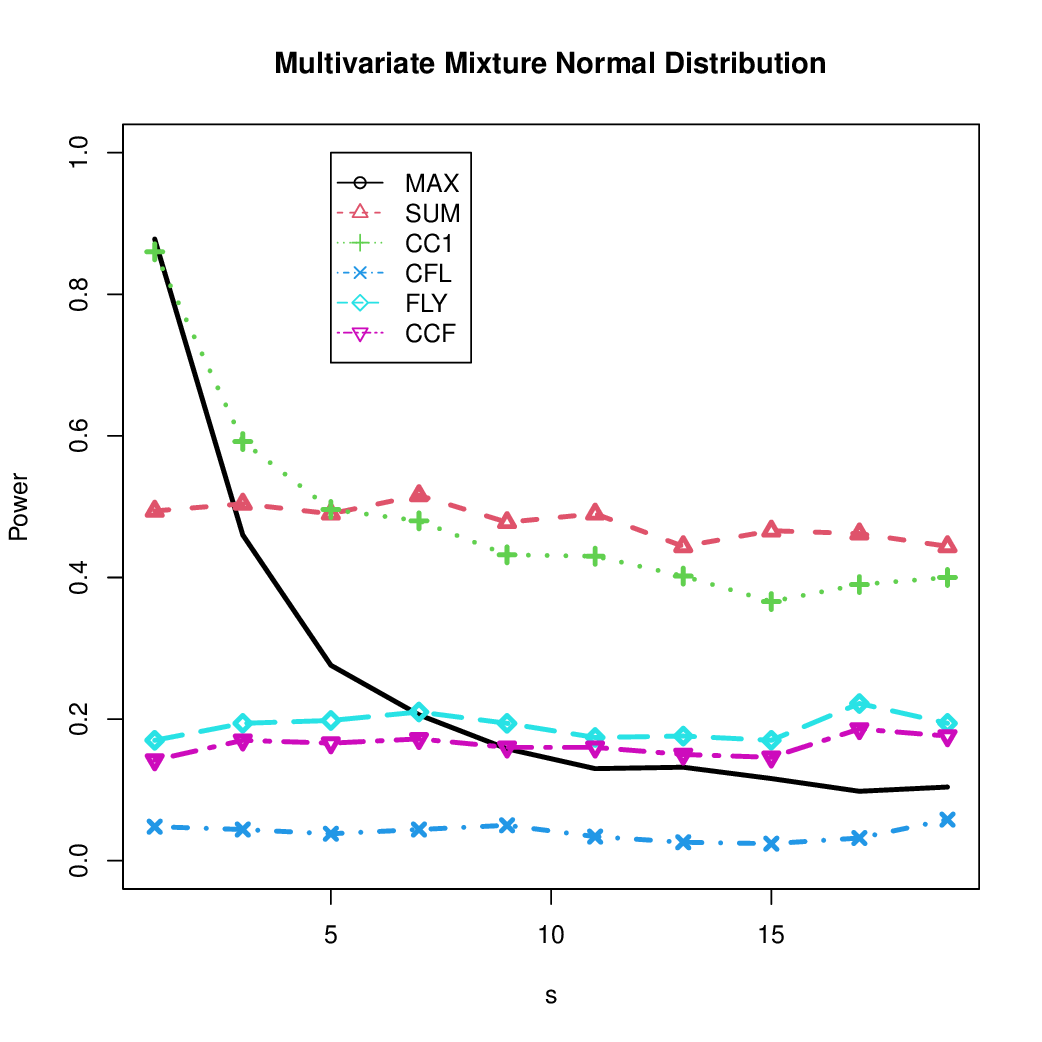}
        \parbox[t]{\textwidth}{\centering (c) Mixture Normal Distribution}
        \label{fig:m}
    \end{minipage}
    \label{fig1}
\end{figure}

Next we restrict our attention to Model II for the covariance matrix with sample size $n = 100$ and dimension $p = 120$. For the alternative hypothesis, we specify $\bmu=\kappa\sqrt{\log p/(ns)}\bms^{1/2}(\mathbf{1}^\top_s,\mathbf{0}^\top_{p - s})^\top$ to guarantee $\mO^{1/2}\bmu=\kappa\sqrt{\log p/(ns)}(\mathbf{1}^\top_s,\mathbf{0}^\top_{p - s})^\top$, where $s$ represents the sparsity parameter of the alternative hypothesis. For normal distribution, we set $\kappa=2$, while for multivariate $t$ distribution, we set $\kappa=1.5$ and for mixture normal distribution, we set $\kappa=0.6$.  
Figure \ref{fig1} presents the power curves of each test under various scenarios. Our findings indicate that, under the normal distribution, the performance of $T_{SUM}$ and $T_{MAX}$ is comparable to that of $T_{FLY}$ and $T_{CFL}$, respectively. However, for non-normal distributions, our proposed robust test procedures, namely $T_{SUM}$, $T_{MAX}$, and $T_{CC1}$, significantly outperform the other three methods, $T_{FLY}$, $T_{CFL}$, and $T_{CCF}$. This clearly demonstrates the robustness and efficiency of our methods when dealing with heavy-tailed distributions. Furthermore, when the sparsity parameter $s$ is small, the max-type test procedures $T_{MAX}$ and $T_{CFL}$ exhibit higher power compared to the sum-type test procedures $T_{SUM}$ and $T_{FLY}$. Conversely, for dense alternatives (i.e., when $s$ is large), the sum-type test procedures outperform the max-type test procedures. The Cauchy combination test procedures $T_{CC1}$ and $T_{CCF}$ consistently perform well across different levels of sparsity. In conclusion, our newly proposed test procedure $T_{CC1}$ not only performs admirably for heavy-tailed distributions but also demonstrates good performance across different levels of sparsity in the alternatives, thus exhibiting double robustness. 

\begin{figure}[htbp]
    \caption{Power curves of three Cauchy Combination tests  with different sparsity and models and $n=100,p=120$.}
    \centering
    \begin{minipage}{0.8\textwidth}
        \includegraphics[width=\textwidth]{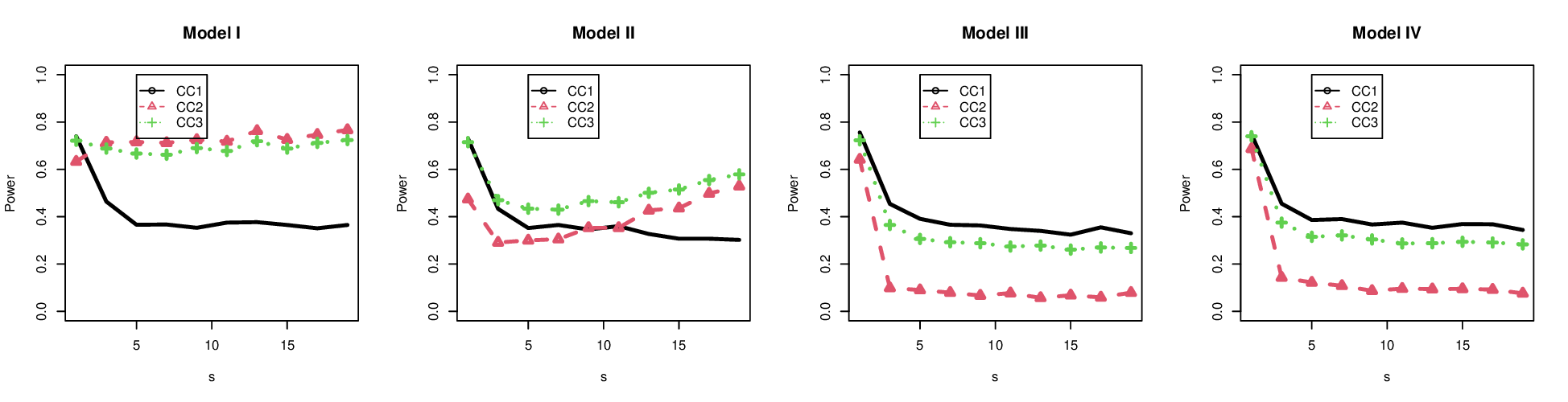}
        \parbox[t]{\textwidth}{\centering (a) Normal Distribution}
        \label{fig:n}
    \end{minipage}
    \begin{minipage}{0.8\textwidth}
        \includegraphics[width=\textwidth]{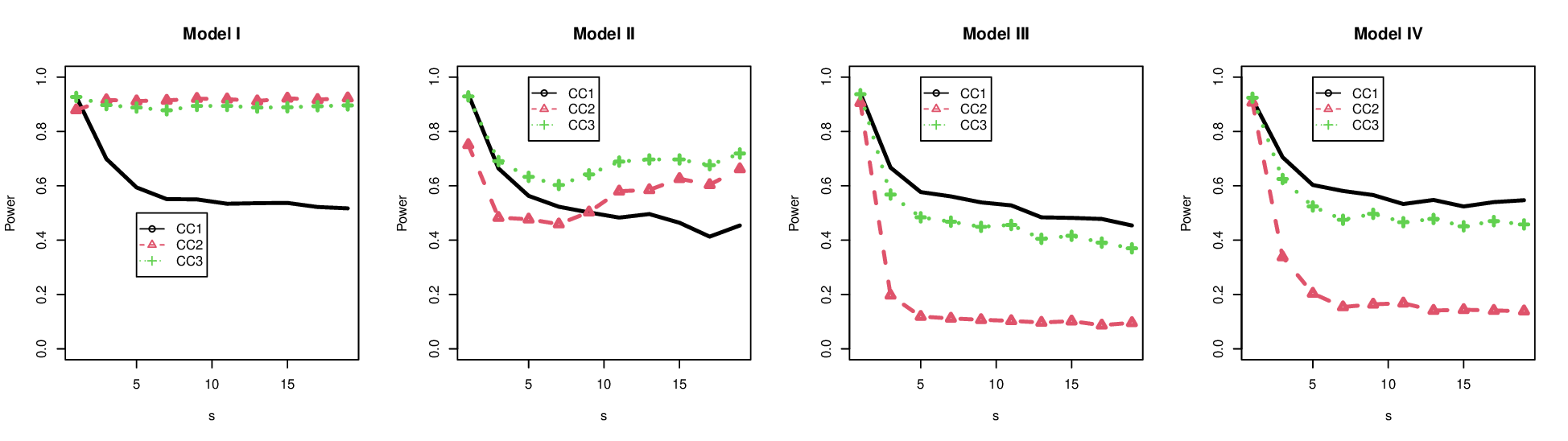}
       \parbox[t]{\textwidth}{\centering (b) Multivariate $t_3$ Distribution}
        \label{fig:t}
    \end{minipage}
    \begin{minipage}{0.8\textwidth}
        \includegraphics[width=\textwidth]{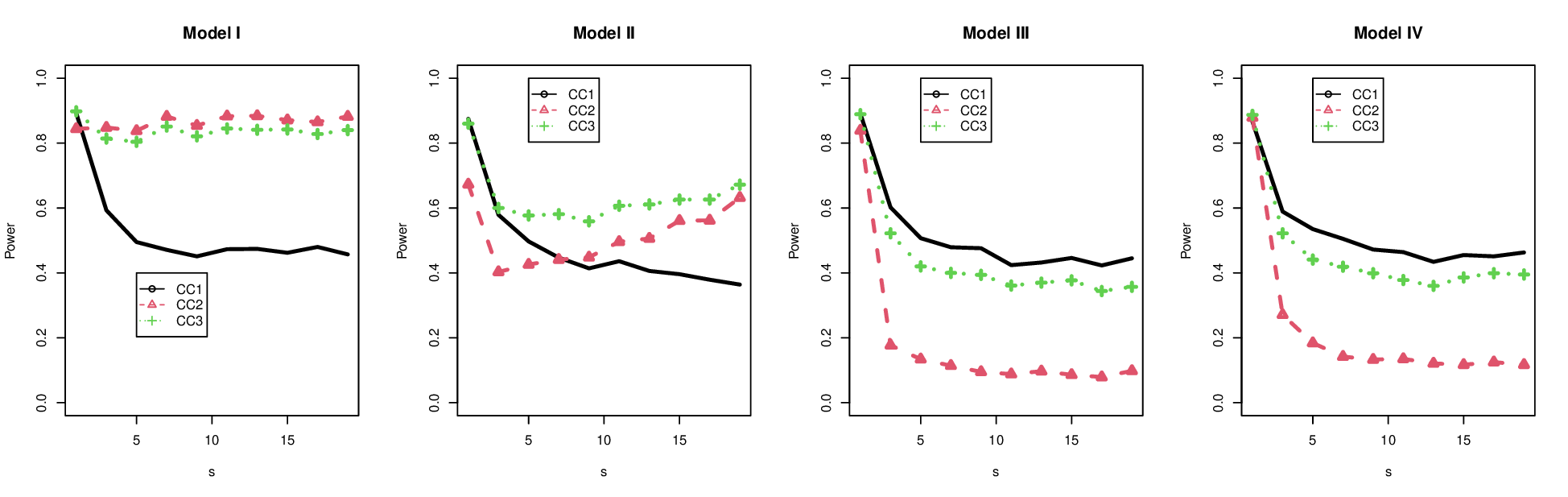}
        \parbox[t]{\textwidth}{\centering (c) Mixture Normal Distribution}
        \label{fig:m}
    \end{minipage}
    \label{fig2}
\end{figure}

Finally, we compare the performance of three Cauchy combination test procedures—$T_{CC1}$, $T_{CC2}$, and $T_{CC3}$—across different models. The alternative hypotheses considered are the same as those in Figure \ref{fig1}. Figure \ref{fig2} presents the power curves of these tests under various model settings and distributions. For Model I, $T_{CC2}$ generally outperforms $T_{CC1}$, whereas for Models III and IV, $T_{CC1}$ tends to be more efficient. In the case of Model II, $T_{CC2}$ exhibits lower power than $T_{CC1}$ when the signal sparsity $s$ is small, but surpasses $T_{CC1}$ as $s$ increases. Overall, the relative performance of $T_{CC1}$ and $T_{CC2}$ is highly sensitive to the underlying model structure.
In contrast, the newly proposed test $T_{CC3}$ demonstrates consistently strong and robust performance across all models and distributions. It often achieves the highest power among the three, making it a preferred choice in practice—not only due to its robustness to varying distributional assumptions, but also its adaptability to different sparsity levels of the alternatives.

\subsection{Quadratic Discriminant Analysis}
We compare our proposed method, HRQDA, with the SQDA method proposed by \cite{Li2015} and the SeQDA method proposed by \cite{Wu2019}. The SQDA method estimates the covariance matrix using the banding method proposed by \cite{bickel2008regularized}, while the SeQDA method estimates the covariance matrix of the transformed sample by simplifying the structure of the covariance matrices.

We consider the following three elliptical distributions:
\begin{itemize}
\item[(i)] Multivariate Normal Distribution: $X_{i1}\sim N(\bmu_1,\bms_1)$, $X_{i2}\sim N(\bmu_2,\bms_2)$;
\item[(ii)] Multivariate $t$-distribution:  $X_{i1}\sim t(\bmu_1,\bms_1,3)/\sqrt{3}$, $X_{i2}\sim t(\bmu_2,\bms_2,3)/\sqrt{3}$;
\item[(iii)] Mixture of multivariate Normal distribution: $X_{i1}\sim MN(\bmu_1,\bms_1,10,0.8)/\sqrt{22.8}$, \\
$X_{i2}\sim MN(\bmu_2,\bms_2,10.0.8)/\sqrt{22.8}$;
\end{itemize}
We consider three models for the covariance matrix:
\begin{itemize}
\item[(I)] $\bms_1=(0.6^{|i-j|})_{1\le i,j\le p}$, $\bms_2=\I_p$;
\item[(II)] $\bms_1=(0.6^{|i-j|})_{1\le i,j\le p}$, $\bms_2=0.5\I_p+0.5\bm 1 \bm 1^\top$;
\item[(III)] $\mO_1=(0.6^{|i-j|})_{1\le i,j\le p}$, $\bms_1=\mO_1^{-1}$, $\bms_2=\mO_1$.
\end{itemize}
The covariance matrices in Model (I) could be approximated by two banded matrix. In model (II), The second covariance matrix $\bms_2$ satisfy the structure assumption in \cite{Wu2019}, while violate the sparse assumption in \cite{Li2015}. For Model (III), The first covariance matrix $\bms_1$ satisfy the sparsity assumption of the precision matrix, while violate both assumptions of \cite{Wu2019} and \cite{Li2015}. For the mean vector, we simply consider $\bmu_1=\bf{0}, \bmu_2=0.1\times\bf{1}$.  In the simulation, we generate $n_1 = n_2 = 100$ training and
test samples of the same size and two dimensions $p=120,240$.

Table \ref{tabqda} presents the average classification rates of the three aforementioned methods. Through observation, it can be seen that the proposed HRQDA generally exhibits the best performance in most scenarios. In Model I, since the covariance matrix can be effectively estimated by the banding method, SQDA demonstrates better performance than SeQDA in most cases. When dealing with normal distributions, the performance of HRQDA is comparable to that of SQDA. However, in the context of non-normal distributions, HRQDA significantly outperforms both SQDA and SeQDA. Model II is more applicable to SdQDA rather than SQDA. Consequently, SQDA shows the poorest performance in this model. Similarly, under normal distributions, the performance of HRQDA is similar to that of SeQDA. Nevertheless, for non-normal distributions, HRQDA outperforms SeQDA. In Model III, HRQDA still maintains the best performance. Collectively, these results indicate that the proposed HRQDA method yields excellent performance across a broad spectrum of distributions and covariance matrix models. 

\begin{table}[htbp]
	\centering
 \caption{The average classification rate ($\%$) and standard deviation (in parenthesis) of each method.}
	\begin{tabular}{c|ccc|ccc} \hline \hline
 &\multicolumn{3}{c}{$p=120$}&\multicolumn{3}{c}{$p=240$}\\ \hline
Distributions& HRQDA & SQDA &SeQDA & HRQDA & SQDA &SeQDA\\ \hline
\multicolumn{7}{c}{Model I}\\ \hline
(i)&0.99(0.01)&0.94(0.08)&0.63(0.03)&1(0)&0.96(0.08)&0.64(0.04)\\
(ii)&0.95(0.06)&0.64(0.11)&0.55(0.04)&0.97(0.08)&0.6(0.09)&0.54(0.05)\\
(iii)&0.92(0.13)&0.55(0.07)&0.52(0.04)&0.96(0.1)&0.55(0.08)&0.51(0.04)\\
\hline
\multicolumn{7}{c}{Model II}\\ \hline
(i)&1(0.01)&0.77(0.1)&0.97(0.01)&1(0)&0.78(0.14)&1(0.01)\\
(ii)&0.99(0.01)&0.55(0.06)&0.68(0.03)&1(0)&0.55(0.05)&0.68(0.03)\\
(iii)&0.99(0.01)&0.53(0.04)&0.54(0.05)&1(0)&0.53(0.04)&0.53(0.03)\\
\hline
\multicolumn{7}{c}{Model III}\\ \hline
(i)&1(0)&1(0.02)&0.76(0.04)&1(0)&1(0.02)&0.76(0.04)\\
(ii)&1(0)&0.82(0.11)&0.66(0.03)&1(0)&0.81(0.12)&0.65(0.03)\\
(iii)&0.77(0.09)&0.6(0.02)&0.52(0.05)&0.78(0.10)&0.60(0.02)&0.51(0.04)\\
\hline \hline
	\end{tabular}
	\label{tabqda}
\end{table}

\section{Real data application}
This study employed the dataset (accession: GSE12288) utilized by \cite{sinnaeve2009gene}, which includes gene expression profiles of 110 coronary artery disease (CAD) patients (selected via Duke CAD index, CADi $>$ 23) and 112 healthy controls. Preprocessing involved applying two-sample $t$-tests to each gene, retaining those with $p$-values $<0.01$
, yielding 297 genes for analysis.

We compared our novel HRQDA method with two established QDA approaches: SQDA and SeQDA. The dataset was randomly partitioned into training (73 CAD patients, 75 controls) and testing subsets (37 CAD patients, 37 controls), reflecting approximately two-thirds and one-third of the data, respectively. This splitting procedure was repeated 200 times to ensure reliability. Accuracy metrics were computed across these iterations.

The performance of classifiers was evaluated using four key metrics:
\begin{itemize}
    \item \textbf{Accuracy} (\( \text{Acc} \)): Proportion of correctly classified samples:
    \[
    \text{Acc} = \frac{TP + TN}{TP + TN + FP + FN}.
    \]
    \item \textbf{Specificity} (\( \text{Spec} \)): Proportion of true negatives correctly identified:
    \[
    \text{Spec} = \frac{TN}{TN + FP}.
    \]
    \item \textbf{Sensitivity} (\( \text{Sens} \)): Proportion of true positives correctly identified:
    \[
    \text{Sens} = \frac{TP}{TP + FN}.
    \]
    \item \textbf{Matthews Correlation Coefficient} (\( \text{MCC} \)): Balanced measure of classification quality:
    \[
    \text{MCC} = \frac{TP \cdot TN - FP \cdot FN}{\sqrt{(TP + FP)(TP + FN)(TN + FP)(TN + FN)}}.
    \]
\end{itemize}
where \( TP \) (true positive), \( TN \) (true negative), \( FP \) (false positive), and \( FN \) (false negative) represent the counts of respective classification outcomes. All metrics range between $0$ and $1$, except MCC which ranges between $-1$ and $1$, with higher values indicating better performance.

Table \ref{tab:metrics_comparison} underscores the superior performance of HRQDA, which achieved the highest mean accuracy ($0.760$) among compared methods, surpassing SQDA ($0.710$) and SeQDA ($0.729$). This improvement is further validated by its highest Matthews Correlation Coefficient ($0.527$), indicating optimal balance between true positive and true negative predictions. HRQDA also demonstrated the strongest sensitivity ($0.821$), reflecting its robust ability to identify positive cases, while maintaining competitive specificity ($0.708$). These results highlight HRQDA's effectiveness in discriminating CAD patients from healthy individuals using gene expression data, particularly in scenarios requiring reliable detection of positive samples without excessive false negatives.

\begin{table}[htbp]
\centering
\caption{Comparison of evaluation metrics and standard deviation (in parenthesis) for each methods}
\label{tab:metrics_comparison}
\begin{tabular}{c|cccc}
\hline\hline
Method       & Accuracy       & Specificity    & Sensitivity    & MCC                \\
\hline
HRQDA        & 0.760 (0.042) & 0.708 (0.088) & 0.821 (0.069) & 0.527 (0.082)      \\
SQDA         & 0.710 (0.051) & 0.707 (0.112) & 0.702 (0.111) & 0.429 (0.103)      \\
SeQDA        & 0.729 (0.051) & 0.685 (0.084) & 0.772 (0.070) & 0.461 (0.102)      \\
\hline\hline
\end{tabular}
\end{table}

\section{Conclusion}
In this paper, we proposed a high-dimensional extension of the Hettmansperger-Randles estimator for simultaneously estimating the spatial median and scatter matrix under elliptically symmetric distributions. We demonstrated the applicability of this estimator to two fundamental problems in high-dimensional statistics: the one-sample location testing problem and quadratic discriminant analysis. Our proposed methods, which leverage the high-dimensional Hettmansperger-Randles estimator, consistently outperform existing approaches that rely on the sample mean and sparse covariance matrix estimators. Simulation studies and theoretical analysis confirm the superior efficiency and robustness of our estimator in high-dimensional settings. Given its strong performance, the high-dimensional Hettmansperger-Randles estimator offers a promising tool for broader applications in high-dimensional data analysis. In particular, it may be fruitfully applied to other important problems such as the two-sample location test \citep{feng2016multivariate} and the high-dimensional linear asset pricing model \citep{feng2022high}. These potential extensions warrant further investigation in future research.

\section{Appendix}
Recall that For $i=1,2,\cdots,n,\boldsymbol{U}_{i}=U(\bm{\varepsilon}_i)=U\{\mathbf{\Omega}^{1/2}(\boldsymbol{X}_{i}-\boldsymbol{\mu})\}$ and $r_{i} = \|\bm{\varepsilon}_i\|=\| \mathbf{\Omega} ^{1/2}( \boldsymbol{X}_{i}-\boldsymbol{\mu }) \|$ as the scale-invariant spatial-sign and radius of $\boldsymbol{X}_{i}-\boldsymbol{\mu}$, where $U(\boldsymbol{X})=\boldsymbol{X}/\|\boldsymbol{X}\|\mathbb{I}(\boldsymbol{X}\neq0)$ is the multivariate sign function of $\boldsymbol{X}$, with $\mathbb{I}(\cdot)$ being the indicator function. The moments of $r_i$ is defined as $\zeta_k=\mathbb{E}\left(r_{i}^{-k}\right)$.We denote the estimated version $\boldsymbol{U}_i$ and $r_i$ as $\hat{r}_i=\|\hat{\mathbf{\Omega}}^{1/2}(\boldsymbol{X}_i\boldsymbol{-\mu})\|$ and $\hat{U}_i=\hat{\mathbf{\Omega}}^{1/2}(\boldsymbol{X}_i\boldsymbol{-}\boldsymbol{\mu})/\|\hat{\mathbf{\Omega}}^{1/2}(\boldsymbol{X}_i\boldsymbol{-}\boldsymbol{\mu})\|$, respectively$,i=1,2,\cdots,n.$ Finally, we denote various positive constants by $C,C_1,C_2,\dots$ without mentioning this explicitly.
\subsection{The lemmas to be used}
The following result is a one-sample special case of Lemma 1 in \cite{feng2016multivariate}.

\begin{lemma}\label{lemma3}
 Under Assumption \ref{assu1}, for any matrix $\mM$, we have $\E[\{U(\bm{\varepsilon}_i)^\top\mM U(\bm{\varepsilon}_i)\}^2]=O\{p^{-2}\mathrm{tr}(\mM^\top\mM)\}$.
\end{lemma}

As it plays a key role in our analysis, we restate Theorem 1 from \cite{lu2025} below.

\begin{lemma}\label{lemma4}
    Under Assumptions \ref{assu1}-\ref{assu4}, $\hat{\mathbf{\Omega}}$ defined in (\ref{sglasso}) satisfies the following property.When $n$,$p$ are sufficiently large, there exist constants $C_{\eta,T}$ and $C$, such that if we pick 
    $$
    \lambda_{n}=T\bigg\{\frac{\sqrt{2}C(8+\eta^{2}C_{\eta,T})}{\eta^{2}}\sqrt{\frac{\log p}{n}}+\frac{C_{\eta,T}}{\sqrt{p}}\bigg\},
    $$ 
    with probability larger than $1-2p^{-2}$, the following inequalities hold:
    \begin{align*}
        &\|\hat{\mathbf{\Omega}}-\mathbf{\Omega}\|_{\infty}\leq4\|\mathbf{\Omega}\|_{L_{1}}\lambda_{n},\\
        &\|\hat{\mathbf{\Omega}}-\mathbf{\Omega}\|_{\mathrm{op}}\leq\|\hat{\mathbf{\Omega}}-\mathbf{\Omega}\|_{L_{1}}\leq C_{4}\lambda_{n}^{1-q}s_{0}(p),\\
        &p^{-1}\|\hat{\mathbf{\Omega}}-\mathbf{\Omega}\|_{F}^{2}\leq C_{5}\lambda_{n}^{2-q}s_{0}(p),
    \end{align*}
    where $C_{4}\leq(1+2^{1-q}+3^{1-q})(4\|\mathbf{\Omega}\|_{L_{1}})^{1-q}$ and $C_{5}\leq4\|\mathbf{\Omega}\|_{L_{1}}C_{4}.$
\end{lemma}
\begin{lemma}\label{lemma5}
    Define a random matrix $\hat{\mathbf{Q}}=n^{-1}\sum_{i=1}^n \hat{r}_i^{-1}\hat{\bm{U}}_i\hat{\bm{U}}^\top_i\in \mathbb{R}^{p\times p}$, and let $\hat{\mathbf{Q}}_{jl}$ denote its $(j,l)$-th element. Assume $\lambda_n^{1-q}s_0(p)(\log p)^{1/2}=o(1)$, and satisfy Assumptions \ref{assu1}–\ref{assu4}. Then we have
    \begin{equation*}
     |\hat{\mathbf{Q}}_{jl}|\lesssim p^{-3/2}\mathbb{I}(j=l)+O_p\left\{n^{-1/2}p^{-3/2}+ \lambda_{n}^{1-q}s_{0}(p) p^{-3/2}\right\}.
\end{equation*}
\end{lemma}

\begin{lemma}\label{lemma6}
    Suppose the Assumptions in Lemma \ref{lemma4} hold, then $\hat{\zeta}_1\overset{p}{\rightarrow}\zeta_1$ as $(n,p)\to\infty$, where $\hat{\zeta}_1=n^{-1}\sum_{i=1}^n \|\hat{\mathbf{\Omega}}^{1/2}(\X_1-\hat\bmu_1)\|^{-1}$.
\end{lemma}
\begin{lemma}\label{lemma7}
Suppose the Assumptions in Lemma \ref{lemma6} hold with $s_0(p)\asymp p^{1-\delta}$ for some positive constant $\delta\le 1/2$ Then, if \,$\log p=o(n^{1/3})$,
\begin{equation}\label{l8}
    \begin{aligned}&(i)\left\|n^{-1}\sum_{i=1}^{n}\zeta_{1}^{-1}\hat{\boldsymbol{U}}_{i}\right\|_{\infty}=O_{p}\left\{n^{-1/2}\log^{1/2}(np)\right\},\\&(ii)\left\|\zeta_{1}^{-1}n^{-1}\sum_{i=1}^{n}\delta_{1,i}\hat{\boldsymbol{U}}_{i}\right\|_{\infty}=O_{p}(n^{-1}).\end{aligned}
\end{equation}
where $\delta_{1,i}$ is defined in the proof in Lemma \ref{lemma1}.
\end{lemma}
Lemma \ref{lemma8} presents the well-known Nazarov's inequality, whose detailed proof is provided in Appendix A of \cite{chernozhukov2017central}.
\begin{lemma}[Nazarov’s inequality]\label{lemma8}
    Let $\bm{Y}_0=(Y_{0,1},Y_{0,2},\cdots,Y_{0,p})^\top$ be a centered Gaussian random vector in $\mathbb{R}^p$ and $\mathbb{E}(Y_{0,j}^2)\ge b$ for all $j=1,2,\cdots,p$ and some constant $b>0$, then for every $y\in \mathbb{R}^p$ and $a >0$,
    $$\mathbb{P}(\boldsymbol{Y}_{0}\leq y+a)-\mathbb{P}(\boldsymbol{Y}_{0}\leq y)\lesssim a\log^{1/2}(p).$$
\end{lemma}
We restate Lemma S9 in \cite{Feng2022AsymptoticIO}.
\begin{lemma}\label{lemma10} For each $d \geq 1$, we have
$$\lim_{p \to \infty} H(d,p) \leq \frac{1}{d!}\pi^{-d/2}e^{-dy/2} ,$$
where $H(d,p) \doteq \sum_{1 \leq i_1 < \cdots < i_d \leq p} \mathbb{P}(B_{i_1} \cdots B_{i_d})$, $B_{i_d}=\{|y_{i_d}|\ge \sqrt{2\log p-
\log \log p+y}\}$, $\bm{Y}=(y_1,\cdots,y_p)^\top\sim N(\bm{0},\mathbf{R}).$
\end{lemma}
\begin{lemma}\label{lemma11}
    Let $\bm{u}\in\mathbb{R}^p$ be a random vector uniformly distributed on the unit sphere $\mathbb{S}^{p-1}$. $\mathbf{A}\in\mathbb{R}^{p\times p}$ is a non-random matrix. Then we have $\mathbb{E}(\bm{u}^\top \mathbf{A}\bm{u})=p^{-1}\tr(\mathbf{A})$ and $\mathrm{Var}(\bm{u}^\top \mathbf{A}\bm{u})\asymp p^{-2}\|\mathbf{A}\|^2_F$ as $p\to \infty$.
\end{lemma}
\begin{lemma}\label{lemma12}
Under Assumption \ref{assu1}, we have 
\begin{equation*}
    \frac{\widehat{\tr (\bm{\Xi})}}{\tr (\bm{\Xi})}-1=O_p(n^{-1/2}).
\end{equation*}
\end{lemma}
We next restate Lemma 8.9 from \cite{shen2025spatial}.
\begin{lemma}\label{lemma13}
For positive matrix $\mathbf{X}$, $\mathbf{Y}$,
$$\log|\mathbf{X}|\leq\log|\mathbf{Y}|+\tr\{\mathbf{Y}^{-1}(\mathbf{X}-\mathbf{Y})\}.$$
\end{lemma}
\subsection{Proof of main lemmas}
\subsubsection{Proof of Lemma \ref{lemma5}}
\begin{proof}
Denote $\hat{\mathbf{I}}=\hat{\mathbf{\Omega}}^{1/2}\mathbf{\Sigma}^{1/2}$. Set $\mathbf{\hat{I}}_i^\top$ and $\mathbf{\Omega}_i^\top$ be the $i$th row of $\mathbf{\hat{I}}$ and $\mathbf{\Omega}$ respectively.
\begin{align*}
    \hat{\mathbf{Q}}_{jl}&=\frac{1}{n}\sum_{i=1}^n \hat{r}_i^{-1}\hat{\bm{U}}_{ij}\hat{\bm{U}}_{il}\\
    &=\frac{1}{n}\sum_{i=1}^n{\hat{r}_i^{-3}}\hat{\mathbf{I}}_j^\top\bm{\varepsilon}_i\hat{\mathbf{I}}_l^\top\bm{\varepsilon}_i\\
    &=\frac{1}{n}\sum_{i=1}^n\|\hat{\mathbf{I}}\bm{\varepsilon}_i\|^{-3}(\mathbf{\hat{I}}_j^T\bm{\varepsilon}_i)(\mathbf{\hat{I}}_l^T\bm{\varepsilon}_i)\\
    &=A_1+A_2+A_3,
\end{align*}
    where $A_1$, $A_2$ and $A_3$ are defined as follows
\begin{align*}
    A_1&=\frac{1}{n}\sum_{i=1}^n\left(\|\hat{\mathbf{I}}\bm{\varepsilon}_i\|^{-3}-\|{\mathbf{I}}\bm{\varepsilon}_i\|^{-3}\right)(\mathbf{\hat{I}}_j^T\bm{\varepsilon}_i)(\mathbf{\hat{I}}_l^T\bm{\varepsilon}_i);\\
   A_2&=\frac{1}{n}\sum_{i=1}^n\left(\|{\mathbf{I}}\bm{\varepsilon}_i\|^{-3}-\zeta_{3}\right)(\mathbf{\hat{I}}_j^T\bm{\varepsilon}_i)(\mathbf{\hat{I}}_l^T\bm{\varepsilon}_i);\\
   A_3&=\frac{1}{n}\sum_{i=1}^n\zeta_{3}(\mathbf{\hat{I}}_j^T\bm{\varepsilon}_i)(\mathbf{\hat{I}}_l^T\bm{\varepsilon}_i).
\end{align*}
Given Lemma \ref{lemma4} and under Assumption \ref{assu2}, we obtain that
\begin{align*}
\|\hat{\mathbf{I}}\boldsymbol{\varepsilon}_{i}\|^{2}
=&\bm{\varepsilon}^T_i\mathbf{\Sigma}^{1/2}(\hat{\mathbf{\Omega}}-\mathbf{\Omega})\mathbf{\Sigma}^{1/2}\bm{\varepsilon}_i+r_i^2\\
\leq& r_i^2+(\bm{\varepsilon}^T_i\mathbf{\Sigma}\bm{\varepsilon}_i)\|\hat{\mathbf{\Omega}}-\mathbf{\Omega}\|_{op} \\
\leq& r_i^2+\eta^{-1}r_i^2\|\hat{\mathbf{\Omega}}-\mathbf{\Omega}\|_{op}\\
\doteq&r_i^{2}\left(1+H\right),
\end{align*}
where $H=\eta^{-1}\|\hat{\mathbf{\Omega}}-\mathbf{\Omega}\|_{op}=O_p\{\lambda_{n}^{1-q}s_{0}(p)\}.$
Therefore, for any integer $k$,
\begin{align}\label{H}
\|\hat{\mathbf{I}}\boldsymbol{\varepsilon}_{i}\|^{k}=&\{\bm{\varepsilon}^T_i\mathbf{\Sigma}^{1/2}(\hat{\mathbf{\Omega}}-\mathbf{\Omega})\mathbf{\Sigma}^{1/2}\bm{\varepsilon}_i+r_i^2\}^{k/2}\notag\\
\leq& r_i^k(1+H)^{k/2}\notag\\
:=&r_i^{k}\left(1+H_k\right),
\end{align}    
where $H_k=(1+H)^{k/2}-1=O_p\{\lambda_{n}^{1-q}s_{0}(p)\}.$

Similar to the proof of Lemma A3 in \cite{cheng2023}, we have
\begin{align*}
    \mathbb{E}(A_1)
    =&\mathbb{E}\left\{\frac{1}{n}\sum_{i=1}^n\left(\|\hat{\mathbf{I}}\bm{\varepsilon}_i\|^{-3}-\|{\mathbf{I}}\bm{\varepsilon}_i\|^{-3}\right)(\mathbf{\hat{I}}_j^\top\bm{\varepsilon}_i)(\mathbf{\hat{I}}_l^\top\bm{\varepsilon}_i)\right\}\\
    =&\mathbb{E}\left\{\frac{1}{n}\sum_{i=1}^n\left(\left\|\bm{\varepsilon}_i\right\|^{-3}H_{-3}\right)(\mathbf{\hat{I}}_j^\top\bm{\varepsilon}_i)(\mathbf{\hat{I}}_l^\top\bm{\varepsilon}_i)\right\}\\
    =&\mathbb{E}\left\{(A_2+A_3)H_{-3}\right\}.
\end{align*}
Firstly, notice that,
\begin{align*}
    \mathbf{\hat{I}}_j^\top\bm{\varepsilon}_i=&(\mathbf{\hat{I}}-\mathbf{I})_j^\top\bm{\varepsilon}_i+{\varepsilon}_{ij}\\
    =&(\mathbf{\hat{\Omega}}^{1/2}-\mathbf{\Omega}^{1/2})_j^\top\mathbf{\Sigma}^{1/2}\bm{\varepsilon}_i+\varepsilon_{ij},
\end{align*}
thus,
\begin{align*}
    \mathbf{\hat{I}}_j^\top\bm{\varepsilon}_i-\varepsilon_{ij}=& \frac{1}{2}\{\mathbf{\Omega}^{-1/2}(\mathbf{\hat{\Omega}}-\mathbf{\Omega})\}_j^\top\mathbf{\Sigma}^{1/2}\bm{\varepsilon}_i+o_p\big[\{\mathbf{\Omega}^{-1/2}(\mathbf{\hat{\Omega}}-\mathbf{\Omega})\}_j^\top\mathbf{\Sigma}^{1/2}\bm{\varepsilon}_i\big]\\
        \lesssim & \|\mathbf{\Omega}^{-1/2}\|_{L_1}\|\mathbf{\hat{\Omega}}-\mathbf{\Omega}\|_{L_1}\|\mathbf{\Sigma}^{1/2}\|_{L_1}\|r_i\mathbf{\Sigma}^{1/2}\bm{U}_i\|_{\infty}\\
    =& O_p\{\lambda_n^{1-q}s_0(p)(\log p)^{1/2}\}=o_p(1).
\end{align*}
In the above equation, the second to last equation  from the following facts: (1) Since $\bms$ is a positive define symmetric matrix, and under the Assumption \ref{assu1}, we have $\|\mathbf{\Omega}^{-1/2}\|_{L_1}\le\{\lambda_{\max}(\bms)\|\bms\|_{L_1}\}^{1/2}=O(1)$. (2) Furthermore, according to the second formula of Lemma \ref{lemma4}, $\|\mathbf{\hat{\Omega}}-\mathbf{\Omega}\|_{L_1}=O_p\{\lambda_{n}^{1-q}s_{0}(p)\}$. (3) As for $\bm{U}_i$ is uniformly distributed on a p-dimensional unit sphere, $\|\bm{U}_i\|_\infty=O_p(\sqrt{\log p/p})$ and $r_i=O_p(\sqrt{p})$, we have 
$\|r_i\mathbf{\Sigma}^{1/2}\bm{U}_i\|_{\infty}=O_p\{(\log p)^{1/2}\}$.

Next, we analyze $A_2$ and $A_3$. Since $\E(r_i^2)=p$,
\begin{align*}
    A_2=&\frac{1}{n}\sum_{i=1}^n\left(\|{\mathbf{I}}\bm{\varepsilon}_i\|^{-3}-\zeta_{3}\right)\{\varepsilon_{ij}+o_p(1)\}\{\varepsilon_{il}+o_p(1)\}\\
    =&\frac{1}{n}\sum_{i=1}^n(r_i^{-1}-\zeta_{3}r_i^2)U_{ij}U_{il}\mathbb{I}(j=l)+o_p(1)\\
=&\zeta_1p^{-1}\mathbb{I}(j=l)+O_p( n^{-1/2}p^{-3/2})
\lesssim p^{-3/2}\mathbb{I}(j=l)+O_p( n^{-1/2}p^{-3/2}).
\end{align*}
and
\begin{align*}
    A_3&=\frac{1}{n}\sum_{i=1}^n\zeta_{3}\{\varepsilon_{ij}+o_p(1)\}\{\varepsilon_{il}+o_p(1)\}\\
    &\lesssim p^{-3/2}\mathbb{I}(j=l)+O_p( n^{-1/2}p^{-3/2}).
\end{align*}
It follows that,
\begin{equation*}
    |\hat{\mathbf{Q}}_{jl}|\lesssim\left\{p^{-3/2}\mathbb{I}(j=l)+O_p(  n^{-1/2}p^{-3/2})\right\}[1+O_p\{\lambda_{n}^{1-q}s_{0}(p)\}].
\end{equation*}
Thus,
\begin{equation*}
     |\hat{\mathbf{Q}}_{jl}|\lesssim p^{-3/2}\mathbb{I}(j=l)+O_p\left\{   n^{-1/2}p^{-3/2}+ \lambda_{n}^{1-q}s_{0}(p) p^{-3/2}\right\}.
\end{equation*}
\end{proof}
\subsubsection{Proof of Lemma \ref{lemma6}}
\begin{proof}
    Denote $\hat{\boldsymbol{\bm{\theta}}}=\hat{\boldsymbol{\mu}}-\boldsymbol{\mu}.$
$$\begin{aligned}\|\hat{\mathbf{\Omega}}^{1/2}(\boldsymbol{X}_{i}-\hat{\boldsymbol{\mu}})\|&=\|\mathbf{\Omega}^{1/2}(\boldsymbol{X}_{i}-\boldsymbol{\mu})\|(1+r_{i}^{-2}\|(\hat{\mathbf{\Omega}}^{1/2}-\mathbf{\Omega}^{1/2})(\boldsymbol{X}_{i}-\boldsymbol{\mu})\|^{2}\\
&+r_{i}^{-2}\|\hat{\mathbf{\Omega}}^{1/2}\hat{\boldsymbol{\bm{\theta}}}\|^{2}+2r_{i}^{-2}\boldsymbol{U}_{i}^{\top}(\hat{\mathbf{\Omega}}^{1/2}-\mathbf{\Omega}^{1/2})\mathbf{\Omega}^{-1/2}\boldsymbol{U}_{i})\\
&-2r_{i}^{-1}\boldsymbol{U}_{i}^{\top}\hat{\mathbf{\Omega}}^{1/2}\hat{\boldsymbol{\bm{\theta}}}-2r_{i}^{-1}\boldsymbol{U}_{i}\mathbf{\Omega}^{-1/2}(\hat{\mathbf{\Omega}}^{1/2}-\mathbf{\Omega}^{1/2})\hat{\mathbf{\Omega}}^{1/2}\hat{\boldsymbol{\bm{\theta}}})^{1/2}.\end{aligned}$$
By combining the third expression in Lemma \ref{lemma4}, the Taylor expansion and Markov's inequality, we obtain $r_i^{-2}\|(\hat{\mathbf{\Omega}}^{1/2}-\mathbf{\Omega}^{1/2})(\boldsymbol{X}_i-\boldsymbol{\mu})\|^{2}=O_{p}\left\{\lambda_n^{2-q}s_0(p)\right\}=o_{p}(1)$. Based on Lemma \ref{lemma1} and under the Assumption \ref{assu1}, we have $r_i^{-2}\|\hat{\mathbf{\Omega}}^{1/2}\hat{\boldsymbol{\bm{\theta}}}\|^{2}=O_{p}(n^{-1})=o_{p}(1)$. Similarly,  by the Cauchy
inequality, the other parts are also $o_p(1).$ So,
$$n^{-1}\sum_{i=1}^n\left\|\hat{\mathbf{\Omega}}^{1/2}\left(\boldsymbol{X}_i-\hat{\boldsymbol{\mu}}\right)\right\|^{-1}=\left\{n^{-1}\sum_{i=1}^n\left\|\mathbf{\Omega}^{1/2}\left(\boldsymbol{X}_i-\boldsymbol{\mu}\right)\right\|^{-1}\right\}\left\{1+o_p(1)\right\}.$$
Obviously,  $\mathbb{E}\left ( n^{- 1}\sum _{i= 1}^nr_i^{-1}\right ) = \zeta _1$ and  $\mathrm{Var}\left ( n^{- 1}\zeta _1^{- 1}\sum _{i= 1}^nr_i^{-1}\right ) = O\left ( n^{- 1}\right ) .$ Finally,  the proof
is completed.
\end{proof}
\subsubsection{Proof of Lemma \ref{lemma7}}
\begin{proof}
From the proof of Lemma \ref{lemma5}, we can see that $\hat \I_j^\top\bm{\varepsilon}_i-\varepsilon_{ij}=O_p\{\lambda_n^{1-q}s_0(p)(\log p)^{1/2}\}$. Moreover, for any integer $k$, we have $\hat{r}_i^{k}\le r_i^{k}(1+H_k)$, where $H_k=O_p\{\lambda_{n}^{1-q}s_{0}(p)\}$.
Recall that $\hat{\boldsymbol{U}}_i=U\{\mathbf{\hat{\Omega}}^{1/2}(\X_i-\bmu)\}$, since $ r_i^{-1}=O_p(p^{-1/2})$, then for any $j\in \{ 1, 2, \cdots , p\} $,
\begin{align*}
\hat{U}_{ij}=\hat{r}_i^{-1}\hat \I_j^\top\bm{\varepsilon}_i\le &r_i^{-1}(1+H_{-1})\varepsilon_{ij}+r_i^{-1}(1+H_{-1})(\hat \I_j^\top\bm{\varepsilon}_i-\varepsilon_{ij})\\
=&(1+H_{-1})U_{ij}+o_p\{(1+H_{-1})U_{ij}\}.
\end{align*}
Therefore, we obtain that $\hat\U_i\le \U_i(1+H_{-1})$ for $i=1,2,\dots,n$ with the assumption $\lambda_n^{1-q}s_0(p)(\log p)^{1/2}=o(1)$. 
According to the Lemma A4 in \cite{cheng2023}, we have $\left\|n^{-1/2} \sum_{i=1}^n \zeta_{1}^{-1} \bm{U}_i\right\|_\infty = O_p\{\log^{1/2}(np)\}$ and $\left\|n^{-1} \sum_{i=1}^n (\zeta_{1}^{-1} \bm{U}_i)^2\right\|_\infty = O_p(1)$ with $\log p = o(n^{1/3})$.



Therefore, we have
\begin{align*}
\left\|n^{-1}\sum_{i=1}^{n}\zeta_{1}^{-1}\hat{\boldsymbol{U}}_{i}\right\|_{\infty}=&\left\|n^{-1}\sum_{i=1}^{n}\zeta_{1}^{-1}(1+H_{-1})\boldsymbol{U}_{i}\right\|_{\infty}\\
\leq&|1+H_{-1}|\cdot\left\|n^{-1}\sum_{i=1}^{n}\zeta_{1}^{-1}\boldsymbol{U}_{i}\right\|_{\infty}=O_{p}\left\{n^{-1/2}\log^{1/2}(np)\right\}.
\end{align*}
Similarly
\begin{align*}
&\left\|\zeta_{1}^{-1}n^{-1}\sum_{i=1}^{n}\delta_{1,i}\hat{\bm{U}}_{i}\right\|_{\infty}\leq|1+H_{-1}|\cdot\left\|\zeta_{1}^{-1}n^{-1}\sum_{i=1}^{n}\delta_{1,i}\bm{U}_{i}\right\|_{\infty}\\
    \leq& O_{p}\{n^{-1}(1+n^{-1/2}\log^{1/2}p)\}=O_{p}(n^{-1}).
\end{align*}
\end{proof}
\subsubsection{Proof of Lemma \ref{lemma11}}
\begin{proof}
Since $\E(\bm{u}\bm{u}^\top)=p^{-1}\I_p$, then $\E(\bm{u}^\top \mathbf{A}\bm{u})=\tr\{\mathbf{A}\E(\bm{u}\bm{u}^\top)\}=p^{-1}\tr (\mathbf{A})$. Let $\mathbf{A}=(a_{ij})_{i,j=1}^p$, $\bm{u}=(u_1,\dots,u_p)^\top$, \begin{align*}
    \mathbb{E}(\bm{u}^\top \mathbf{A}\bm{u})^2=&\mathbb{E}\left(\sum_{i=1}^p a_{ii}u_i^2+\sum_{1\le i\neq j\le p}a_{ij}u_iu_j\right)^2\\
    =&\mathbb{E}\left\{\sum_{i=1}^p a^2_{ii}u_i^4+\sum_{1\le i\neq j\le p}(a_{ij}^2+a_{ii}a_{jj})u^2_iu^2_j\right\}\\
    =&\frac{3}{p(p+2)}\sum_{i=1}^p a^2_{ii}+\frac{1}{p(p+2)}\sum_{1\le i\neq j\le p}a_{ij}^2+a_{ii}a_{jj},
\end{align*}
where the last equality because that $(u_1^2,\dots,u_p^2)^\top$ follow a Dirichlet distribution $D_p(1/2,\dots,1/2)$\citep{oja2010multivariate}. As a consequence, we have $\E(u_i^4)=3/\{p(p+2)\}$ and $\E(u_i^2u_j^2)=1/\{p(p+2)\}$ for any  $i\neq j$. Combining the two results above and after some straightforward calculations, we obtain $\var(\bm{u}^\top \mathbf{A}\bm{u})\asymp p^{-2}\|\mathbf{A}\|^2_F$.
\end{proof}
\subsubsection{Proof of Lemma \ref{lemma12}}
\begin{proof}
Recall that $\widehat{\tr(\mathbf{\Xi}_i)}$ is defined as in Section \ref{sec:QDA}. Notice that
\begin{align*}
\widehat{\tr(\mathbf{\Xi}_i)}=&\frac{1}{n-1}\sum_{j=1}^n\bm{X}_j^T\bm{X}_j-\frac{n}{n-1}\bar{\bm{X}}^T\bar{\bm{X}}\\
=&\frac{\sum_{j=1}^n\bm{X}_j^T\bm{X}_j-\sum_{i,j}\bm{X}_i^T\bm{X}_j} {n(n-1)}\\
=&\frac{\sum_{i\neq j}-\bm{X}_i^T\bm{X}_j+\bm{X}_j^T\bm{X}_j}{n(n-1)}\\
=&\frac{\sum_{i\neq j\neq k}-\bm{X}_i^T\bm{X}_j+\bm{X}_j^T\bm{X}_j}{n(n-1)(n-2)}\\
=&\frac{\sum_{i\neq j\neq k}\bm{X}_i^T\bm{X}_k-\bm{X}_j^T\bm{X}_k-\bm{X}_i^T\bm{X}_j+\bm{X}_j^T\bm{X}_j}{n(n-1)(n-2)}\\
=&\frac{\sum_{i\neq j\neq k}(\bm{X}_i-\bm{X}_j)^T(\bm{X}_k-\bm{X}_j)}{n(n-1)(n-2)},
\end{align*}
which implies that our estimate of $\tr(\bm{\Xi})$ is the same as that of \cite{shen2025spatial}. Thus, we complete the proof according to Lemma 8.4 of \cite{shen2025spatial}.
\end{proof}
\subsubsection{Proof of Lemma \ref{lemma1}}
\begin{proof}
As $\boldsymbol{\mu}$ is a location parameter, we assume $\boldsymbol{\mu}=0$ without loss of generality.
Note that given $\hat{\boldsymbol{\Omega}}$, the estimator $\hat{\boldsymbol{\mu}}$ satisfies
$$\sum_{i=1}^n U\{\hat{\mathbf{\Omega}}^{1/2}(\mathbf{X}_{i}-\boldsymbol{\hat{\mu}})\}=0,$$
Therefore, the estimator $\hat{\boldsymbol{\mu}}$ is defined as the minimizer of the following objective function:
\begin{equation}
L(\boldsymbol{\bm{\theta}})=\sum_{i=1}^n\left\|\hat{\mathbf{\Omega}}^{1/2}(\mathbf{X}_{i}-\boldsymbol{\bm{\theta}})\right\|.
\end{equation}
 Our goal is find $b_{n,p}$ such that $\|\bm{\hat{\mu}}\|=O_p(b_{n,p})$.  The existence of a $b_{n,p}^{-1}$-consistent local minimizer is implied by the fact that for an arbitrarily small $\varepsilon>0$, there exist a sufficiently large constant $C$, which does no depend on $n$ or $p$, such that
\begin{equation}\label{l1.1}
    \lim\inf_n\mathbb{P}\left\{ \inf_{\bm{u}\in\mathbb{R}^p,\|\bm{u}\|=C} L(b_{n,p}\bm{u})>L(\bm{0})\right\}>1-\varepsilon.
\end{equation}
Firstly, we prove Equation (\ref{l1.1}) holds when  $b_{n,p}=p^{1/2}n^{-1/2}$. Consider the expansion of $\|\hat{\mathbf{\Omega}}^{1/2}(\mathbf{X}_{i}-b_{n,p}\bm{u})\|$:
\begin{align*}
   \| \hat{\mathbf{\Omega}}^{1/2}(\mathbf{X}_{i}-b_{n,p}\bm{u})\|
    =\| \hat{\mathbf{\Omega}}^{1/2}\mathbf{X}_{i}\| \left(1-2b_{n,p}\hat{r}_i^{-1}\bm{u}^\top \hat{\mathbf{\Omega}}^{1/2}\hat{\bm{U}_i}+b_{n,p}^2\hat{r}_i^{-2}\bm{u}^\top\hat{\mathbf{\Omega}}\bm{u}\right)^{1/2}
\end{align*}
Note that $b_{n,p}\hat{r}_i^{-1}\bm{u}^\top \hat{\mathbf{\Omega}}^{1/2}\hat{\bm{U}_i}=O_p(n^{-1/2})$ and $b_{n,p}^2\hat{r}^{-2}\bm{u}^\top\hat{\mathbf{\Omega}}\bm{u}=O_p(n^{-1})$. These orders follow from the following argument.
Since we already know that $\hat r_i^k\le r_i^k(1+H_k)$ and $\hat \U_i\le \U_i(1+H_{-1})$ with $H_k=O_p\{\lambda_{n}^{1-q}s_{0}(p)\}$ for any integer $k$, thus, 
\begin{align*}
b_{n,p}\hat{r}_i^{-1}\bm{u}^\top \hat{\mathbf{\Omega}}^{1/2}\hat{\bm{U}_i}\le &b_{n,p}(1+H_k)^2r_i^{-1}\bm{u}^\top \mathbf{\Omega}^{1/2}\bm{U}_i+b_{n,p}(1+H_k)^2r_i^{-1}\bm{u}^\top(\hat{\mathbf{\Omega}}^{1/2}-\mathbf{\Omega}^{1/2})\bm{U}_i.
\end{align*}
For the first term, by independence between $r_i$ and $\U_i$, we have
 $\E\{(r_i^{-1}\bm{u}^\top \mathbf{\Omega}^{1/2}\bm{U}_i)^2\}=\E(r_i^{-2})\E\{(\bm{u}^\top \mathbf{\Omega}^{1/2}\bm{U}_i)^2\}=\zeta_2p^{-1}\tr(\mathbf{\Omega})$, which implies that $\hat{r}_i^{-1}\bm{u}^\top \hat{\mathbf{\Omega}}^{1/2}\hat{\bm{U}_i}=O_p(p^{-1/2})$. Similarly, for the second term, applying Taylor expansion and Lemma \ref{lemma4} yields: 
$$
\E[\{r_i^{-1}\bm{u}^\top(\hat{\mathbf{\Omega}}^{1/2}-\mathbf{\Omega}^{1/2})\bm{U}_i\}^2]\lesssim p^{-2}\|(\hat{\mathbf{\Omega}}-\mathbf{\Omega})^2\|_{op}\le p^{-2}\|\hat{\mathbf{\Omega}}-\mathbf{\Omega}\|_{op}^2\lesssim p^{-2}\lambda_n^{2-2q}s^2_0(p),
$$
which implies that $r_i^{-1}\bm{u}^\top(\hat{\mathbf{\Omega}}^{1/2}-\mathbf{\Omega}^{1/2})\bm{U}_i=O_p\{p^{-1}\lambda_n^{1-q}s_0(p)\}$. Hence 
\begin{align*}
b_{n,p}\hat{r}_i^{-1}\bm{u}^\top \hat{\mathbf{\Omega}}^{1/2}\hat{\bm{U}_i}=&O_p\{b_{n,p}p^{-1/2}+b_{n,p}p^{-1}\lambda_n^{1-q}s_0(p)\}\\
=&O_p(n^{-1/2}).
\end{align*}
As the same way, we have $b_{n,p}^2\hat{r}^{-2}\bm{u}^\top\hat{\mathbf{\Omega}}\bm{u}=O_p(n^{-1})$.
Then we have
\begin{align*}
   \| \hat{\mathbf{\Omega}}^{1/2}(\mathbf{X}_{i}-b_{n,p}\bm{u})\| 
    =&\|\hat{\mathbf{\Omega}}^{1/2}\mathbf{X}_{i}\| -b_{n,p}\bm{u}^\top \hat{\mathbf{\Omega}}^{1/2}\hat{\bm{U}}_i\\
    &+\frac{ 1}{2}b_{n,p}^2\hat{r}_i^{-1}\bm{u} \hat{\mathbf{\Omega}}^{1/2}\left( \mathbf{I}_p-\hat{\bm{U}}_i\hat{\bm{U}}_i^\top \right) \hat{\mathbf{\Omega}}^{1/2}\bm{u}+O_p(p^{1/2}n^{-3/2}).
\end{align*}

So, it can be easily seen
\begin{equation}\label{l1.2}
\begin{aligned}
    &p^{-1/2}\left\{ L(b_{n,p}\bm{u})-L(\bm{0}) \right\}\\
    =&-n^{-1/2}\bm{u}^\top\hat{\mathbf{\Omega}}^{1/2}\sum_{i=1}^n\hat{\bm{U}}_i\\
    &+2^{-1}p^{1/2}\bm{u}\hat{\mathbf{\Omega}}^{1/2}\left\{n^{-1}\sum_{i=1}^n\left( \hat{r}_i^{-1}\mathbf{I}_p-\hat{r}_i^{-1}\hat{\bm{U}}_i\hat{\bm{U}}_i^\top  \right)\right\}\hat{\mathbf{\Omega}}^{1/2}\bm{u}+O_p(n^{-1/2}).
\end{aligned}
\end{equation}
Notice that $\mathbb{E}\left( \|n^{-1/2}\sum_{i=1}^n\hat{\bm{U}}_i\|^2 \right)=O(1)$ and $\mathrm{Var}\left( \|n^{-1/2}\sum_{i=1}^n\hat{\bm{U}}_i\|^2 \right)=O(1)$. Accordingly
\begin{equation*}
    \left|-n^{-1/2}\bm{u}^\top\hat{\mathbf{\Omega}}^{1/2}\sum_{i=1}^n\hat{\bm{U}}_i\right|\le \left\|\hat{\mathbf{\Omega}}^{1/2}\bm{u}\right\|\left\|n^{-1/2}\sum_{i=1}^n\hat{\bm{U}}_i\right\|=O_p(1).
\end{equation*}
Recall the definition $\mathbf{\hat{Q}}=n^{-1}\sum_{i=1}^n\hat{r}_i^{-1}\hat{\bm{U}}_i\hat{\bm{U}}_i^\top$ in Lemma \ref{lemma5}. After some tedious calculation, we can obtain that $\mathbb{E}\{\mathrm{tr}(\mathbf{\hat{Q}}^2)\}=O\{p^{-2}+n^{-1}p^{-1}+\lambda_n^{2-2q}s_0^2(p)p^{-1}\}$. Then $\mathbb{E}( \bm{u}^\top\hat{\mathbf{\Omega}}^{1/2}\mathbf{\hat{Q}}\hat{\mathbf{\Omega}}^{1/2}\bm{u} )^2\le \mathbb{E}\left\{( \bm{u}^\top\hat{\mathbf{\Omega}}\bm{u} )^2\mathrm{tr}( \mathbf{\hat{Q}}^2 ) \right\}=O\{p^{-2}+n^{-1}p^{-1}+\lambda_n^{2-2q}s_0^2(p)p^{-1}\}$, which leads to $\bm{u}^\top\hat{\mathbf{\Omega}}^{1/2}\mathbf{\hat{Q}}\hat{\mathbf{\Omega}}^{1/2}\bm{u}=O_p\{p^{-1}+n^{-1/2}p^{-1/2}+\lambda_n^{1-q}s_0(p)p^{-1/2}\}$. Thus we have
\begin{align*}
    &p^{1/2}\bm{u}\hat{\mathbf{\Omega}}^{1/2}\left\{ \frac{1}{n}\sum_{i=1}^n\left( \hat{r}_i^{-1}\mathbf{I}_p-\hat{r}_i^{-1}\hat{\bm{U}}_i\hat{\bm{U}}_i^\top  \right)\right\}\hat{\mathbf{\Omega}}^{1/2}\bm{u}\\
    =&p^{1/2}n^{-1}\sum_{i=1}^n \hat{r}_i^{-1}\bm{u}\hat{\mathbf{\Omega}}\bm{u}+o_p(1),
\end{align*}
where we use the fact that $n^{-1}\sum_{i=1}^n\hat{r}_i^{-1}=\zeta_{1}+O_p\{n^{-1/2}p^{-1/2}+\lambda_n^{1-q}s_0(p)p^{-1/2}\}$. By choosing a sufficient large $C$, the second term in (\ref{l1.2}) dominates the first term uniformly in $\|\bm{u}\|=C$. Hence, (\ref{l1.2}) holds and accordingly $\hat{\bm{\mu}}=O_p(b_{n,p})$. The estimator $\hat{\boldsymbol{\mu}}$ satisfies $\sum_{i=1}^nU\{\hat{\mathbf{\Omega}}^{1/2}(\boldsymbol{X}_i{-\hat{\boldsymbol{\mu}}})\}=0$, which is is equivalent to
\begin{equation*}
    n^{-1}\sum_{i=1}^{n}(\hat{\boldsymbol{U}}_{i}-\hat{r}_{i}^{-1}\hat{\mathbf{\Omega}}^{1/2}\hat{\boldsymbol{\mu}})(1-2\hat{r}_{i}^{-1}\hat{\boldsymbol{U}}_{i}^{\top}\hat{\mathbf{\Omega}}^{1/2}\hat{\boldsymbol{\mu}}+\hat{r}_{i}^{-2}\hat{\boldsymbol{\mu}}^{\top}\hat{\mathbf{\Omega}}\hat{\boldsymbol{\mu}})^{-1/2}=0.
\end{equation*}
 By the first-Taylor expansion, the above equation can be rewritten as:
\begin{equation*}
    n^{-1}\sum_{i=1}^{n}\left(\hat{\boldsymbol{U}}_{i}-\hat{r}_{i}^{-1}\hat{\mathbf{\Omega}}^{1/2}\hat{\boldsymbol{\mu}}\right)\left(1+\hat{r}_{i}^{-1}\hat{\boldsymbol{U}}_{i}^{\top}\hat{\mathbf{\Omega}}^{1/2}\hat{\boldsymbol{\mu}}-2^{-1}\hat{r}_{i}^{-2}\left\|\hat{\mathbf{\Omega}}^{1/2}\hat{\boldsymbol{\mu}}\right\|^{2}+\delta_{1,i}\right)=0,
\end{equation*}
where $\delta_{1,i}=O_{p}\big\{(\hat{r}_{i}^{-1}\hat{\bm{U}}_{i}^{\top}\hat{\mathbf{\Omega}}^{1/2}\hat{\boldsymbol{\mu}}-2^{-1}\hat{r}_{i}^{-2}\|\hat{\mathbf{\Omega}}^{1/2}\hat{\boldsymbol{\mu}}\|^{2})^{2}\big\}=O_{p} (n^{-1})$, which implies
\begin{equation}\label{eq3}
    \begin{aligned}
    &\frac{1}{n}\sum_{i=1}^{n}(1-2^{-1}\hat{r}_{i}^{-2}\hat{\boldsymbol{\mu}}^{\top}\hat{\mathbf{\Omega}}\hat{\boldsymbol{\mu}}+\delta_{1,i})\hat{\boldsymbol{U}}_{i}+\frac{1}{n}\sum_{i=1}^{n}\hat{r}_{i}^{-1}(\hat{\boldsymbol{U}}_{i}^{\top}\hat{\mathbf{\Omega}}^{1/2}\hat{\boldsymbol{\mu}})\hat{\boldsymbol{U}}_{i}\\
    =&\frac{1}{n}\sum_{i=1}^{n}(1+\delta_{1,i}+\delta_{2,i})\hat{r}_{i}^{-1}\hat{\mathbf{\Omega}}^{1/2}\hat{\boldsymbol{\mu}},
\end{aligned}
\end{equation}
where $\delta_{2,i}=O_{p}(\hat{r}_{i}^{-1}\hat{\boldsymbol{U}}_{i}^{\top}\hat{\mathbf{\Omega}}^{1/2}\hat{\boldsymbol{\mu}}-2^{-1}\hat{r}_{i}^{-2}\|\hat{\mathbf{\Omega}}^{1/2}\hat{\boldsymbol{\mu}}\|^{2})=O_{p}(\delta_{1,i}^{1/2})$. By Assumption \ref{assu1} and Markov inequality, we have that: $\max r_i^{-2}=O_p(p^{-1}n^{1/2})$, $\max\delta_{1,i}=O_p\left( \|\hat{\mathbf{\Omega}}^{1/2}\bm{\hat{\mu}}\|^2\max \hat{r}^{-2}_i \right)=O_p(n^{-1/2})$ and $\max \delta_{2,i}=O_p(n^{-1/4})$.
Considering the second term in Equation (\ref{eq3}),
\begin{equation*}
    \frac{1}{n}\sum_{i=1}^{n}\hat{r}_{i}^{-1}(\hat{\bm{U}}_{i}^{\top}\hat{\mathbf{\Omega}}^{1/2}\hat{\boldsymbol{\mu}})\hat{\bm{U}}_{i}=\frac{1}{n}\sum_{i=1}^{n}\hat{r}_{i}^{-1}(\hat{\bm{U}}_{i}\hat{\bm{U}}_{i}^{\top}\hat{\mathbf{\Omega}}^{1/2})\hat{\boldsymbol{\mu}}=\hat{\mathbf{Q}}\hat{\mathbf{\Omega}}^{1/2}\hat{\boldsymbol{\mu}}.
\end{equation*}
From Lemma \ref{lemma5} we acquire
\begin{equation*}
     |\hat{\mathbf{Q}}_{jl}|\lesssim p^{-3/2}\mathbb{I}(j=l)+O_p\left\{  n^{-1/2}p^{-3/2}+ \lambda_{n}^{1-q}s_{0}(p) p^{-3/2}\right\},
\end{equation*}
and this implies that,
\begin{equation}\label{eq13}
\begin{aligned}
    &\|\hat{\mathbf{Q}}\hat{\mathbf{\Omega}}^{1/2}\hat{\bm{\mu}}\|_\infty\\
    \le&\|\hat{\mathbf{Q}}\|_1\|\hat{\mathbf{\Omega}}^{1/2}\hat{\bm{\mu}}\|_\infty\\
    =& O_p\left\{n^{-1/2}p^{-1/2}+ \lambda_{n}^{1-q}s_{0}(p) p^{-1/2}\right\}\|\hat{\mathbf{\Omega}}^{1/2}\hat{\bm{\mu}}\|_\infty\\
\end{aligned}
\end{equation}
According to Lemma \ref{lemma7}, we obtain
\begin{align*}
&\left\|\zeta_{1}^{-1}n^{-1}\sum_{i=1}^{n}\hat{r}_{i}^{-2}\|\hat{\mathbf{\Omega}}^{1/2}\hat{\boldsymbol{\mu}}\|^{2}\hat{\boldsymbol{U}}_{i}\right\|_{\infty}\leq|1+H_{u}|\cdot\left\|\zeta_{1}^{-1}n^{-1}\sum_{i=1}^{n}\hat{r}_{i}^{-2}\|\hat{\mathbf{\Omega}}^{1/2}\hat{\boldsymbol{\mu}}\|^{2}\boldsymbol{U}_{i}\right\|_{\infty}\\=& O_{p}(n^{-1})\big[1+O_{p}\{\lambda_n^{1-q}s_0(p)(\log p)^{1/2}\}\big]=O_{p}(n^{-1}).
\end{align*}
Using the fact that $\zeta_{1}^{-1}n^{-1}\sum_{i=1}^n r_i^{m-1}=1+O_p(n^{-1/2})$ and Equation (\ref{H}), we have
\begin{align*}
    \frac{1}{n}\zeta_{1}^{-1}\sum_{i=1}^{n}\hat{r}_{i}^{-1}=&\frac{1}{n}\zeta_{1}^{-1}\sum_{i=1}^{n}r_{i}^{-1}[1+O_{p}\{\lambda_n^{1-q}s_0(p)\}]\\=&\left\{1+O_{p}(n^{-1/2})\right\}\left[1+O_{p}\{\lambda_n^{1-q}s_0(p)\}\right]\\=&1+O_{p}\{\lambda_n^{1-q}s_0(p)\}.
\end{align*}
We final obtain:
\begin{equation*}
\begin{aligned}\left\|\hat{\mathbf{\Omega}}^{1/2}\hat{\boldsymbol{\mu}}\right\|_{\infty}&\lesssim\left\|\zeta_{1}^{-1}n^{-1}\sum_{i=1}^{n}\hat{\bm{U}}_{i}\right\|_{\infty}+\zeta_{1}^{-1}\left\|\mathbf{\hat{Q}}\hat{\mathbf{\Omega}}^{1/2}\hat{\bm\mu}\right\|_{\infty}\\&\lesssim p^{-1}\left\|\hat{\mathbf{\Omega}}^{1/2}\hat{\boldsymbol{\mu}}\right\|_{\infty}+O_{p}\left\{n^{-1/2}\log^{1/2}(np)\right\}\\&+ O_p\left\{  n^{-1/2}p^{-1}+ \lambda_{n}^{1-q}s_{0}(p) p^{-1}\right\}\|\hat{\mathbf{\Omega}}^{1/2}\hat{\bm{\mu}}\|_\infty.\end{aligned}
\end{equation*}
Thus we conclude that:
$$\left\|\hat{\mathbf{\Omega}}^{1/2}\hat{\boldsymbol{\mu}}\right\|_{\infty}=O_{p}\{n^{-1/2}\log^{1/2}(np)\},$$
as $s_0(p)\asymp p^{1-\delta}.$
In addition, by equation (\ref{eq13}) we have
$$\begin{aligned}
\left\|\zeta_{1}^{-1}\mathbf{\hat{Q}}\hat{\mathbf{\Omega}}^{1/2}\hat{\boldsymbol{\mu}}\right\|_{\infty}&=O_{p}\left[p^{1/2}\{n^{-1/2}p^{-1/2}+\lambda_n^{1-q}s_0(p)p^{-1/2}\}n^{-1/2}\log^{1/2}(np)\right]\\
&=O_p\left\{n^{-1}\log^{1/2}(np)+n^{-1/2}\lambda_n^{1-q}s_0(p)\log^{1/2}(np)\right\},
\end{aligned}$$
and
$$\begin{aligned}&n^{-1}\sum_{i=1}^{n}\hat{r}_{i}^{-1}\left(1+\delta_{1,i}+\delta_{2,i}\right)\\=&\zeta_{1}\left\{1+O_{p}\left(n^{-1/4}\right)\right\}\left[1+O_{p}\{\lambda_n^{1-q}s_0(p)\}\right]\\=&\zeta_{1}\left[1+O_{p}\{n^{-1/4}+\lambda_n^{1-q}s_0(p)\}\right].\end{aligned}$$
Finally, we can write
$$n^{1/2}\hat{\mathbf{\Omega}}^{1/2}(\hat{\boldsymbol{\mu}}-\boldsymbol{\mu})=n^{-1/2}\zeta_{1}^{-1}\sum_{i=1}^n\boldsymbol{U}_i+C_n,$$
where 
\begin{align*}
C_n=&\zeta_1^{-1}\big\{\big( -2^{-1}n^{-1/2}\sum_{i=1}^{n}\hat{r}_i^{-2}\hat{\bm{U}}_i     \big) \hat{\bm{\mu}}\hat{\mathbf{\Omega}}\hat{\bm{\mu} }\big\} +\zeta _1^{-1}\left ( n^{-1/2}\sum_{i=1}^{n} \delta_{1,i} \hat{\bm{U}}_i \right )  +\zeta _1^{-1}n^{1/2} \hat{\mathbf{Q}}\hat{\mathbf{\Omega}}^{1/2}\hat{\bm{\mu} } \\
&+n^{-1/2} \sum_{i=1}^{n} (\delta _{1,i}+\delta _{2,i}) \hat{r}_i^{-1}\hat{\mathbf{\Omega}}^{1/2}\hat{\bm{\mu}}.
\end{align*}
By previous discussion, we have
$$\begin{aligned}\|C_{n}\|_{\infty}= & O_p\big[n^{-1/2}+n^{-1/2}+n^{-1}\log^{1/2}(np)+n^{-1/2}\lambda_n^{1-q}s_0(p)\log^{1/2}(np)+\big\{n^{-1/4}+\lambda_n^{1-q}s_0(p)\big\}\log^{1/2}(np)\big]\\
=&O_{p}\big\{n^{-1/4}\log^{1/2}(np)+\lambda_n^{1-q}s_0(p)\log^{1/2}(np)\big\}.\end{aligned}$$
Then we complete the proof.
\end{proof}
\subsubsection{Proof of Lemma \ref{lemma2}}
\begin{proof}
     Let $L_{n, p}= n^{-1/4}\log^{1/2}(np)+\lambda_n^{1-q}s_0(p)\log^{1/2}(np)$, according to Lemma \ref{lemma1}, for any sequence $\eta_n\to\infty$ and any $t\in\mathbb{R}^p$,
$$\begin{aligned}\mathbb{P}\{n^{1/2}\hat{\mathbf{\Omega}}^{1/2}(\hat{\boldsymbol{\mu}}-\boldsymbol{\mu})\leq t\}&=\mathbb{P
}\big(n^{-1/2}\zeta_{1}^{-1}\sum_{i=1}^{n}\bm{U}_{i}+C_{n}\leq t\big)\\&\leq\mathbb{P}\big(n^{-1/2}\zeta_{1}^{-1}\sum_{i=1}^{n}\bm{U}_{i}\leq t+\eta_{n}L_{n,p}\big)+\mathbb{P}(\|C_{n}\|_{\infty}>\eta_{n}L_{n,p}).\end{aligned}$$
According to Lemma A4. in \cite{cheng2023} and $\mathbb{E}\{(\zeta_{1}^{-1}U_{i,j})^{4}\}\lesssim 3$ and $\mathbb{E}\{(\zeta_{1}^{-1}U_{i,j})^{2}\}\gtrsim\bar{B}^{-2}$ uniformly for all $i=1,2,\cdots,n$, $j=1,2,\cdots,p$,
the Gaussian approximation for independent partial sums in \cite{koike2021notes} yields:
$$\begin{aligned}\mathbb{P}\big(n^{1/2}\zeta_{1}^{-1}\sum_{i=1}^{n}\bm{U}_{i}\leq t+\eta_{n}L_{n,p}\big)&\leq\mathbb{P}(\bm{Z}\leq t+\eta_{n}L_{n,p})+O[\{n^{-1}\log^{5}(np)\}^{1/6}]\\&\leq\mathbb{P}(\bm{Z}\leq t)+O\{\eta_{n}L_{n,p}\log^{1/2}(p)\}+O[\{n^{-1}\log^{5}(np)\}^{1/6}],\end{aligned}$$
where $\bm{Z}\sim N\left(0,p^{-1}\zeta_{1}^{-2}\mathbf{I}_p\right)$, and the second inequality follows from Nazarov’s inequality (Lemma \ref{lemma8}). Thus,
$$\begin{aligned}\mathbb{P}\{n^{1/2}\hat{\mathbf{\Omega}}^{1/2}(\hat{\boldsymbol{\mu}}-\boldsymbol{\mu})\leq t\}\leq&\mathbb{P}(\boldsymbol{Z}\leq t)+O\{\eta_{n}L_{n,p}\log^{1/2}(p)\}+O(\{n^{-1}\log^{5}(np)\}^{1/6})\\
&+\mathbb{P}(|C_{n}|_{\infty}>\eta_{n}l_{n,p}).\end{aligned}$$
On the other hand, we have
$$\mathbb{P}\{n^{1/2}\hat{\mathbf{\Omega}}^{1/2}(\hat{\boldsymbol{\mu}}-\boldsymbol{\mu})\leq t\}\geq\mathbb{P}(\boldsymbol{Z}\leq t)-O\{\eta_{n}L_{n,p}\log^{1/2}(p)\}-O(\{n^{-1}\log^{5}(np)\}^{1/6})\\-\mathbb{P}(\|C_{n}\|_{\infty}>\eta_{n}l_{n,p}),$$
where $\mathbb{P}(\|C_n\|_\infty>\eta_nl_{n,p})\to0$ as $n\to\infty$ by Lemma \ref{lemma1}.
Then we have that, if $\log p=o(n^{1/5})$ ,
$$\sup\limits_{t\in\mathbb{R}^p}|\mathbb{P}\{n^{1/2}\hat{\mathbf{\Omega}}^{1/2}(\hat{\boldsymbol{\mu}}-\boldsymbol{\mu})\leq t\}-\mathbb{P}(\boldsymbol{Z}\leq t)|\to0.$$
Furthermore, by Corollary 3.1 in \cite{chernozhukov2017central}, we have
$$\rho_{n}(\mathcal{A}^{si})=\sup_{A\in\mathcal{A}^{si}}|\mathbb{P}\{n^{1/2}\hat{\mathbf{\Omega}}^{1/2}(\hat{\boldsymbol{\mu}}-\boldsymbol{\mu})\in A\}-\mathbb{P}(\boldsymbol{Z}\in A)|\to0.$$
The proof is thus complete.
\end{proof}
\subsection{Proof of main results}
\subsubsection{Proof of Theorem \ref{thm1}}
\begin{proof}
Recall that $\bm{Z}\sim N(0,p\zeta_1^2\mathbf{I}_p)$. Under the null hypothesis, Theorem 1 in  \cite{CLX14} establishes that as $p \to \infty$, we have
$$\mathbb{P}\left(p\zeta_1^2 \max_{1 \leq i \leq p} Z_i^2-2 \log p+\log \log p \leq x\right) \to F(x) = \exp \left(-\frac{1}{\sqrt{\pi}} e^{-x/2} \right),$$
for any $x \in \mathbb{R}$.
Thus, by applying the triangle inequality, using Lemma~\ref{lemma6} and Corollary~\ref{cor1}, we obtain that under the null hypothesis,
\begin{align*}
    &\left| \mathbb{P}\left(n\left\|\hat{\mathbf{\Omega}}^{1 / 2} \hat{\boldsymbol{\mu}}\right\|_{\infty}^2  \hat{\zeta}_{1}^2p-2 \log p+\log \log p \leq x\right) - F(x) \right|\\
    \leq& \left| \mathbb{P}\left( n\left\|\hat{\mathbf{\Omega}}^{1 / 2} \hat{\boldsymbol{\mu}}\right\|_{\infty}^2  {\zeta}_{1}^2p-2 \log p+\log \log p \leq x \right) - F(x) \right| + o(1)\\
    \leq& \left| \mathbb{P}\left( n\left\|\hat{\mathbf{\Omega}}^{1 / 2} \hat{\boldsymbol{\mu}}\right\|_{\infty}^2  {\zeta}_{1}^2p-2 \log p+\log \log p \leq x \right) - \mathbb{P}\left( p\zeta_1^2 \max_{1 \leq i \leq p} Z_i^2 - 2 \log p + \log \log p \leq x \right) \right|\\
    &+\left| \mathbb{P}\left( p\zeta_1^2 \max_{1 \leq i \leq p} Z_i^2 - 2 \log p + \log \log p \leq x \right) - F(x) \right| + o(1) \to 0,
\end{align*}
for any $x \in \mathbb{R}$.
\end{proof}
\subsubsection{Proof of Theorem \ref{thm2}}
\begin{proof}
Under alternative hypothesis for small $\alpha$, we have
\begin{align*}
   \mathbb{P}&\left(T_{MAX}>q_{1-\alpha}\Big|H_1\right) \\    =\mathbb{P}&\left(n\|\hat{\mathbf{\Omega}}^{1/2}\hat{\bm{\mu}}\|_{\infty}^2\hat{\zeta}_1^2p-2\log p+\log\log p>q_{1-\alpha}\Big|H_1\right)\\   =\mathbb{P}&\left(n^{1/2}\|\hat{\mathbf{\Omega}}^{1/2}\hat{\bm{\mu}}\|_{\infty}\hat{\zeta}_1p^{1/2}>(2\log p+\log\log p+q_{1-\alpha})^{1/2}\Big|H_1\right)\\
   \geq \mathbb{P}&\left(n^{1/2}\|\hat{\mathbf{\Omega}}^{1/2}{\bm{\mu}}\|_{\infty}\hat{\zeta}_1p^{1/2}-n^{1/2}\|\hat{\mathbf{\Omega}}^{1/2}(\hat{\bm{\mu}}-\bm{\mu})\|_{\infty}\hat{\zeta}_1p^{1/2}>(2\log p+\log\log p+q_{1-\alpha})^{1/2}\Big|H_1\right)\\
  =\mathbb{P}&\Big(n\|\hat{\mathbf{\Omega}}^{1/2}(\hat{\bm{\mu}}-\bm{\mu})\|_\infty^2\hat{\zeta}_1^2p-2\log p+\log\log p\\
   &\le n\|\hat{\mathbf{\Omega}}^{1/2}\bm{\mu}\|_{\infty}^2\hat{\zeta}_1^2p-2(2\log p+\log\log p+q_{1-\alpha})^{1/2}n^{1/2}\|\hat{\mathbf{\Omega}}^{1/2}\bm{\mu}\|_{\infty}\hat{\zeta}_1p^{1/2}+q_{1-\alpha}\Big|H_1\Big).
\end{align*}
By Lemma \ref{lemma4}, Lemma \ref{lemma6} and Theorem \ref{thm1}, we have
\begin{align*}
     \mathbb{P}&\left(T_{MAX}>q_{1-\alpha}\Big|H_1\right) \\ 
     \le \mathbb{P}&\Big(n\|\hat{\mathbf{\Omega}}^{1/2}(\hat{\bm{\mu}}-\bm{\mu})\|_\infty^2\hat{\zeta}_1^2p-2\log p+\log\log p\\
   &\ge n\|\hat{\mathbf{\Omega}}^{1/2}\bm{\mu}\|_{\infty}^2\hat{\zeta}_1^2p-2(2\log p+\log\log p+q_{1-\alpha})^{1/2}n^{1/2}\|\hat{\mathbf{\Omega}}^{1/2}\bm{\mu}\|_{\infty}\hat{\zeta}_1p^{1/2}+q_{1-\alpha}\Big|H_1\Big)\\
   =\mathbb{P}&\Big(n\|\hat{\mathbf{\Omega}}^{1/2}(\hat{\bm{\mu}}-\bm{\mu})\|_\infty^2{\zeta}_1^2p-2\log p+\log\log p\\
   &\le n\|{\mathbf{\Omega}}^{1/2}\bm{\mu}\|_{\infty}^2{\zeta}_1^2p-2(2\log p+\log\log p+q_{1-\alpha})^{1/2}n^{1/2}\|{\mathbf{\Omega}}^{1/2}\bm{\mu}\|_{\infty}{\zeta}_1p^{1/2}+q_{1-\alpha}+o(1)\Big|H_1\Big)\\
   =F&\left(n\|{\mathbf{\Omega}}^{1/2}\bm{\mu}\|_{\infty}^2{\zeta}_1^2p-2(2\log p+\log\log p+q_{1-\alpha})^{1/2}n^{1/2}\|{\mathbf{\Omega}}^{1/2}\bm{\mu}\|_{\infty}{\zeta}_1p^{1/2}+q_{1-\alpha}+o(1)\right)+o(1)\rightarrow 1,
\end{align*}
when $\|\mathbf{\Omega}^{1/2}\bm{\mu}\|_\infty\ge \widetilde{C}n^{-1/2}\{\log p-2\log\log(1-\alpha)^{-1}\}^{1/2}$.
\end{proof}
\subsubsection{Proof of Theorem \ref{thm3}}
\begin{proof}
By Lemma \ref{lemma2}, we have the Gaussian approximation
\begin{align*}
    \sup _{A \in \mathcal{A}^{\mathrm{re}}}\left|\mathbb{P}\left(n^{1 / 2}p^{1/2}\zeta_1\hat{\mathbf \Omega}^{1/2}\hat{\boldsymbol{\mu}} \in A\right)-\mathbb{P}\left(\boldsymbol{G}+ n^{1 / 2}p^{1/2}\zeta_1\hat{\mathbf \Omega}^{1/2}\bmu\in A\right)\right|\to 0,
\end{align*}
  where $\boldsymbol{G} := p^{1/2} \zeta_1 \boldsymbol{Z} \sim N(0, \I_p)$. Then
    \begin{align*}
    &\sup _{t \in \mathbb{R}}\left|\mathbb{P}\left(\left\|n^{1 / 2}p^{1/2}\zeta_1\hat{\mathbf \Omega}^{1/2}\hat{\boldsymbol{\mu}} \right\|^2\le t\right)-\mathbb{P}\left(\left\|\boldsymbol{G}+ n^{1 / 2}p^{1/2}\zeta_1\hat{\mathbf \Omega}^{1/2}{\boldsymbol{\mu}}\right\|^2\le t\right)\right|\\
    =&\sup _{t \in \mathbb{R}}\left|\mathbb{P}\left(\left\|n^{1 / 2}p^{1/2}\zeta_1\hat{\mathbf \Omega}^{1/2}\hat{\boldsymbol{\mu}} \right\|^2\le t\right)-\mathbb{P}\left\{\chi^2\left(p,\left\| n^{1 / 2}p^{1/2}\zeta_1\hat{\mathbf \Omega}^{1/2}{\boldsymbol{\mu}}\right\|^2\right)\le t\right\}\right|\\
    \to&\sup _{t \in \mathbb{R}}\left|\mathbb{P}\left(\left\|n^{1 / 2}p^{1/2}\zeta_1\hat{\mathbf \Omega}^{1/2}\hat{\boldsymbol{\mu}} \right\|^2\le t\right)-\mathbb{P}\left\{\chi^2\left(p,\left\| n^{1 / 2}p^{1/2}\zeta_1{\mathbf \Omega}^{1/2}{\boldsymbol{\mu}}\right\|^2\right)\le t\right\}\right|\\
    =&\sup _{t \in \mathbb{R}}\bigg|\mathbb{P}\left(\frac{\left\|n^{1 / 2}p^{1/2}\zeta_1\hat{\mathbf \Omega}^{1/2}\hat{\boldsymbol{\mu}} \right\|^2-p-\left\| n^{1 / 2}p^{1/2}\zeta_1{\mathbf \Omega}^{1/2}{\boldsymbol{\mu}}\right\|^2}{\sqrt{2p+4\left\| n^{1 / 2}p^{1/2}\zeta_1{\mathbf \Omega}^{1/2}{\boldsymbol{\mu}}\right\|^2}}\le t\right)\\
    &-\mathbb{P}\left\{\frac{\chi^2\left(p,\left\| n^{1 / 2}p^{1/2}\zeta_1\hat{\mathbf \Omega}^{1/2}{\boldsymbol{\mu}}\right\|^2\right)-p-\left\| n^{1 / 2}p^{1/2}\zeta_1{\mathbf \Omega}^{1/2}{\boldsymbol{\mu}}\right\|^2}{\sqrt{2p+4\left\| n^{1 / 2}p^{1/2}\zeta_1{\mathbf \Omega}^{1/2}{\boldsymbol{\mu}}\right\|^2}}\le t\right\}\bigg|\\
     =&\sup _{t \in \mathbb{R}}\left|\mathbb{P}\left(\frac{\left\|n^{1 / 2}p^{1/2}\zeta_1\hat{\mathbf \Omega}^{1/2}\hat{\boldsymbol{\mu}} \right\|^2-p-\left\| n^{1 / 2}p^{1/2}\zeta_1{\mathbf \Omega}^{1/2}{\boldsymbol{\mu}}\right\|^2}{\sqrt{2p+4\left\| n^{1 / 2}p^{1/2}\zeta_1{\mathbf \Omega}^{1/2}{\boldsymbol{\mu}}\right\|^2}}\le t\right)-\Phi(t)\right|\to0,
\end{align*}
as $(n,p)\to\infty$.
 Therefore, under the null hypothesis, we have $T_{SUM}\cd N(0,1)$; Under the alternative hypothesis, assuming $\left\| n^{1 / 2}p^{1/2}\zeta_1{\mathbf \Omega}^{1/2}{\boldsymbol{\mu}}\right\|^2=o(p)$, we have 
$$T_{SUM}-2^{-1/2}np^{1/2}\zeta_1^2\bm{\mu}^\top\mathbf{\Omega}\bm{\mu}\cd N(0,1).$$ Then we complete the proof.
\end{proof}
\subsubsection{Proof of Theorem \ref{thm5}}
\begin{proof}
    Recall that Corollary \ref{cor1}, as $n\to\infty$, we have
    $$\tilde\rho_{n,comb}=\sup _{t_1,t_2 \in \mathbb{R}}\left|\mathbb{P}\left(n^{1 / 2}\Vert\hat{\mathbf \Omega}^{1/2}(\hat{\boldsymbol{\mu}}-\boldsymbol{\mu})\Vert_{\infty} \leqslant t_1, n^{1 / 2}\Vert\hat{\mathbf \Omega}^{1/2}(\hat{\boldsymbol{\mu}}-\boldsymbol{\mu})\Vert \leqslant t_2\right)-\mathbb{P}\left(\Vert \boldsymbol Z\Vert_{\infty} \leqslant t_1, \Vert \boldsymbol Z\Vert \leqslant t_2\right)\right| \rightarrow 0.$$
    By $\|\bm{\mu}\|_\infty=o(n^{-1/2})$, $\|\bm{\mu}\|=o(p^{1/4}n^{-1/2})$, Assumption \ref{assu3} and Lemma \ref{lemma4} we have 
    \begin{align*}
        \sup _{t_1,t_2 \in \mathbb{R}}\Big|\mathbb{P}\left(n^{1 / 2}p^{1/2}\zeta_1\Vert\hat{\mathbf \Omega}^{1/2}\hat{\boldsymbol{\mu}}\Vert_{\infty}+o(1) \leqslant t_1, n^{1 / 2}p^{1/2}\zeta_1\Vert\hat{\mathbf \Omega}^{1/2}\hat{\boldsymbol{\mu}}\Vert+o(p^{1/2}) \leqslant t_2\right)\\
        -\mathbb{P}\left(p^{1/2}\zeta_1\Vert \boldsymbol Z\Vert_{\infty} \leqslant t_1, p^{1/2}\zeta_1\Vert \boldsymbol Z\Vert \leqslant t_2\right)\Big| \rightarrow 0.
    \end{align*}
    Hence, applying the continuous mapping theorem, we obtain that
     \begin{align*}
     \sup _{t_1,t_2 \in \mathbb{R}}\bigg|&\mathbb{P}\left(T_{MAX}+o(1) \leqslant t_1, T_{SUM}+o(1)\leqslant t_2\right)\\
     &-\mathbb{P}\left(p\zeta_1^2\Vert \boldsymbol Z\Vert_{\infty}^2-2\log p+\log\log p \leqslant t_1,  (2p)^{-1/2}(p\zeta_1^2\Vert\boldsymbol Z\Vert^2-p) \leqslant t_2\right)\bigg| \rightarrow 0.
     \end{align*}
     By Theorem 3 in \cite{Feng2022AsymptoticIO}, we have $p^{1/2}\zeta_1\Vert \boldsymbol Z\Vert_{\infty}^2-2\log p+\log\log p$ and $(2p)^{-1/2}(p\zeta_1^2\Vert\boldsymbol Z\Vert^2-p)$ are asymptotic independent as $p\to\infty$, so we have $T_{MAX}$ and $T_{SUM}$ are asymptotic independent as $n,p\to\infty$.
\end{proof}
\subsubsection{Proof of Theorem \ref{thm6}}
\begin{proof}
    From the proof of Theorem 7 in \cite{liu2024spatial} we can find that
    $$T_{SUM2}=\frac{2}{n(n-1)}\sum\sum_{i<j}\bm{\tilde{U}}_{i}^{\top}\bm{\tilde{U}}_{j}+\zeta_{1}^{2}\boldsymbol{\mu}^{\top}\mathbf{D}^{-1}\boldsymbol{\mu}+o_{p}(\sigma_{n}),$$
    with $\sigma^2_{n}=2/\{n(n-1)p\}+o(n^{-3}),$ and
    $$n^{1/2}\mathbf{D}^{-1/2}\hat{\boldsymbol{\mu}}=n^{-1/2}\zeta_1^{-1}\sum_{i=1}^n\boldsymbol{\tilde{U}}_i+n^{1/2}\mathbf{D}^{-1/2}\boldsymbol{\mu}+C_n,$$
    where $\bm{\tilde{U}}_{i}:=U(\mathbf{D}^{-1/2}\mathbf{\Sigma}^{1/2}\bm{U}_i)=(\mathbf{R}^{1/2}\bm{U}_i)/\|\mathbf{R}^{1/2}\bm{U}_i\|$, $\mathbf{R}=\mathbf{D}^{-1/2}\bm{\Sigma}\mathbf{D}^{-1/2}$ Thus, we have
    \begin{equation*}
        \frac{T_{SUM2}}{\sigma_n}=\frac{p}{n\sqrt{2\tr(\mathbf{R}^2)}}\left(\bigg\|\sum_{i=1}^n \bm{\tilde{U}}_i\bigg\|^2-n\right)+O(1)=\frac{\|p^{1/2}n^{-1/2}\sum_{i=1}^n\bm{\tilde{U}}_i\|^2-p}{\sqrt{2\mathrm{tr}(\mathbf{R}^2)}}+O(1),
    \end{equation*}
    and
    \begin{equation*}
        T_{MAX2}-2\log p+\log\log p=\bigg\|\sqrt{\frac{p}{n}}\sum_{i=1}^n \bm{\tilde{U}}_i\bigg\|_\infty^2-2\log p+\log\log p.
    \end{equation*}
    Notice that $p^{1/2}n^{-1/2}\sum_{i=1}^n \bm{\tilde{U}}_i=p^{1/2}n^{-1/2}\sum_{i=1}^n \mathbf{R}^{1/2}\bm{U}_i+p^{1/2}n^{-1/2}\sum_{i=1}^n\mathbf{R}^{1/2}\bm{U}_i(1/\|\mathbf{R}^{1/2}\bm{U}_i\|-1)$. Denote $v_i=1/\|\mathbf{R}^{1/2}\bm{U}_i\|-1$, $\mathrm{Var}(v_i)=\sigma_v^2$. We have
   \begin{equation*}
        \|\sqrt{\frac{p}{n}}\sum_{i=1}^n v_i\mathbf{R}^{1/2}\bm{U}_i\|_\infty=\sqrt{\frac{p}{n}}\max_{i\le j\le p}\big|\sum_{i=1}^nv_iU_{ij}\big|\le \frac{1}{\sqrt{n}}\sum_{i=1}^n \big|v_i\big|\max_{i\le j\le p}\sqrt{p}\big|U_{ij}\big|=O_p(\sigma_v\log p). \end{equation*}
    For a random variable $X$, denote $X^{\star}=(X-\mathbb{E}X)/\sqrt{\mathrm{Var}(X)}$. Thus the original proposition is equivalent to proving that $\|\sum_{i=1}^p\bm{U}_i\|_\infty^\star$ is asymptotically independent with $\|\mathbf{R}^{1/2}\sum_{i=1}^p\bm{U}_i\|^\star$ and $\|\sum_{i=1}^p\bm{U}_i\|^\star$ is asymptotically independent with $\|\mathbf{R}^{1/2}\sum_{i=1}^p\bm{U}_i\|_\infty^\star$. 
    Then for any sequence $\eta_{n,p}\to \infty$ and any $t\in\mathbb{R}^p$
    \begin{align*}
        \mathbb{P}\left(\sqrt{\frac{p}{n}}\sum_{i=1}^n \bm{\tilde{U}}_i\le t\right)=&\mathbb{P}\left(\sqrt{\frac{p}{n}}\sum_{i=1}^n \mathbf{R}^{1/2}\bm{{U}}_i+\sqrt{\frac{p}{n}}\sum_{i=1}^n v_i\mathbf{R}^{1/2}\bm{{U}}_i\le t\right)\\
        \le &  \mathbb{P}\left(\sqrt{\frac{p}{n}}\sum_{i=1}^n \mathbf{R}^{1/2}\bm{{U}}_i\le t+\eta_{n,p}\sigma_v\log p\right)+\mathbb{P}\left(\bigg\|\sqrt{\frac{p}{n}}\sum_{i=1}^n v_i\mathbf{R}^{1/2}\bm{{U}}_i\bigg\|_\infty>\eta_{n,p}\sigma_v\log p\right)\\
        \le &\mathbb{P}\left(\sqrt{\frac{p}{n}}\sum_{i=1}^n \mathbf{R}^{1/2}\bm{{U}}_i\le t\right)+o(1)
    \end{align*}
    Similarly, we have $ \mathbb{P}\left(\sqrt{\frac{p}{n}}\sum_{i=1}^n \bm{\tilde{U}}_i\le t\right)\ge\mathbb{P}\left(\sqrt{\frac{p}{n}}\sum_{i=1}^n \mathbf{R}^{1/2}\bm{{U}}_i\le t\right)+o(1) $. We have 
    \begin{equation*}
        \sup_{t\in \mathbb{R}^p}\left|\mathbb{P}\left(\sqrt{\frac{p}{n}}\sum_{i=1}^n \bm{\tilde{U}}_i\le t\right)-\mathbb{P}\left(\sqrt{\frac{p}{n}}\sum_{i=1}^n \mathbf{R}^{1/2}\bm{{U}}_i\le t\right)\right|\to 0.
    \end{equation*}
    Further, 
    \begin{equation*}
    \sup_{A\in\mathcal{A}^{si}}\left|\mathbb{P}\left(\sqrt{\frac{p}{n}}\sum_{i=1}^n \bm{\tilde{U}}_i\in A\right)-\mathbb{P}\left(\sqrt{\frac{p}{n}}\sum_{i=1}^n \mathbf{R}^{1/2}\bm{{U}}_i\in A\right)\right|\to 0.
    \end{equation*}
    From the proof of Lemma \ref{lemma2} we have 
     \begin{equation*}
    \sup_{A\in\mathcal{A}^{si}}\left|\mathbb{P}\left(\sqrt{\frac{p}{n}}\sum_{i=1}^n \bm{{U}}_i\in A\right)-\mathbb{P}\left(\bm{Z}\in A\right)\right|\to 0,
    \end{equation*}
    where $\bm{Z}\sim N(0,\mathbf{I}_p)$. Thus 
     \begin{equation*}
    \sup_{A_1,A_2\in\mathcal{A}^{si}}\left|\mathbb{P}\left(\sqrt{\frac{p}{n}}\sum_{i=1}^n \mathbf{R}^{1/2}\bm{{U}}_i\in A_1, \sqrt{\frac{p}{n}}\sum_{i=1}^n \bm{{U}}_i\in A_2\right)-\mathbb{P}\left(\mathbf{R}^{1/2}\bm{Z}\in A_1,\bm{Z}\in A_2\right)\right|\to 0,
    \end{equation*}
     Thus the original proposition is equivalent to proving that $\|\bm{Z}\|_\infty^\star$ is asymptotically independent with $\|\mathbf{R}^{1/2}\bm{Z}\|^\star$ and $\|\mathbf{R}^{1/2}\bm{Z}\|_\infty^\star$ is asymptotically independent with $\|\bm{Z}\|^\star$. From Theorem 2.2 in \cite{chen2024asymptotic} we have $\|\bm{Z}\|_\infty^\star$ is asymptotically independent with $\|\mathbf{R}^{1/2}\bm{Z}\|^\star$. Consider that $\mathbf{R}^{1/2}\bm{Z}\sim N(\bm{0},\mathbf{R})$. Next we prove $\|\bm{Y}\|_\infty^\star$ is asymptotically independent with $\|\mathbf{R}^{-1/2}\bm{Y}\|^\star$ where $\bm{Y}=\mathbf{R}^{1/2}\bm{Z}\sim N(\bm{0},\mathbf{R})$.
     
 Define $\bm{Y}=(\bm{Y}_1^\top,\bm{Y}_2^\top)^\top\in\mathbb{R}^p$ where $\bm{Y}_1=(Y_1,\ldots,Y_d)^\top$ and $\bm{Y}_2=(Y_{d+1},\ldots,Y_p)^\top.$ And
 $$\mathbf{R}=\left(\begin{array}{cc}\mathbf{R}_{1}&\mathbf{R}_{12}\\\\\mathbf{R}_{21}&\mathbf{R}_{2}\end{array}\right).$$
$$\mathbf{R}^{-1}:=\mathbf{P}=\left(\begin{array}{cc}\mathbf{P}_{1}&\mathbf{P}_{12}\\\\\mathbf{P}_{21}&\mathbf{P}_{2}\end{array}\right).$$
$$\mathbf{K}=\left(\begin{array}{cc}\mathbf{K}_{1}&\mathbf{K}_{12}\\\\\mathbf{K}_{21}&\mathbf{K}_{2}\end{array}\right):=\left(\begin{array}{cc}\mathbf{R}_1^{1/2}\mathbf{P}_{1}\mathbf{R}_1^{1/2}&\mathbf{R}_1^{1/2}\mathbf{P}_{12}\mathbf{R}_2^{1/2}\\\\\mathbf{R}_2^{1/2}\mathbf{P}_{21}\mathbf{R}_1^{1/2}&\mathbf{R}_2^{1/2}\mathbf{P}_{2}\mathbf{R}_2^{1/2}\end{array}\right).$$

So,

$$\bm{Y}^\top \mathbf{P}\bm{Y}=\bm{Y}_1^\top \mathbf{P}_1\bm{Y}_1+2\bm{Y}_1^\top \mathbf{P}_{12}\bm{Y}_2+\bm{Y}_2^\top \mathbf{P}_2\bm{Y}_2.$$


For $\epsilon>0,$ set $z_i$ are i.i.d. Gaussian random variables. Define $\bm{z}_1=(z_1,\cdots,z_d)^\top\in\mathbb{R}^d$, $\bm{z}_1=(z_{d+1},\cdots,z_p)^\top\in\mathbb{R}^{p-d}$. Then there exist $\eta>0$ and $K>0$ such that $\E\{\exp(\eta z_i^2)\}\leq K.$ According to Assumptions of Theorem 7 in \cite{liu2024spatial}, we can get $\lambda_{\max}(\mathbf{K}_1)\leq\lambda_{\max}(\mathbf{K})<c_1$ for a constant $c_1 > 0$.

$$\begin{aligned}\pr(\bm{Y}_{1}^{\top}\mathbf{P}_{1}\bm{Y}_{1}>\sqrt{2p}\epsilon)&\leq \pr(c_1\bm{z}_{1}^{\top}\bm{z}_{1}>\sqrt{2p}\epsilon)\\
&=\pr\biggl(\eta\sum_{i=1}^{d}{z}_{i}^{2}>\sqrt{2p}\epsilon c_1^{-1}\eta\epsilon\biggr)\\
&\leq\exp(-\sqrt{2p}\epsilon c_1^{-1}\eta\epsilon)\E(\mathrm{e}^{\eta\sum_{i=1}^{d}z_{i}^{2}})\\&=\exp(-\sqrt{2p}\epsilon c_1^{-1}\eta\epsilon)\{E(\mathrm{e}^{\eta z_{i}^{2}})\}^{d}\\&\leq K^{d}\exp(-\sqrt{2p}\epsilon c_1^{-1}\eta\epsilon).\end{aligned}$$

Define $\mathbf{K}=\mathbf{O}^{\top}\mathbf{\Lambda} \mathbf{O}$ where $\mathbf{O}=(q_{ij})_{1\leq i,j\leq p}$ is an orthogonal matrix and $\mathbf{\Lambda}=\operatorname{diag}\{\lambda_{1},\ldots,\lambda_{p}\},\lambda_{i}$, $i=1,\ldots,p$ are the eigenvalues of $\mathbf{K}.$ Note that $\sum_{1\leq j\leq p}k_{ij}^2$ is the $i$-th diagonal element of $\mathbf{K}^{2}=\mathbf{O}^{\top}\mathbf{\Lambda}^2 \mathbf{O}.$ We have$\sum_{1\leq j\leq p}k_{ij}^{2}=\sum_{l=1}^{p}q_{li}^{2}\lambda_{l}^{2}\leq c_1^{2}$ .Next,  define $\theta = \sqrt {(2\eta)/(dc_1^2)}.$ We have 
\begin{align*}
  \pr(\bm{Y}_{1}^{\top}\mathbf{P}_{12}\bm{Y}_{2}\geq\sqrt{2p}\epsilon)&\leq\exp(-\sqrt{2p}\theta\epsilon)\E(\exp(\theta \bm{z}_{1}^{\top}\mathbf{K}_{12}\bm{z}_{2}) \\
 & =\exp(-\sqrt{2p}\theta\epsilon)\E(\mathrm{e}^{\theta\sum_{i=1}^{d}\sum_{j=d+1}^{p}k_{ij}z_{i}z_{j}}) \\
 & =\exp(-\sqrt{2p}\theta\epsilon)\E\{\E(\mathrm{e}^{\theta\sum_{j=d+1}^{p}(\sum_{i=1}^{d}k_{ij}z_{i})z_{j}}|\bm{z}_1)\} \\
 & =\exp(-\sqrt{2p}\theta\epsilon)\E\left[\prod_{j=d+1}^p\E\{\mathrm{e}^{(\theta\sum_{i=1}^dk_{ij}z_i)z_j}|\bm{z}_1\}\right] \\
 & \leq\exp(-\sqrt{2p}\theta\epsilon)\E\left[\prod_{j=d+1}^{p}\exp\left\{\frac{\theta^{2}}{2}\left(\sum_{i=1}^{d}k_{ij}z_{i}\right)^{2}\right\}\right] \\
 & =\exp(-\sqrt{2p}\theta\epsilon)\E\left[\exp\left\{\frac{\theta^2}{2}\sum_{j=d+1}^p\left(\sum_{i=1}^dk_{ij}z_i\right)^2\right\}\right]\\
 & \leq\exp(-\sqrt{2p}\theta\epsilon)\E\left\{\exp\left(\frac{d\theta^2}{2}\sum_{j=d+1}^p\sum_{i=1}^dk_{ij}^2z_i^2\right)\right\} \\
 & \leq\exp(-\sqrt{2p}\theta\epsilon)\E\left\{\exp\left(\frac{dc_1^2\theta^2}{2}\sum_{i=1}^dz_i^2\right)\right\} \\
 & =\exp(-\sqrt{2p}\theta\epsilon)\E\left\{\exp\left(\eta\sum_{i=1}^dz_i^2\right)\right\} \\
 & \leq K^d\exp(-\sqrt{2p}\theta\epsilon).
 \end{align*}

 So

$$\pr(\bm{Y}_1^\top \mathbf{P}_{12}\bm{Y}_2\ge\sqrt{2p}\epsilon)\le K^d\exp\bigg(-\sqrt{\frac{4\eta}{dc_1^4}}\epsilon p^{1/2}\bigg).$$ 

Similarly, we also can prove that

$$\pr\big\{(-\bm{Y}_1)^\top \mathbf{P}_{12}\bm{Y}_2\ge\sqrt{2p}\epsilon\big\}\le K^d\exp\bigg(-\sqrt{\frac{4\eta}{dc_1^4}}\epsilon p^{1/2}\bigg).$$

Let $\Theta _{p}= \bm{Y}_{1}^{\top }\mathbf{P}_{1}\bm{Y}_{1}+ 2\bm{Y}_{1}^{\top }\mathbf{P}_{12}\bm{Y}_{2}.$

\begin{align*}\pr(|\Theta_{p}|>\sqrt{2p}\epsilon)&\leq \pr(\bm{Y}_{1}^{\top}\mathbf{P}_{1}\bm{Y}_{1}>\sqrt{2p}\epsilon/2)+\pr(|\bm{Y}_{1}^{\top}\mathbf{P}_{12}\bm{Y}_{2}|>\sqrt{2p}\epsilon/4)\\&\leq \pr(\bm{Y}_{1}^{\top}\mathbf{P}_{1}\bm{Y}_{1}>\sqrt{2p}\epsilon/2)+\pr(\bm{Y}_{1}^{\top}\mathbf{P}_{12}\bm{Y}_{2}>\sqrt{2p}\epsilon/8)+\pr(-\bm{Y}_{1}^{\top}\mathbf{P}_{12}\bm{Y}_{2}>\sqrt{2p}\epsilon/8).\end{align*}
Denote $A_p(x)=\left\{\frac{\bm{Y}\mathbf{R}^{-1}\bm{Y}-p}{\sqrt{2p}}\right\}\le x$, $B_i=\left\{|Y_1|\ge \sqrt{2\log p-\log\log p}\right\}$, so there exist a constant $c_\epsilon>0$
\begin{align*}
 & \pr(|\Theta_{p}|>\sqrt{2p}\epsilon)\leq K^{d}\exp(-c_{\epsilon}p^{1/2}), \\
 & \pr(A_p(x)B_1\cdots B_d) \\
 & =\pr\left(\frac{\bm{Y}_{2}^{\top}\mathbf{P}_{2}\bm{Y}_{2}-p+\Theta_{p}}{\sqrt{2p}}\leq x,B_{1}\cdots B_{d}\right) \\
 & \leq \pr\left(\frac{\bm{Y}_{2}^{\top}\mathbf{P}_{2}\bm{Y}_{2}-p+\Theta_{p}}{\sqrt{2p}}\leq x,|\Theta_{p}|\leq\sqrt{2p}\epsilon,B_{1}\cdots B_{d}\right)+\pr(|\Theta_{p}|>\sqrt{2p}\epsilon) \\
 & \leq \pr\left(\frac{\bm{Y}_{2}^{\top}\mathbf{P}_{2}\bm{Y}_{2}-p}{\sqrt{2p}}\leq x+\epsilon,B_{1}\cdots B_{d}\right)+K^{d}\exp(-c_{\epsilon}p^{1/2}) \\
 &  =\pr\left(\frac{\bm{Y}_{2}^{\top}\mathbf{P}_{2}\bm{Y}_{2}-p}{\sqrt{2p}}\leq x+\epsilon\right)P(B_{1}\cdots B_{d})+K^{d}\exp(-c_{\epsilon}p^{1/2}) \\
 & \leq\left\{\pr\left(\frac{\bm{Y}_{2}^{\top}\mathbf{P}_{2}\bm{Y}_{2}-p}{\sqrt{2p}}\leq x+\epsilon,|\Theta_{p}|\leq\sqrt{2p}\epsilon\right)+\pr(|\Theta_{p}|>\sqrt{2p}\epsilon)\right\}P(B_{1}\cdots B_{d}) \\
 & +K^d\exp(-c_\epsilon p^{1/2}) \\
 &\leq \pr\left(\frac{\bm{Y}_{2}^{\top}\mathbf{P}_{2}\bm{Y}_{2}-p+\Theta_{p}}{\sqrt{2p}}\leq x+2\epsilon\right)\pr(B_{1}\cdots B_{d})+2K^{d}\exp(-c_{\epsilon}p^{1/2}) \\
 & =\pr\{A_p(x+2\epsilon)\}\pr(B_{1}\cdots B_{d})+2K^{d}\exp(-c_{\epsilon}p^{1/2}).
\end{align*}
Similarly, we can prove that

$$\pr(A_p(x)B_1\cdots B_d)\geq \pr\{A_p(x-2\epsilon)\}\pr(B_1\cdots B_d)-2K^d\exp(-c_\epsilon p^{1/2}).$$

So, we have

$$|\pr(A_{p}(x)B_{1}\cdots B_{d})-\pr\{A_{p}(x)\}\cdot \pr(B_{1}\cdots B_{d})|\\\leq\Delta_{p,\epsilon}\cdot \pr(B_{1}\cdots B_{d})+2K^{d}\exp(-c_{\epsilon}p^{1/2}),$$

where

$$\begin{aligned}\Delta_{p,\epsilon}&=|\pr\{A_{p}(x)\}-\pr(A_{p}(x+2\epsilon))|+|\pr\{A_{p}(x)\}-\pr\{A_{p}(x-2\epsilon)\}|\\&=\pr\{A_{p}(x+2\epsilon)\}-\pr\{A_{p}(x-2\epsilon)\}.\end{aligned}$$

Obviously, the equation discussed above holds for all $i_1,\ldots,i_d.$ Thus,

$$\begin{aligned}&\sum_{1\leq i_{1}<\cdots<i_{d}\leq p}|\pr(A_{p}(x)B_{i_{1}}\cdots B_{i_{d}})-\pr\{A_{p}(x)\}\cdot \pr(B_{i_{1}}\cdots B_{i_{d}})|\\&\leq\sum_{1\leq i_{1}<\cdots<i_{d}\leq p}[\Delta_{p,\epsilon}\cdot \pr(B_{i_{1}}\cdots B_{i_{d}})+2K^{d}\exp(-c_{\epsilon}p^{1/2})]\\&\leq\Delta_{p,\epsilon}\cdot H(d,p)+\left(\begin{array}{c}p\\d\end{array}\right)\cdot2K^{d}\exp(-c_{\epsilon}p^{1/2}).\end{aligned}$$

Because $\pr\{A_p(x)\}\rightarrow \Phi(x)$ as $p \rightarrow \infty$. So $\Delta_{p,\epsilon} \rightarrow \Phi(x+2\epsilon) - \Phi(x-2\epsilon)$. By letting $\epsilon \rightarrow 0$, we have $\Delta_{p,\epsilon} \rightarrow 0$. By Lemma \ref{lemma10} as $p\to \infty$ we have
\begin{equation*}
   \sum_{1 \leq i_1 < \cdots < i_d \leq p} |\pr(A_p(x)B_{i_1}\cdots B_{i_d}) - \pr\{A_p(x)\}\cdot \pr(B_{i_1}\cdots B_{i_d})| \to 0
\end{equation*}

Then, repeat the procedure in proof of Theorem 2.2 in \cite{chen2024asymptotic} we have
\begin{align*}
\limsup_{p\to\infty} \pr\left(\|\mathbf{R}^{-1/2}\bm{Y}\|^{2*}\leq x,\|\bm{Y}\|_\infty^{*2}\leq y\right)&=
    \limsup_{p\to\infty} \pr\left(\frac{\bm{Y}^\top \mathbf{R}^{-1}\bm{Y}-p}{\sqrt{2p}}\leq x,\max_{1\leq i\leq p}|Y_i|>l_p\right)\\
    &\leq\Phi(x)\cdot[1-F(y)]+\lim_{p\to\infty}H(p,2k+1),
\end{align*}
\begin{align*}
\liminf_{p\to\infty} \pr\left(\|\mathbf{R}^{-1/2}\bm{Y}\|^{2*}\leq x,\|\bm{Y}\|_\infty^{*2}\leq y\right)&=
    \liminf_{p\to\infty}\pr\left(\frac{\bm{Y}^\top \mathbf{R}^{-1}\bm{Y}-p}{\sqrt{2p}}\leq x,\max_{1\leq i\leq p}|Y_i|>l_p\right)\\
    &\leq\Phi(x)\cdot[1-F(y)]-\lim_{p\to\infty}H(p,2k+1).
\end{align*}

Then we can get $\|\bm{Y}\|_\infty^\star$ is asymptotically independent with $\|\mathbf{R}^{-1/2}\bm{Y}\|^\star$ by sending $p\to \infty$ and then sending  $k\to \infty$.
\end{proof}
\subsubsection{Proof of Theorem \ref{thm7}}
\begin{proof}
    Set $Q(\bm{x})=\Delta_d^2(\bm{x})-c\varsigma_p$ and $\hat{Q}(\bm{x})=\hat{\Delta}_d^2(\bm{x})-c\hat{\varsigma}_p$. Thus we have
\begin{align*}
    R_{HRQDA}-R_{QDA}=&\int _{\hat{Q}< 0 }\frac{1}{2}f_1(\bm{x})\mathrm{d}\bm{x}+  \int _{\hat{Q}\ge  0 }\frac{1}{2}f_2(\bm{x})\mathrm{d}\bm{x}-\left ( \int _{{Q}< 0 }\frac{1}{2}f_1(\bm{x})\mathrm{d}\bm{x}+  \int _{{Q}\ge  0 }\frac{1}{2}f_2(\bm{x})\mathrm{d}\bm{x} \right ) \\
    =& \int_{Q(\bm{x})\ge 0}\frac{1}{2}\left\{f_1(\bm{x})-f_2(\bm{x})\right\}\mathrm{d}\bm{x}+\int_{\hat{Q}(\bm{x})< 0}\frac{1}{2}\left\{f_1(\bm{x})-f_2(\bm{x})\right\}\mathrm{d}\bm{x}.
\end{align*}

Notice that $\int \frac{1}{2}\{f_1(\bm{x})-f_2(\bm{x})\}\mathrm{d}\bm{x}=0$, we have
\begin{equation}\label{eq:hrqdaorder}
    \begin{aligned}
    |R_{HRQDA}-R_{QDA}|&=\Big|\int_{Q(\bm{x})\ge 0, \hat{Q}(\bm{x})<0}\frac{1}{2}\left\{f_1(\bm{x})-f_2(\bm{x})\right\}\mathrm{d}\bm{x}\Big|\\
    &\le \frac{1}{2}\mathbb{E}_{\bm{x}\sim f_1}\bm{1}\{0\le Q(\bm{x})< Q(\bm{x})- \hat{Q}(\bm{x})\}+\frac{1}{2}\mathbb{E}_{\bm{x}\sim f_2}\bm{1}\{0\le Q(\bm{x})< Q(\bm{x})- \hat{Q}(\bm{x})\}\\
    &=\frac{1}{2}\mathbb{P}_{x\sim f_1}\left\{0\le \frac{1}{p}Q(\bm{x})<\frac{1}{p}M(\bm{x})\right\}+\frac{1}{2}\mathbb{P}_{x\sim f_2}\left\{0\le \frac{1}{p}Q(\bm{x})<\frac{1}{p}M(\bm{x})\right\},
\end{aligned}
\end{equation}
where $M(\bm{x}):=Q(\bm{x})- \hat{Q}(\bm{x})$. By calculations, we can get \begin{align*}
    M(\bm{x})=&(\bm{x}-\bm{\mu}_1)^\top\{\bm{\Xi}^{-1}_2-\bm{\Xi}^{-1}_1-(\tilde{\bm{\Omega}}_2-\tilde{\bm{\Omega}}_1)\}(\bm{x}-\bm{\mu}_1)-2(\bm{\mu}_1-\hat{\bm{\mu}})^\top(\tilde{\bm{\Omega}}_2-\tilde{\bm{\Omega}}_1)(\bm{x}-\bm{\mu}_1)+2(\bm{\delta}^\top\bm{\Xi}^{-1}_2-\bm{\hat{\delta}}^\top\bm{\tilde{\Omega}}_2)(\bm{x}-\bm{\mu}_1)\\
    &+(\bm{\mu}_1-\hat{\bm{\mu}}_1)^\top(\tilde{\bm{\Omega}}_2-\tilde{\bm{\Omega}}_1)(\bm{\mu}_1-\hat{\bm{\mu}}_1)-2\hat{\bm{\delta}}^\top\tilde{\bm{\Omega}}_2(\bm{\mu}_1-\hat{\bm{\mu}}_1)+\bm{\delta}^\top\bm{\Xi}^{-1}_2\bm{\delta}-\hat{\bm{\delta}}^\top\bm{\tilde{\Omega}}_2\bm{\hat{\delta}}-c(\varsigma_p-\hat{\varsigma}_p).
\end{align*}

Next we calculate the variance of $\frac{1}{p}Q(\bm{x})$
\begin{align*}
    \mathrm{Var}_{\bm{x}\sim f_1}\left\{\frac{1}{p}Q(\bm{x})\right\}=&\mathrm{Var}_{\bm{x}\sim f_1}\left\{\frac{1}{p}(\bm{x}-\bm{\mu}_1)^\top(\bm{\Xi}^{-1}_2-\bm{\Xi}^{-1}_1)(\bm{x}-\bm{\mu}_1)-\frac{1}{p}2\bm{\delta}^\top\bm{\Xi}^{-1}_2(\bm{x}-\bm{\mu}_1)+\frac{1}{p}\bm{\delta}^\top\bm{\Xi}^{-1}_2\bm{\delta}-\frac{c}{p}\varsigma_p\right\}\\
    =&\mathrm{Var}_{\bm{x}\sim f_1}\left\{\frac{1}{p}(\bm{x}-\bm{\mu}_1)^\top(\bm{\Xi}^{-1}_2-\bm{\Xi}^{-1}_1)(\bm{x}-\bm{\mu}_1)-\frac{1}{p}2\bm{\delta}^\top\bm{\Xi}^{-1}_2(\bm{x}-\bm{\mu}_1)\right\}\\
    =&\mathbb{E}_{\bm{x}\sim f_1}\Bigg[\left\{\frac{1}{p}(\bm{x}-\bm{\mu}_1)^\top(\bm{\Xi}^{-1}_2-\bm{\Xi}^{-1}_1)(\bm{x}-\bm{\mu}_1)\right\}^2+\left\{\frac{1}{p}2\bm{\delta}^\top\bm{\Xi}^{-1}_2(\bm{x}-\bm{\mu}_1)\right\}^2\\
    &-2\left\{\frac{1}{p}(\bm{x}-\bm{\mu}_1)^\top(\bm{\Xi}^{-1}_2-\bm{\Xi}^{-1}_1)(\bm{x}-\bm{\mu}_1)\right\}\left\{\frac{1}{p}2\bm{\delta}^\top\bm{\Xi}^{-1}_2(\bm{x}-\bm{\mu}_1)\right\}\Bigg]\\
    &-\left[\mathbb{E}_{\bm{x}\sim f_1}\left\{\frac{1}{p}(\bm{x}-\bm{\mu}_1)^\top(\bm{\Xi}^{-1}_2-\bm{\Xi}^{-1}_1)(\bm{x}-\bm{\mu}_1)\right\}\right]^2\\
    =&\mathbb{E}_{\bm{x}\sim f_1}\Bigg[\left\{\frac{1}{p}(\bm{x}-\bm{\mu}_1)^\top(\bm{\Xi}^{-1}_2-\bm{\Xi}^{-1}_1)(\bm{x}-\bm{\mu}_1)\right\}^2+\left\{\frac{1}{p}2\bm{\delta}^\top\bm{\Xi}^{-1}_2(\bm{x}-\bm{\mu}_1)\right\}^2\Bigg]\\
    &-\left[\mathbb{E}_{\bm{x}\sim f_1}\left\{\frac{1}{p}(\bm{x}-\bm{\mu}_1)^\top(\bm{\Xi}^{-1}_2-\bm{\Xi}^{-1}_1)(\bm{x}-\bm{\mu}_1)\right\}\right]^2\\
    =&\mathrm{Var}_{x\sim f_1}\left\{\frac{1}{p}(\bm{x}-\bm{\mu}_1)^\top(\bm{\Xi}^{-1}_2-\bm{\Xi}^{-1}_1)(\bm{x}-\bm{\mu}_1)\right\}+\mathbb{E}_{\bm{x}\sim f_1}\left\{\frac{1}{p}2\bm{\delta}^\top\bm{\Xi}^{-1}_2(\bm{x}-\bm{\mu}_1)\right\}^2\\
    =&\mathrm{Var}\left\{\frac{r^2}{p}\bm{U}^\top\mathbf{\Xi}^{1/2}_1(\bm{\Xi}^{-1}_2-\bm{\Xi}^{-1}_1)\mathbf{\Xi}^{1/2}_1\bm{U}\right\}+\mathrm{Var}\left\{\frac{2r}{p}\bm{\delta}^\top\bm{\Xi}^{-1}_2\mathbf{\Xi}^{1/2}_1\bm{U}\right\}\\
    =&\mathbb{E}\left(\frac{r^4}{p^2}\right)\mathrm{Var}\left\{\bm{U}^\top\mathbf{\Xi}^{1/2}_1(\bm{\Xi}^{-1}_2-\bm{\Xi}^{-1}_1)\mathbf{\Xi}^{1/2}_1\bm{U}\right\}+\mathbb{E}\left(\frac{4r^2}{p}\right)\mathrm{Var}\left(\frac{1}{\sqrt{p}}\bm{\delta}^\top\bm{\Xi}^{-1}_2\mathbf{\Xi}^{1/2}_1\bm{U}\right)\\
    &+\mathrm{Var}\left(\frac{r^2}{p}\right)\left[\mathbb{E}\left\{\bm{U}^\top\mathbf{\Xi}^{1/2}_1(\bm{\Xi}^{-1}_2-\bm{\Xi}^{-1}_1)\mathbf{\Xi}^{1/2}_1\bm{U}\right\}\right]^2+\mathrm{Var}\left(\frac{2r}{\sqrt{p}}\right)\left\{\mathbb{E}\left(\frac{1}{\sqrt{p}}\bm{\delta}^\top\bm{\Xi}^{-1}_2\mathbf{\Xi}^{1/2}_1\bm{U}\right)\right\}^2.
\end{align*}
By Assumptions \ref{assu1}, \ref{assu6} and Lemma \ref{lemma11} we have
$$\mathrm{Var}_{\bm{x}\sim f_1}\left\{\frac{1}{p}Q(\bm{x})\right\}\asymp \frac{1}{p^2}\left\{\|\mathbf{\Xi}^{1/2}_1(\bm{\Xi}^{-1}_2-\bm{\Xi}^{-1}_1)\mathbf{\Xi}^{1/2}_1\|_F^2+\|\bm{\Xi}^{-1}_2\mathbf{\Xi}^{1/2}_1\bm{\delta}\|^2\right\}.$$
By Assumptions \ref{assu2} and \ref{assu3} we have $\|\mathbf{\Xi}^{1/2}_1(\bm{\Xi}^{-1}_2-\bm{\Xi}^{-1}_1)\mathbf{\Xi}^{1/2}_1\|_F\asymp\|\mathbf{\Xi}^{1/2}_2(\bm{\Xi}^{-1}_1-\bm{\Xi}^{-1}_2)\mathbf{\Xi}^{1/2}_2\|_F\asymp\|\mathbf{\Sigma}^{1/2}_1(\bm{\Omega}_2-\bm{\Omega}_1)\mathbf{\Sigma}^{1/2}_1\|_F\asymp\|\mathbf{\Sigma}^{1/2}_2(\bm{\Omega}_1-\bm{\Omega}_2)\mathbf{\Sigma}^{1/2}_2\|_F$ and $\|\bm{\Xi}^{-1}_2\mathbf{\Xi}^{1/2}_1\bm{\delta}\|\asymp\sqrt{\frac{p}{t_0(p)}}\|\bm{\delta}\|$. Thus, we have $\mathrm{Var}_{\bm{x}\sim f_1}\big\{\frac{1}{p}Q(\bm{x})\big\}\asymp \frac{1}{p^2}\sigma_Q^2(p)\asymp 1$. Similarly, $\mathrm{Var}_{\bm{x}\sim f_2}\big\{\frac{1}{p}Q(\bm{x})\big\}\asymp \frac{1}{p^2}\sigma_Q^2(p)\asymp 1$.

Next, we bound the discrepancy between $\tilde{\bm{\Omega}}_k$ and ${\bm{\Xi}}^{-1}_k$ by Lemma \ref{lemma4} and \ref{lemma12}.
$$\begin{aligned}
    \|\tilde{\bm{\Omega}}_k-{\bm{\Xi}}^{-1}_k\|_\infty=&\left\|\frac{p}{\widehat{\tr(\bm{\Xi}_k)}}\hat{\bm{\Omega}}_k-\frac{p}{\tr(\bm{\Xi}_k)}{\bm{\Omega}}_k\right\|_\infty\\
    \le &\left(\frac{p}{\widehat{\tr(\bm{\Xi}_k)}}\left|\frac{\widehat{\tr (\bm{\Xi}_k)}}{\tr (\bm{\Xi}_k)}-1\right|\right)\|\hat{\bm{\Omega}}_k\|_\infty+\|\hat{\bm{\Omega}}_k-{\bm{\Omega}}_k\|_\infty\\
    =& O_p(\lambda_n+n^{-1/2})=O_p(\lambda_n).
\end{aligned}$$
Similarly, we have $ \|\tilde{\bm{\Omega}}_k-{\bm{\Xi}}^{-1}_k\|_{op}\le\|\tilde{\bm{\Omega}}_k-{\bm{\Xi}}^{-1}_k\|_{L_1}=O_p(\lambda_n^{1-q}s_0(p))$. And $\frac{1}{p}\|\tilde{\bm{\Omega}}_k-{\bm{\Xi}}^{-1}_k\|_{F}^2\le \|\tilde{\bm{\Omega}}_k-{\bm{\Xi}}^{-1}_k\|_{L_1}$, $\|\tilde{\bm{\Omega}}_k-{\bm{\Xi}}^{-1}_k\|_{\infty}=O_p\{\lambda_n^{2-q}s_0(p)\}$. By the proof of Lemma \ref{lemma1} we have $\|\bm{\mu}-\hat{\bm{\mu}}\|=O_p(p^{1/2}n^{-1/2})$ and $\|\bm{\mu}-\hat{\bm{\mu}}\|_\infty=O_p\{n^{-1/2}\log^{1/2}(np)\}$. Then we bound the $p^{-1}M(\bm{x})$ under $\bm{x}\sim f_1$.
$$\begin{aligned}
    \frac{1}{p}(\bm{x}-\bm{\mu}_1)^\top\big\{{\bm{\Xi}}^{-1}_2-{\bm{\Xi}}^{-1}_1-(\tilde{\bm{\Omega}}_2-\tilde{\bm{\Omega}}_1)\big\}(\bm{x}-\bm{\mu}_1)=&\frac{r^2}{p}\bm{U}^\top\mathbf{\Sigma}^{1/2}_1\big\{{\bm{\Xi}}^{-1}_2-{\bm{\Xi}}^{-1}_1-(\tilde{\bm{\Omega}}_2-\tilde{\bm{\Omega}}_1)\big\}\mathbf{\Sigma}^{1/2}_1\bm{U}\\
    \le& \frac{r^2}{p}\|\mathbf{\Sigma}^{1/2}_1({\bm{\Xi}}^{-1}_2-{\bm{\Xi}}^{-1}_1-(\tilde{\bm{\Omega}}_2-\tilde{\bm{\Omega}}_1))\mathbf{\Sigma}^{1/2}_1\|_{op}\\
    =&O_p\{\lambda_n^{1-q}s_0(p)\}, 
\end{aligned}$$
$$\begin{aligned}
    \frac{1}{p}(\bm{\mu}_1-\hat{\bm{\mu}})^\top(\tilde{\bm{\Omega}}_2-\tilde{\bm{\Omega}}_1)(\bm{x}-\bm{\mu}_1)\le& \frac{1}{\sqrt{p}}\|\bm{\mu}_1-\bm{\hat{\mu}}_1\|\frac{r}{\sqrt{p}}\|(\tilde{\bm{\Omega}}_2-\tilde{\bm{\Omega}}_1)\mathbf{\Sigma}^{1/2}_1\bm{U}\|\\
    =&O_p(n^{-1/2}),\end{aligned}$$
$$\begin{aligned}
    \frac{1}{p}(\bm{\delta}^\top{\bm{\Xi}}^{-1}_2-\bm{\hat{\delta}}^\top\bm{\tilde{\Omega}}_2)(\bm{x}-\bm{\mu}_1)\le &\frac{1}{p}(\|\bm{\delta}{\bm{\Xi}}^{-1}_2-\bm{\delta}\tilde{\bm{\Omega}}_2\|+\|\bm{\delta}\bm{\tilde{\Omega}}_2-\bm{\hat{\delta}}\tilde{\bm{\Omega}}_2\|)\|\bm{x}-\bm{\mu}_1\|\\
    \le &\frac{1}{p}(\|\bm{\delta}\|\|{\bm{\Xi}}^{-1}_2-\tilde{\bm{\Omega}}_2\|_{op}+\|\bm{\delta}-\hat{\bm{\delta}}\|\|\tilde{\bm{\Omega}}\|_{op})\|\bm{x}-\bm{\mu}_1\|\\
    =& O_p(\lambda_n^{1-q}s_0(p)+n^{-1/2})= O_p\{\lambda_n^{1-q}s_0(p)\},\end{aligned}$$
$$\begin{aligned}
    \frac{1}{p}(\bm{\mu}_1-\hat{\bm{\mu}}_1)^\top(\tilde{\bm{\Omega}}_2-\tilde{\bm{\Omega}}_1)(\bm{\mu}_1-\hat{\bm{\mu}}_1)\le&\frac{1}{p}\|\bm{\mu}_1-\hat{\bm{\mu}}_1\|^2\|\tilde{\bm{\Omega}}_2-\tilde{\bm{\Omega}}_1\|_{op}=O_p(n^{-1}),\end{aligned}$$
$$\begin{aligned}
\frac{1}{p}\hat{\bm{\delta}}^\top\tilde{\bm{\Omega}}_2(\bm{\mu}_1-\hat{\bm{\mu}}_1)\le \frac{1}{p}\|\bm{\hat{\delta}}\|\|\Tilde{\bm{\Omega}}_2\|_{op}\|\bm{\mu}_1-\hat{\bm{\mu}}_1\|=O_p(n^{-1/2}),\end{aligned}$$
$$\begin{aligned}
\frac{1}{p}(\bm{\delta}^\top{\bm{\Xi}}^{-1}_2\bm{\delta}-\hat{\bm{\delta}}^\top\bm{\tilde{\Omega}}_2\bm{\hat{\delta}})\le& \frac{1}{p}(\bm{\delta}^\top{\bm{\Xi}}^{-1}_2\bm{\delta}-\bm{\delta}^\top\bm{\tilde{\Omega}}_2\bm{\delta}+\bm{\delta}^\top\bm{\tilde{\Omega}}_2\bm{\delta}-\bm{\delta}^\top\bm{\tilde{\Omega}}_2\bm{\hat{\delta}}+\bm{\delta}^\top\bm{\tilde{\Omega}}_2\bm{\hat{\delta}}-\bm{\hat{\delta}}^\top\bm{\tilde{\Omega}}_2\bm{\hat{\delta}})\\
=&O_p\{\lambda_n^{1-q}s_0(p)+n^{-1/2}\}=O_p\{\lambda_n^{1-q}s_0(p)\}.\end{aligned}$$
Denote $\mathbf{D}_{\Omega}=\bm{\Xi}_2^{-1}-\bm{\Xi}_1^{-1}$, $\Tilde{\mathbf{D}}_{\Omega}=\Tilde{\bm{\Omega}}_2-\Tilde{\bm{\Omega}}_1$,
$$\begin{aligned}
(\hat{{\varsigma}}_p-{\varsigma}_p) &= \log|\tilde{\mathbf{D}}_\Omega\tilde{\bm{\Omega}}^{-1}_{1}+\mathbf{I}_{p}|-\log|\mathbf{D}_\Omega\bm{\Xi}_{1}+\mathbf{I}_{p}| \\
 & \leq\mathrm{tr}\big\{(\mathbf{D}_\Omega\bm{\Xi}_{1}+\mathbf{I}_{p})^{-1}(\tilde{\mathbf{D}}_\Omega\tilde{\bm{\Omega}}^{-1}_{1}-\mathbf{D}_\Omega\bm{\Xi}_{1})\big\} \\
 & =\mathrm{tr}\big\{(-\mathbf{D}_\Omega\bm{\Xi}_{2}+\mathbf{I}_{p})(\tilde{\mathbf{D}}_\Omega\tilde{\bm{\Omega}}^{-1}_{1}-\mathbf{D}_\Omega\bm{\Xi}_{1})\big\} \\
 & =\mathrm{tr}\big\{(-\mathbf{D}_\Omega\bm{\Xi}_{2})(\tilde{\mathbf{D}}_\Omega\tilde{\bm{\Omega}}^{-1}_{1}-\mathbf{D}_\Omega\bm{\Xi}_{1})\big\}+\mathrm{tr}(\tilde{\mathbf{D}}_\Omega\tilde{\bm{\Omega}}^{-1}_{1}-\mathbf{D}_\Omega\bm{\Xi}_{1}) \\
 & \leq\|\mathbf{D}_\Omega\bm{\Xi}_{2}\|_{F}\cdot\|\tilde{\mathbf{D}}_\Omega\tilde{\bm{\Omega}}^{-1}_{1}-\mathbf{D}_\Omega\bm{\Xi}_{1}\|_{F}+\mathrm{tr}(\tilde{\mathbf{D}}_\Omega\tilde{\bm{\Omega}}^{-1}_{1}-\mathbf{D}_\Omega\bm{\Xi}_{1}) \\
 & \lesssim\|\mathbf{D}_\Omega\|_{F}\|\bm{\Xi}_{2}\|_{op}\cdot\|\tilde{\mathbf{D}}_\Omega\tilde{\bm{\Omega}}^{-1}_{1}-\mathbf{D}_\Omega\bm{\Xi}_{1}\|_{F}, \\
 \end{aligned}$$
 where
$$\begin{aligned}
 & \left\|\mathbf{D}_\Omega\bm{\Xi}_1-\tilde{\mathbf{D}}_\Omega\tilde{\bm{\Omega}}^{-1}_1\right\|_F \\
 & \leq\left\|\mathbf{D}_\Omega\bm{\Xi}_{1}-\tilde{\mathbf{D}}_\Omega\bm{\Xi}_{1}\right\|_{F}+\left\|\tilde{\mathbf{D}}_\Omega(\bm{\Xi}_{1}-\tilde{\bm{\Omega}}^{-1}_{1})\right\|_{F} \\
 & \leq\|\mathbf{D}_\Omega-\tilde{\mathbf{D}}_\Omega\|_{F}\|\bm{\Xi}_{1}\|_{op}+\|\tilde{\mathbf{D}}_\Omega\|_{F}\|\bm{\Xi}_{1}-\tilde{\bm{\Omega}}^{-1}_{1}\|_{op}.
\end{aligned}$$
Then, we can find $p^{-1}(\hat{{\varsigma}}_d-{\varsigma}_d)=O_p \{\lambda_n^{1-q/2}s_0(p)^{1/2}+\lambda_n^{1-q}s_0(p)\}$. Thus, under $\bm{x}\sim f_1$ we have $\frac{1}{p}M(\bm{x})=O_p \{\lambda_n^{1-q/2}s_0(p)^{1/2}+\lambda_n^{1-q}s_0(p)\}$. Similarly, we can get the same solution under $\bm{x}\sim f_2$. From the previous discussion, we know that $\frac{1}{p}Q(\bm{x})$ is non-degenerate. Recall (\ref{eq:hrqdaorder}) we have
\begin{equation}
    \begin{aligned}
    |R_{HRQDA}-R_{QDA}|&\le\frac{1}{2}\mathbb{P}_{x\sim f_1}\left\{0\le \frac{1}{p}Q(\bm{x})<\frac{1}{p}M(\bm{x})\right\}+\frac{1}{2}\mathbb{P}_{x\sim f_2}\left\{0\le \frac{1}{p}Q(\bm{x})<\frac{1}{p}M(\bm{x})\right\}\\
    &=O_p \{\lambda_n^{1-q/2}s_0^{1/2}(p)+\lambda_n^{1-q}s_0(p)\}.
\end{aligned}
\end{equation}
\end{proof}
\bibliographystyle{chicago}
\bibliography{ref.bib}

\end{document}